\documentclass[a4paper,11pt]{article}

\usepackage{jheppub}
\usepackage{xpatch}

\makeatletter
\xpatchcmd{\maketitle}
  {\noindent{\small\scshape\@fpheader}\@preprint\par}
  {\noindent{\small\scshape\hphantom{Prepared for submission to JHEP}}\@preprint\par}
  {}{}
\makeatother

\usepackage[T1]{fontenc}
\usepackage[utf8]{inputenc} 
\usepackage[english]{babel}
\usepackage{csquotes}

\usepackage{mathtools} 
\usepackage{amssymb}
\usepackage{mathrsfs}
\usepackage{bm}

\usepackage{graphicx}
\usepackage{booktabs}
\usepackage{multirow}
\usepackage{array} 
\newcolumntype{C}[1]{>{\centering\let\newline\\\arraybackslash\hspace{0pt}}m{#1}}
\newcolumntype{L}[1]{>{\let\newline\\\arraybackslash\hspace{0pt}}m{#1}}
\usepackage{placeins}

\usepackage[dvipsnames]{xcolor}
\usepackage{xspace}
\usepackage{enumitem}
\usepackage{changepage}
\usepackage{marginnote}
\usepackage{siunitx}

\usepackage[colorlinks=true,citecolor=blue,linkcolor=blue]{hyperref}
\usepackage[capitalize]{cleveref}

\definecolor{Morange}{rgb}{0.880722,0.611041,0.142051}
\definecolor{Mgreen}{rgb}{0.560181,0.691569,0.194885}
\definecolor{Mred}{rgb}{0.922526,0.385626,0.209179}
\definecolor{dfa}{rgb}{0., 0., 0.}
\definecolor{dfb}{rgb}{0.8862745098039215, 0.2901960784313725, 0.2}
\definecolor{dfc}{rgb}{0.20392156862745098, 0.5411764705882353, 0.7411764705882353}
\definecolor{dfd}{rgb}{0.596078431372549, 0.5568627450980392, 0.8352941176470589}
\definecolor{dfe}{rgb}{0.4666666666666667, 0.4666666666666667, 0.4666666666666667}
\definecolor{dff}{rgb}{0.984313725490196, 0.7568627450980392, 0.3686274509803922}
\definecolor{dfg}{rgb}{0.5568627450980392, 0.7294117647058823, 0.2588235294117647}
\definecolor{dfh}{rgb}{1., 0.7098039215686275, 0.7215686274509804}

\definecolor{cborange}{HTML}{e69f00}
\definecolor{cbgreen}{HTML}{009e73}
\definecolor{cbyellow}{HTML}{f1dd42}
\definecolor{cblblue}{HTML}{56b4e9}
\definecolor{cbblue}{HTML}{0072b2}
\definecolor{defgrey}{HTML}{9f9f9f}
\definecolor{defgreen}{HTML}{8eba42}

\usepackage{xpatch}
\makeatletter
\patchcmd{\@ssect@ltx}
    {\addcontentsline{toc}{#1}{\protect\numberline{}#8}}
    {}
    {}
    {}

\DeclareSIUnit\year{yr}
\DeclareSIUnit\years{yrs}

\makeatletter
\providecommand*{\diff}%
	{\@ifnextchar^{\DIfF}{\DIfF^{}}}
\def\DIfF^#1{%
	\mathop{\mathrm{\mathstrut d}}%
		\nolimits^{#1}\gobblespace}
\def\gobblespace{%
	\futurelet\diffarg\opspace}
\def\opspace{%
	\let\DiffSpace\!%
	\ifx\diffarg(%
		\let\DiffSpace\relax
	\else
		\ifx\diffarg[%
			\let\DiffSpace\relax
		\else
  			\ifx\diffarg\{%
				\let\DiffSpace\relax
			\fi\fi\fi\DiffSpace}

\newcommand{\Lumgeom}{\mathscr{L}_\text{geom}}
\newcommand{\Lumint}{\mathscr{L}_\text{int}}

\newcommand{\Ltot}{\mathscr{L}^{0.95}_\text{int}}
\newcommand{\Lx}[1]{\mathscr{L}^{#1}_\text{int}}           
\newcommand{\fx}[1]{f_{#1}}                     

\definecolor{Red}{rgb}{1.,0.,0.}
\newcommand{\ATAP}[1]{{\color{defgreen}}}
\newcommand{\ATLAS}[1]{{\color{purple}}}

\usepackage[nolist]{acronym}
\begin{acronym}
	\acro{SM}{Standard Model}
	\acro{BSM}{beyond the Standard Model}
	\acro{COM}{center-of-mass}
	\acro{QED}{Quantum Electrodynamics}
	\acro{EFT}{Effective Field Theory}
	\acro{RHS}{right hand side}
	\acro{LHS}{left hand side}
	\acro{EM}{Electromagnetism}
	\acro{SNR}{signal to noise ratio}
	\acro{CL}{confidence level}
	\acro{VEV}{vacuum expectation value}
	\acro{EWSB}{Electroweak Symmetry Breaking}
	\acro{PDF}{parton distribution function}
	\acro{MLE}{maximum likelihood estimator}
	\acro{NDF}{number density function}
	\acro{MC}{Monte Carlo}
	\acro{NP}{New Physics}
	\acro{LHC}{Large Hadron Collider}
    \acro{HL-LHC}{High Luminosity LHC}
    \acro{BR}{branching ratio}
    \acro{DOF}{degrees of freedom}
    \acro{MIP}{minimally ionizing particle}
    \acro{SUSY}{Supersymmetry}
    \acro{MSSM}{minimally supersymmetric Standard Model}
    \acro{HSCP}{heavy stable charged particle}
    \acro{pCM}{parton-center-of-mass energy}
    \acro{RF}{radio frequency}
    \acro{LWFA}{laser-wakefield acceleration}
    \acro{PWFA}{particle-beam wakefield acceleration}
    \acro{IP}{interaction point}
\end{acronym}

\newcommand\myshade{80}
\colorlet{mylinkcolor}{ForestGreen}
\colorlet{mycitecolor}{Red}
\colorlet{myurlcolor}{violet}

\hypersetup{
  linkcolor  = mylinkcolor!\myshade!black,
  citecolor  = mycitecolor!\myshade!black,
  urlcolor   = myurlcolor!\myshade!black,
  colorlinks = true
}

\title{Searches for electroweak states at future plasma wakefield colliders}

\author[a]{So Chigusa,}
\author[b,c]{Simon Knapen,}
\author[d,e]{Toby Opferkuch,}
\author[b,c]{Inbar Savoray,}
\author[b,c]{Christiane Scherb,}
\author[f,g]{and Weishuang Linda Xu}

\affiliation[a]{Center for Theoretical Physics -- a Leinweber Institute, Massachusetts Institute of Technology,\\
77 Massachusetts Avenue, Cambridge, MA 02139, USA}
\affiliation[b]{Theoretical Physics Group, Lawrence Berkeley National Laboratory, Berkeley, CA 94720, USA}
\affiliation[c]{Leinweber Institute for Theoretical Physics, Department of Physics, University of California, \mbox{Berkeley}, CA 94720, USA}
\affiliation[d]{SISSA International School for Advanced Studies, Via Bonomea 265, 34136, Trieste, Italy}
\affiliation[e]{INFN, Sezione di Trieste, Via Bonomea 265, 34136, Trieste, Italy}
\affiliation[f]{Kavli Institute for Particle Astrophysics \& Cosmology, Stanford University, \\Stanford, CA 94305, USA}
\affiliation[g]{Particle Theory Group, SLAC National Accelerator Laboratory, Stanford, CA 94305, USA}

\emailAdd{schigusa@mit.edu}
\emailAdd{smknapen@lbl.gov}
\emailAdd{toby.opferkuch@sissa.it}
\emailAdd{inbarsavoray@berkeley.edu}
\emailAdd{cscherb@lbl.gov}
\emailAdd{wlxu@stanford.edu}

\abstract{
We quantify the discovery potential of future multi-TeV plasma wakefield colliders for new electroweak multiplets. 
We include beam--beam effects through realistic luminosity spectra, comparing five collider configurations: 
$e^+e^-$ and $e^-e^-$ machines with round- and flat-beams, and a $\gamma\gamma$ collider. 
The beam--beam effects qualitatively change search strategies relative to idealized mono-energetic lepton colliders, highlighting the importance of the low-energy part of the luminosity spectrum and additional beam-induced initial-state channels.
Our results have implications for accelerator R\&D priorities, since key electroweak targets may remain accessible even if efficient positron acceleration and flat-beam delivery prove technically challenging at the multi-TeV scale.
}

\date{\today}
\preprint{MIT-CTP/5981}

\begin{document}
\maketitle

\section{Introduction}
\label{sec:intro}

The quest to uncover the fundamental laws of Nature has continually driven innovation in particle accelerator technology. 
Following the success of the LHC, the community is exploring complementary avenues towards the next major leap in energy. 
The FCC proton-collider roadmap aims for \SI{100}{\TeV} through an ambitious scaling of proven concepts, while muon and plasma wakefield colliders are being pursued through targeted R\&D as less mature but potentially transformative technologies for multi-TeV lepton collisions. 
This diversified, staged approach to the energy frontier is now explicitly reflected in the major international planning efforts. 
Both the 2020 European Strategy for Particle Physics \cite{Adolphsen:2022ibf} and the 2023 P5 Report \cite{P5:2023wyd} identify this challenge as central to the future of particle physics.
To extend the energy reach of the LHC by roughly an order of magnitude, the next energy frontier collider needs to achieve collisions with a \ac{pCM} of about \SI{10}{\TeV}. This requires a \SI{10}{\TeV} electron or muon collider or a roughly \SI{100}{\TeV} proton collider.
All three approaches face formidable challenges, yet recent advances in accelerator science inspire optimism that these obstacles can be overcome. 

A central challenge is achieving an acceptable facility footprint and cost, particularly for high energy electron colliders.
Wakefield accelerators aim to achieve this by using collective plasma oscillations to provide extremely large accelerating gradients, well beyond what can be achieved with \ac{RF} cavities. 
The key idea is that an intense driver---either a laser pulse or a high-current particle bunch---perturbs the plasma and generates a trailing longitudinal electric field. 
This field acts like the ``wake'' behind a moving object and can efficiently accelerate a trailing bunch of charged particles. 
These extremely large gradients make \SI{10}{\TeV}-scale linear electron collisions a plausible target.
This particle acceleration method has been demonstrated experimentally, but requires additional R\&D to enable a compact, high energy implementation (see \cref{sec:wakefield}).
Recently, the \SI{10}{\TeV} \ac{pCM} Wakefield Collider Design study was launched \cite{Gessner:2025acq}, which aims to chart this R\&D path towards a wakefield collider that can reach the \SI{10}{\TeV} frontier. The study aims to deliver an end-to-end design concept, with self-consistent parameters and an overall costing.
This will require choosing among competing technologies to focus and direct future R\&D efforts.

Direct input from theoretical and experimental particle physics is essential, since these decisions on the accelerator side are strongly intertwined with the physics potential of a future machine.
This is because strong beam--beam interactions are expected to play a qualitative role at a \SI{10}{\TeV} electron collider relying on plasma wakefield acceleration. 
These interactions disrupt the beams during the collision, smearing out the energy spectra of the beams and generating large fluxes of secondary $\gamma$ and $e^\pm$ particles~\cite{YokoyaChen:1992}.
These secondary particles strongly affect the detector design, as well as the rates for both signal and background processes.
It is currently an open question to what extent these beam--beam interactions must be mitigated through the accelerator design, or whether they can be leveraged in a high energy discovery machine (see e.g.~\cite{Barklow:2023iav}).

Moreover, given the technological challenges associated with efficiently accelerating $e^+$ in wakefield colliders, one should understand and quantify how the physics case is affected for $e^-e^-$ machines.
In these machines, unless lepton number is violated or carried by the target particles, the final state of any process must include electrons or neutrinos in addition to the particles one seeks to produce.
This greatly reduced the accessible cross sections, for instance $e^-e^-\to \chi^+\chi^- e^-e^-$ as opposed to $e^+e^-\to \chi^+\chi^-$ for the production of a heavy charged particle $\chi$.
This problem can in principle be circumvented in two ways:
Firstly, beam-beam interactions will generate a large flux of secondary $e^+$ and $\gamma$, which will contribute an appreciable luminosity in high energy $e^+e^-$ and $\gamma\gamma$ collisions, even in an $e^-e^-$ collider. 
Secondly, one could convert the $e^-e^-$ collider into a $\gamma\gamma$ collider by Compton scattering the $e^-$ against the photons in a high-intensity laser.
This further increases the luminosity in both $e^+e^-$ and $\gamma\gamma$ collisions.
In this work we study both options, and ask how they compare with an $e^+e^-$ collider.

The answers to these questions will qualitatively impact the direction of the plasma wakefield R\&D program for a future high energy collider.
To this end, a joint effort was established that brings together expertise from accelerator physics, particle experiment, and particle theory.
This paper is the first to emerge from this collaboration and begins to address how the physics targets are affected by choices in accelerator design.
Forthcoming works will address the accelerator optimization \cite{acceleratorpaper}, the modeling of the beam--beam interactions \cite{simulationpaper} \ATAP{please advise on what you would like for this citation, title, authors etc.} and the constraints those beam--beam interactions place on a realistic detector design \cite{detectorpaper}. \ATLAS{please advise on what you would like for this citation, title, authors etc.}

The science targets for a plasma wakefield collider are the same as those for a high energy muon collider, and overlap with those for \SI{100}{\TeV} $pp$ colliders.
There are, however, important qualitative differences in the reach and search strategies, which require thorough and comprehensive studies.
In this paper we compare the discovery potential of five distinct plasma wakefield colliders for new electroweak particles with masses in the multi-TeV range.
Other theory work is forthcoming which studies extended Higgs sectors \cite{scalarpaper}, heavy resonances \cite{Cipressi:pwfa-resonances}, Higgs precision measurements \cite{triplehiggs} and lepton-flavor violating processes \cite{LVF}. 

The remainder of this paper is organized as follows: In \cref{sec:wakefield} we review the main features and challenges associated with plasma wakefield colliders, and preview the preliminary results that will be presented in \cite{acceleratorpaper} and \cite{simulationpaper}, to ensure the present paper is self-contained. We motivate and define our benchmark models in \cref{sec:model}, followed by a detailed analysis of distinct experimental signatures in \cref{sec:signatures}. We compare with other future collider options in~\cref{sec:discuss} and conclude in \cref{sec:conclusion}.

\section{Wakefield colliders}
\label{sec:wakefield}
\subsection{Overview}
A wakefield accelerator relies on the collective motion of a plasma to generate extremely strong accelerating fields, far exceeding those attainable with \ac{RF}-based accelerator technology. 
By driving a plasma wave---essentially a coherent oscillation of electrons---one can create a large electric field that trails behind the driver.
This can be used to accelerate charged particles with high efficiency if injected in phase with the plasma wave.
In \ac{LWFA}, the driver is a short, high-intensity laser pulse which displaces electrons and excites a relativistic plasma wave. 
In \ac{PWFA}, a dense, ultra-relativistic bunch of electrons, positrons, or protons plays the same role: its space-charge field drives the plasma oscillation. 
In both schemes, the amplitude of the plasma wave---and thus the achievable electric fields---is set by the plasma density and can be orders of magnitude larger than in \ac{RF}-based accelerators.
Concretely, the electric fields that can be generated in the wake are roughly  
\begin{align}
E_0 \approx \SI{96}{\giga\volt\per\meter}\sqrt{\frac{n_0}{10^{18}\,\mathrm{cm}^{-3}}}\,,
\end{align}
with $n_0$ the density of the plasma, which quantifies the maximal single-stage acceleration gradient \cite{Tajima:1979,Chen:1985,Esarey:2009,Adli:2019}. 
In comparison, the \ac{RF} technology in the CLIC design would be capable of a gradient of roughly $\SI{100}{\MeV\per\meter}$ \cite{CLICdp:2018cto}.

This high gradient exemplifies the promise of wakefield concepts for multi-TeV colliders, potentially offering a markedly more compact size compared to \ac{RF}-based designs.
To realize this vision, one must address challenges such as the (i) preservation of beam quality (sub-$\mu$m emittance, percent-level energy spread), (ii) efficiency (high wall-plug to beam efficiency and strong drive-to-witness transfer via beam loading), (iii) multi-stage operation with synchronization and tolerances compatible with emittance preservation, (iv) high repetition rate and reliability, and (v) control of final-focus and beam--beam limits at the \ac{IP} \cite{Esarey:2009,Adli:2019,Diederichs:2024zir}. These considerations determine the choice of plasma density, focusing optics, operating regime (linear/quasi-linear versus the nonlinear `blowout' typically used in electron-driven \ac{PWFA}), and staging architecture.

\ATAP{Any other developments you would like us to highlight?}
Recent experimental progress in wakefield acceleration techniques has been substantial. Notably, \ac{LWFA} at BELLA (LBNL) \cite{Esarey:2009} has demonstrated \SI{10}{\GeV} electron beams over \SI{30}{cm} with high beam quality using an optically formed plasma channel \cite{Picksley:2024cdd,Miao:2021ekl}.
Beam-driven \ac{PWFA} at SLAC’s FACET-II~\cite{DiMitri:2024} has achieved highly uniform acceleration with an energy gain of about \SI{6}{\GeV} in a \SI{40}{\centi\meter} lithium plasma using an electron driver~\cite{Hogan:EAAC2025:FACETII}. On the proton-driven front, CERN’s AWAKE collaboration has demonstrated the first GeV-class acceleration of externally injected electrons to \SI{2}{\GeV} using a \SI{400}{\GeV} proton bunch train from the SPS~\cite{Adli:2018}. AWAKE’s Run-2b has demonstrated more uniform acceleration using a plasma density step, with energies up to \SI{1.7}{\GeV} over \SI{10.3}{\meter} and higher gradients \cite{AWAKE:SPSC:SR356:2024}, and is preparing Run-2c aiming at \SIrange{4}{10}{\GeV} with emittance control once operations resume post-LS3 (physics start 2029) \cite{Gschwendtner:EAAC2025:AWAKE}. For prospective $\gamma\gamma$ colliders, intense laser Compton backscattering of a primary electron beam is a well-established and experimentally tested concept~\cite{Ginzburg:1983,Telnov:1990}.
These gradients suggest that the \emph{acceleration length} for a \SI{10}{\TeV}-class machine could be remarkably compact, while the full facility footprint depends on the staging and beam-transport designs \cite{Lindstrom:2020pzp}.

In the nonlinear blowout regime the drive bunch expels plasma electrons, leaving an ion column that focuses electrons but defocuses positrons. 
This severely complicates the \textbf{acceleration of positrons} \cite{PhysRevE.64.045501,Hogan:2003bs}, and has spurred an intensive R\&D program; see \cite{PhysRevAccelBeams.27.034801} for a recent review and references.
High positron acceleration gradients have been demonstrated experimentally in both homogeneous plasmas and in hollow plasma channels. However, in both cases maintaining a low emittance remains a challenge.
Several new concepts have been developed in simulation; the most promising can achieve a theoretical luminosity/power about an order of magnitude worse than \ac{RF} technology (i.e.~CLIC) \cite{PhysRevAccelBeams.27.034801}. 
While impressive progress has been made on the positron problem in the last few decades, there is still a substantial R\&D road ahead to demonstrate its feasibility with a beam quality suitable for a high energy collider.

A second challenge relates to \textbf{beamstrahlung}:
The beams encountering each other at the \ac{IP} give rise to extreme electromagnetic fields due to the beams' high charge and energy \cite{Esberg:2014zia, RevModPhys.94.045001}. 
As a result, large amounts of photons are radiated, which smears the electron energies downwards.\footnote{Recently a set of analytic approximations has been developed that describe the beam--beam interactions in this regime \cite{He:2025jmz}.}
The amount of beamstrahlung that is generated is a function of $1/(\sigma_x + \sigma_y)$, where $\sigma_{x,y}$ are the beam sizes in the transverse directions \cite{Schroeder_2022}, while the luminosity scales as $\sim 1/(\sigma_x \sigma_y)$. For the same luminosity, “flat” beams ($\sigma_x \gg \sigma_y$) therefore suffer less from beamstrahlung than “round” beams ($\sigma_x \approx \sigma_y$), making flat beams preferable when precise control of the \ac{COM} energy is required, e.g.~in Higgs factories.
In plasma wakefield accelerators, however, nonlinear effects can perturb the transverse wakefields, leading to emittance mixing \cite{Diederichs:2024zir}. As a result, the emittance in the horizontal direction decreases while the emittance in the vertical direction increases, causing the geometric mean emittance to rise. Effectively, the initially flat beam becomes rounder and its luminosity decreases, negating the advantages of using a flat beam in the first place.
Substantial research and development is still required to resolve this issue.

Because of these challenges, it is important to compare the physics reach of $e^{+}e^{-}$, $e^{-}e^{-}$, and $\gamma\gamma$ modes, for round and flat beams. This allows us to gauge the relative merit of the various collider options in order to aid the accelerator community with prioritizing their R\&D efforts. Concretely:
\begin{itemize}
\item Can one sidestep the $e^+$ acceleration problem and achieve our physics goals with an $e^-e^-$ or $\gamma\gamma$ collider? 
\item Can we reach our physics goals with round beams, or is it essential to find a way to deliver flat beams to the \ac{IP}?
\end{itemize}
The answer to these questions may well differ depending on the physics targets one considers. 
It also depends on the luminosity that can realistically be achieved for each machine. 
To this end, we will quantify the minimum luminosities required to discover a number of well-motivated benchmark models, as a starting point for a set of comprehensive studies.
In other words, rather than asking which is the ``best'' collider, we seek to quantify \emph{which trade-offs are acceptable.} If certain R\&D challenges can be by-passed without compromising our physics goals, this could greatly accelerate the path to a new, energy frontier collider. 

\subsection{Beam parameters and luminosity spectra}
\label{subsec:beam-parameters-lumi}
\ATAP{In this section we describe your preliminary work on the accelerator design and simulation studies. 
We will adjust the content and language in this section in any way you like, so that the science and credits are presented appropriately. Please don't be shy suggesting edits here, we want everyone to be fully happy with this!
}

There is an intrinsic tension in any high–energy linear collider between achieving high luminosity and keeping beamstrahlung and beam disruption  under control. 
An \ac{RF} machine such as CLIC addresses this by accelerating flat beams with extremely small spot sizes and high bunch charge: at the \ac{COM} energy of $\SI{3}{\TeV}$, the design foresees bunches with $\mathcal{O}(10^{9})$ particles, nanometre–scale vertical beam sizes and tens–of–nanometre horizontal sizes at the \ac{IP}, together with sub–$\SI{100}{\micro\meter}$ bunch lengths. This already leads to substantial beamstrahlung and copious coherent and incoherent pair production~\cite{Linssen:2012hp,CLICdp:2018cto}. 
Plasma–wakefield linear colliders are fundamentally different in how they generate the accelerating field, but they do not escape this basic trade–off.
Collider–relevant wakefield accelerator designs require \SI{}{\giga\volt\per\meter}–class gradients in dense plasmas, which in turn demand very short, high–charge witness bunches with transverse sizes of only a few tens of nanometres and normalized emittances at the $\sim\!\!\SI{10}{\nano\meter}$ level in order to reach CLIC–like luminosities \cite{Adolphsen:2022ibf}.
So far, only the acceleration of round beams has been demonstrated with emittance preservation \cite{Liu:2024ymd,Lindstrom:2024zbo}.
In general, \SI{10}{\TeV} electron beams will inevitably experience strong pinching and large beamstrahlung effects at the \ac{IP}, so a significant flux of secondary particles is unavoidable.
However, this effect is larger for round beams, and it is therefore essential to quantify its impact on the discovery potential of the collider. 

\begin{figure}
    \centering
    \includegraphics[width=\linewidth]{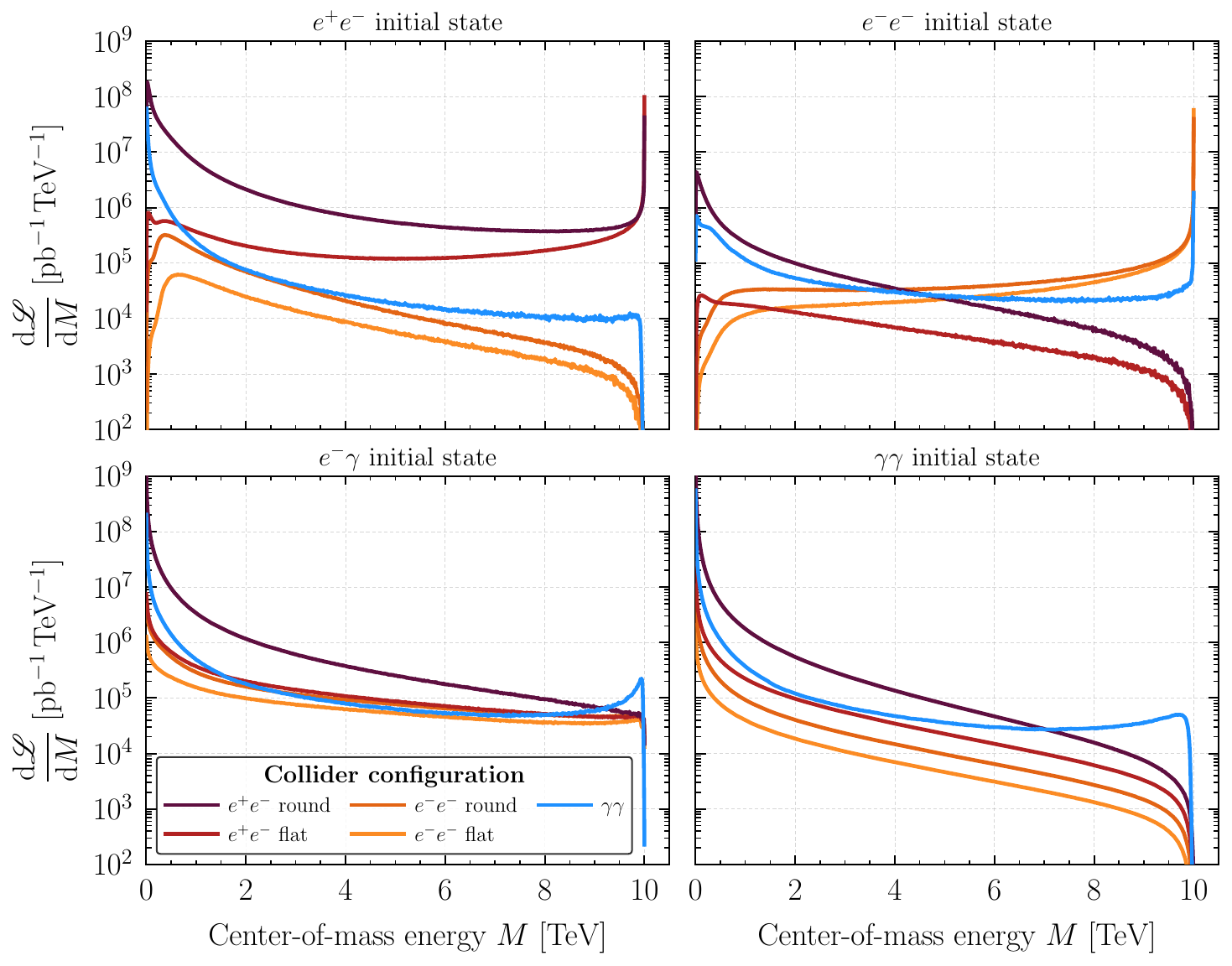}
    \caption{Integrated luminosity spectra $\mathrm{d}\Lumint/\mathrm{d}M$ for four initial states (panels) of Ref.~\cite{simulationpaper}: $e^+e^-$, $e^-e^-$, $e^-\gamma$, and $\gamma\gamma$. Each curve represents one of the five collider configurations (colors). All collider configurations are normalized such that $\Lumgeom = \SI{1}{\per\atto\barn}$. The relative values and the hardness of the spectra reflect the beam--beam dynamics: $e^+e^-$ exhibits stronger pinch and beamstrahlung (more weight at lower $M$), while $e^-e^-$ shows anti-pinch and a more pronounced endpoint spike at $M=2E_{\rm beam}$. Finally, only the $e^-\gamma$ initial state is shown; $e^+\gamma$ is similar for the $e^+e^-$ collider configurations and is otherwise negligible since no primary positron beam exists for the $e^-e^-$ and $\gamma\gamma$ colliders.}
    \label{fig:initial_state_lumis}
\end{figure}

\begin{figure}
    \centering
    \includegraphics[width=\linewidth]{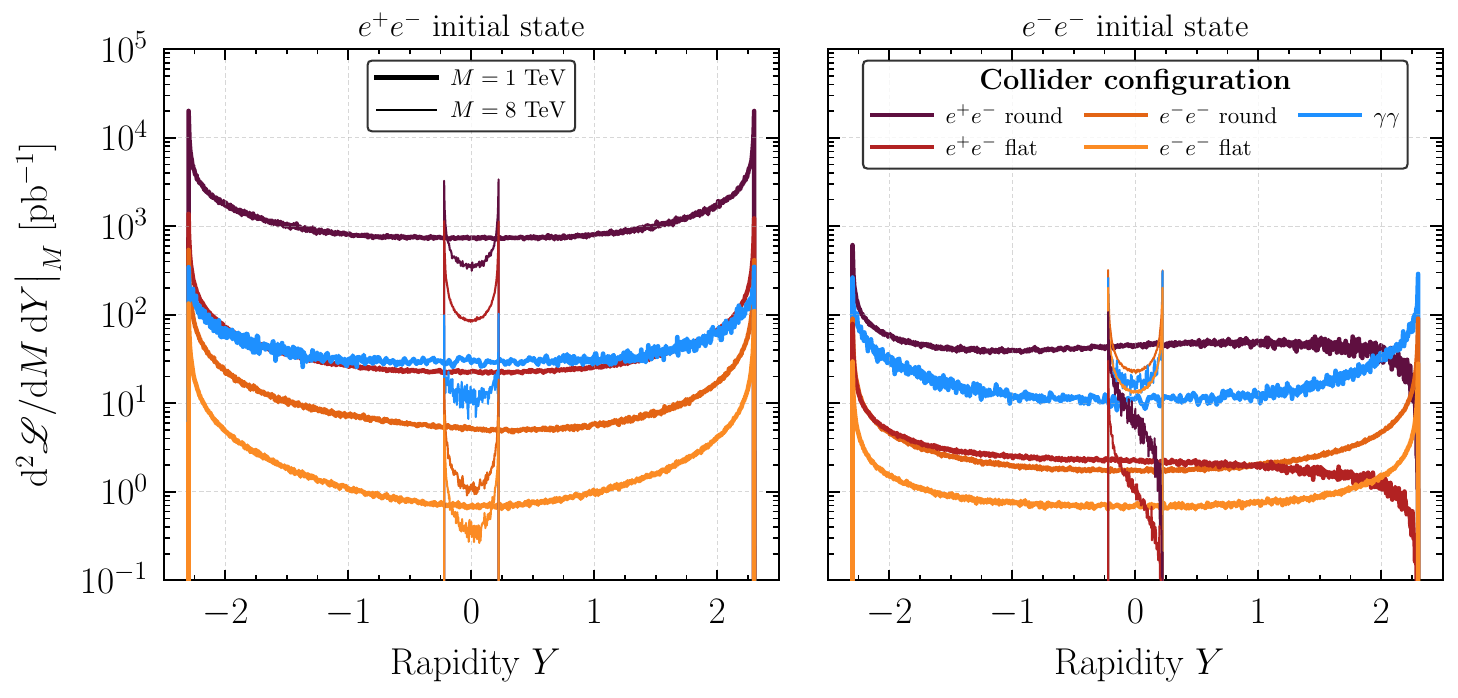}
    \caption{Double-differential integrated luminosity spectra $\left.\mathrm{d}^2\Lumint/(\mathrm{d}M\,\mathrm{d}Y)\right|_{M}$ as a function of rapidity $Y$ for representative masses $M$ (line thickness), across collider configurations (colors) as in \cref{fig:initial_state_lumis}. These were derived from the simulation data in \cite{simulationpaper}, as explained in the text.  For primary--primary channels ($e^\pm e^\mp$ and $e^-e^-$) the distributions approach the kinematic limits $Y=\pm Y_{\max}(M)$. Increasing $M$ narrows the allowed rapidity range $|Y|\le Y_{\max}(M)$ and raises the endpoint weight.}
    \label{fig:initial_state_rapidities}
\end{figure}

In this work we study five colliders: two $e^+e^-$ colliders, with flat and round beams, two $e^-e^-$ colliders, again with flat and round beams, and a $\gamma\gamma$ collider. The corresponding parameter choices are the result of a preliminary optimization of the accelerator parameters, performed by physicists in the LBNL BELLA group \cite{acceleratorpaper,Schroeder:2022xdu}, as summarized in \cref{tab:beam-parameters}. 
These parameters are used as input for detailed simulations of the interaction region, performed by physicists in the LBNL Advanced Modeling Program \cite{simulationpaper}. 
The bunch charge is comparable to that in CLIC, while the transverse beam sizes are somewhat smaller, as CLIC is assuming $\sigma_x \approx \SI{45}{\nano\meter}$ and $\sigma_y \approx \SI{1}{\nano\meter}$ for the \SI{3}{\TeV} benchmark \cite{CLICdp:2018vnx}.
The $e^+e^-$ and $e^-e^-$ simulations are carried out with 
\texttt{WarpX} \cite{warpx}. 
For the $\gamma\gamma$ collider, the $e^-e^-$ beams approaching the interaction region are assumed to be the same as in the $e^-e^-$ collider with round beams.
The effects of the laser Compton conversion and photon collisions were simulated with the \texttt{CAIN} software package \cite{Chen:1994jt}.
Our analysis framework is designed to permit efficient updates as the accelerator parameters and simulation codes continue to evolve. 
The output of these simulations is a set of luminosity spectra, which are summarized in \cref{tab:beam-parameters}, \cref{fig:initial_state_lumis} and \cref{fig:initial_state_rapidities}. 

\begin{table}
\centering
\begin{tabular}{l cc ccccc}\toprule
Collider & \multicolumn{2}{c}{Bunch size [\SI{}{\nano \meter}]} & \multicolumn{5}{c}{Integrated luminosity [\SI{}{\per\pico\barn}]}   \\
configuration  & $\sigma_x$ & $\sigma_y$  &  $\Ltot$ 
& $\Lx{0.20}$ & $\fx{0.20}$
& $\Lx{0.01}$ & $\fx{0.01}$ \\  \midrule
     $e^+e^-$ round   & \num{1.55} & \num{1.55}  & \num{1.5e+07} & \num{1.2e+06} & 8.0\%  & \num{3.6e+05} & 2.4\% \\
     $e^+e^-$ flat    & \num{6}    & \num{0.4}   & \num{2.5e+06} & \num{1.2e+06} & 46.9\% & \num{6.5e+05} & 25.9\% \\ 
     $e^-e^-$ round   & \num{1.55} & \num{1.55}  & \num{7.4e+05} & \num{4.6e+05} & 62.3\% & \num{2.6e+05} & 35.4\% \\
     $e^-e^-$ flat    & \num{6}    & \num{0.4}   & \num{6.8e+05} & \num{5.2e+05} & 75.5\% & \num{3.5e+05} & 50.8\% \\\midrule
     $\gamma\gamma$ collider & \multicolumn{7}{c}{ [round $e^-e^-$ initial beams with laser back-scattering]} \\
     $\,\,\qquad e^-e^-$   & \num{1.55} & \num{1.55}  & \num{4.0e+05} & \num{6.8e+04} & 16.8\% & \num{2.0e+04} & 5.0\% \\
     $\,\,\qquad \gamma\gamma$   &            &             & \num{8.7e+05} & \num{7.1e+04} & 8.1\%  & \num{5.0e+02} & 0.06\%  \\ 
     \toprule
\end{tabular}
   \caption{
   Machine parameters \cite{acceleratorpaper} and integrated luminosities for the primary beams \cite{simulationpaper} for the collider configurations studied. All results are normalized to an integrated geometric luminosity of $\SI{1}{\per\atto\barn}$, which provides a common reference exposure for comparing collider performance.  For all machine configurations listed above, the following apply: beam energy $E_\text{beam}=\SI{5}{\TeV}$,
   number of particles per bunch $N_e=\num{1.2e9}$, bunch length $\sigma_z=\SI{8.5}{\micro\meter}$. 
   For a repetition rate of $\nu_{\rm rep}=\SI{50}{\kilo\hertz}$, this corresponds to an instantaneous geometric luminosity of $\SI{7.52}{\per\atto\barn\per\year}$. 
   For the $\gamma\gamma$ collider, the parameters for the primary $e^-e^-$ beams are the same as those for the round $e^-e^-$ collider.
   The laser frequency for the $\gamma\gamma$-collider was \SI{2.5}{\eV}$/\hbar$. We refer to \cite{acceleratorpaper} for additional details.
   We define $\Ltot$ as the luminosity integrated over the upper 95\% of the kinematic range, with collision \ac{COM} energy $M\in[\SI{0.5}{\TeV},\SI{10}{\TeV}]$, serving as an infrared-safe proxy for the total luminosity. For any top-fraction window $x\in(0,1)$, $\Lx{x}$ denotes the integrated luminosity contained in the highest $x$ fraction of the kinematic range, i.e.~for $M\in[(1-x)\SI{10}{\TeV},\SI{10}{\TeV}]$, and $\fx{x}\equiv \Lx{x}/\Ltot$ is the corresponding fraction of the reference luminosity. Thus, $\Lx{0.20}$ and $\fx{0.20}$ quantify the luminosity (and its fraction) within the top 20\% of the spectrum, while $\Lx{0.01}$ and $\fx{0.01}$ isolate the top 1\%. Smaller $\fx{x}$ indicates stronger beamstrahlung, corresponding to a larger share of luminosity radiated away from the endpoint into the low-$M$ tail. In the $\gamma\gamma$ block, \emph{both} rows correspond to the laser Compton-conversion configuration: the first row `$e^-e^-$' gives the residual electron-electron luminosity \emph{after} the laser interaction, and the second row `$\gamma\gamma$' gives the photon-photon luminosity.
   Not all of the $e^-$ beam energy is converted into high energy photons. Consequently, the resulting $\Lx{0.95}$, $\Lx{0.2}$, and $\Lx{0.05}$ for the residual $e^-e^-$ beams differ from those of the round $e^-e^-$ collider without Compton backscattering. 
  }
    \label{tab:beam-parameters}
\end{table}

We use a derived quantity, the {\it integrated geometric luminosity}, as a common basis for comparing the collider designs. The geometric luminosity is the luminosity the collider would have if beam--beam interactions could somehow be switched off. In other words, it is the luminosity the accelerator delivers to the \ac{IP}. In that sense it is conceptually similar to the luminosity one would refer to in the context of muon and hadron colliders. It is defined by
\begin{align}
    \Lumgeom &\equiv \frac{N_e^2 \nu_\text{rep} T}{4\pi \sigma_x \sigma_y}\,,
\end{align}
where $N_e$ is the number of electrons (or positrons) per bunch, $\nu_\text{rep}$ is the repetition rate, $\sigma_{x,y}$ are the transverse bunch sizes and $T$ total run time of the collider. 
For the machine parameters in \cref{tab:beam-parameters}, all five colliders achieve an instantaneous geometric luminosity of $\SI{7.52}{\per\atto\barn\per\year}\approx 2.4\times 10^{35}\, \mathrm{cm}^{-2}\,  \mathrm{s}^{-1}$ if we assume a $\nu_{\rm rep}=\SI{50}{\kilo\hertz}$ repetition rate. 
As we will see, this is likely somewhat overly aggressive in some cases, as our physics goals can often already be achieved with a lower luminosity, e.g.~by lowering the repetition rate or by allowing for larger bunch sizes.

The beam--beam interactions however disrupt the beams as they cross, as the $e^\pm$ radiate hard photons, which in turn produce additional $e^+e^-$ pairs through the Breit-Wheeler process. This implies that the geometric luminosity is not the right quantity to calculate particle production rates; instead we must use the specific luminosity spectra which take these effects into account. 
They are analogous to parton distribution functions at hadron and muon colliders, but with a few key differences: all particles described by the luminosity spectra are real rather than virtual\footnote{We verified that the contribution from electroweak parton distribution functions is a negligible correction compared to the beam--beam interactions, see \cref{app:pdf}.} and the spectra cannot be factorized into a product of distribution functions for each beam.
In addition, the shape of the spectra depends strongly on the collective beam parameters, such as $\sigma_{x,y}$, $N_e$ etc.
It is through this dependence that the physics case and accelerator R\&D efforts are closely connected.

These luminosity spectra are calculated \cite{simulationpaper} using the \texttt{WarpX} \cite{warpx} and \texttt{CAIN} \cite{Chen:1994jt} codes. 
The former has recently been benchmarked against GUINEA-PIG in ultra-tight collisions and ILC scenarios
\cite{Nguyen:2024coo}.
The \texttt{WarpX} and \texttt{CAIN} outputs are binned in a grid of energies $(E_1,E_2)$ for all combinations of primary and secondary beam particles $\{e^-,e^+, \gamma\}$. We perform the coordinate transformation 
\begin{align}
M \equiv 2\sqrt{E_1 E_2},\qquad
Y \equiv \tfrac12\ln\left(\frac{E_1}{E_2}\right),
\end{align}
with inverse $E_1=(M/2)e^{Y}$, $E_2=(M/2)e^{-Y}$ where $M$ is the \ac{COM} energy and $Y$ is the rapidity for the \ac{COM} frame relative to the lab frame. Thus, the double-differential luminosity spectra for each combination of particles can be expressed as
\begin{align}\label{eq:double-diff-lumi}
\frac{\mathrm{d}^2\Lumint}{\mathrm{d}M\,\mathrm{d}Y}
=\frac{M}{2}\,
\frac{\mathrm{d}^2\Lumint}{\mathrm{d}E_1\,\mathrm{d}E_2}\,,
\quad |Y|\le Y_{\max}(M)=\ln\bigl(2E_{\rm beam}/M\bigr)\,.
\end{align}
We start by showing the single-differential luminosity spectra in \cref{fig:initial_state_lumis} which is obtained by integrating over rapidity
\begin{align}
\frac{\mathrm{d}\Lumint}{\mathrm{d}M}=\int_{-Y_{\max}(M)}^{Y_{\max}(M)}
\frac{\mathrm{d}^2\Lumint}{\mathrm{d}M\,\mathrm{d}Y}\,\mathrm{d}Y\,.~\label{eq:dL_dM}
\end{align}
Each panel of \cref{fig:initial_state_lumis} shows a different set of particles at the \ac{IP}, $e^+e^-$ (top-left), $e^-e^-$ (top-right), $e^-\gamma$ (bottom-left) and $\gamma\gamma$ (bottom-right). The different colors correspond to the different colliders, dark (light) red lines show the luminosity spectra for an $e^+e^-$ machine with round (flat) beams, dark (light) orange for an $e^-e^-$ machine with round (flat) beams and blue for a $\gamma\gamma$ collider. As expected, the $e^+e^-$ ($e^-e^-$) spectra of the $e^+e^-$ ($e^-e^-$) colliders contain a sharp peak at the \SI{10}{\TeV} endpoint, since a fraction of the $e^\pm$ did not radiate a hard photon. 
Similarly, for the $\gamma\gamma$ collider there is a sharp peak in the $e^-e^-$ spectrum, as many $e^-$ neither radiate nor scatter with a photon in the laser.
The $\gamma\gamma$ spectrum (bottom right panel) is moreover enhanced at high $M$, relative to the $\gamma\gamma$ spectrum at the $e^+e^-$ and $e^-e^-$ colliders. 
This can be traced to the different photon spectra from laser backscattering, governed by linear Compton scattering, and from beamstrahlung, which is shaped by strong-field radiation effects.

To perform a realistic physics study, we must always integrate over the spectra in \cref{fig:initial_state_lumis}, folded against the appropriate particle production cross sections.
Before we head there, we can however already gain some qualitative insights by comparing some simpler quantities. First, we define $\Lx{0.95}$ (see \cref{tab:beam-parameters}) by integrating the luminosity spectra over $M\in [\SI{0.5}{\TeV}, \SI{10}{\TeV}]$. (The infrared cut on $M$ is chosen to make the observable more robust against soft radiation.)
We see that $\Lx{0.95}$ is larger for $e^+e^-$ colliders because the opposite charges attract and \emph{pinch} the bunch during the collision, while same charges in the $e^-e^-$ machines repel and \emph{anti-pinch} beams (see e.g.~\cite{YokoyaChen:1992}).  
The larger beamstrahlung in the round-beam configurations also helps explain the behavior of $\Lx{0.95}$, since a greater share of the beam energy is radiated into low-energy particles.

In practice, $\Lx{0.95}$ is not particularly useful for physics studies, as we are often interested in the highest energy collisions only. 
For this purpose we define $\Lx{0.20}$ ($\fx{0.20}\equiv \Lx{0.20}/\Lx{0.95}$) which quantifies the luminosity (fraction) in the highest 20\% of the kinematic range, while $\Lx{0.01}$ ($\fx{0.01}\equiv \Lx{0.01}/\Lx{0.95}$) isolates the top 1\%. Smaller $\fx{x}$ indicates stronger beamstrahlung, i.e.~more luminosity at lower energies.
The fraction of luminosity at \SI{10}{\TeV} is markedly different for machines with $e^+e^-$ bunches versus $e^-e^-$ bunches.
The $e^+e^-$ pinch during the collision, which increases the on-axis electromagnetic field, the beamstrahlung parameter $\Upsilon$, and the average photon emission.
This transfers a larger share of luminosity into the low-$M$ tail, hence a \emph{smaller} luminosity at $M=\SI{10}{\TeV}$. The $e^-e^-$ beams anti-pinch, thus reducing the beamstrahlung, so a \emph{larger} fraction of the luminosity remains at $M\simeq\SI{10}{\TeV}$. 
The same trend appears within a given charge configuration when comparing round versus flat beam configurations: flatter beams ($\sigma_x \gg \sigma_y $) mitigate disruption and beamstrahlung, raising the endpoint fraction relative to round beams (see e.g.~\cite{YokoyaChen:1992}).

Finally, to give an impression of the various rapidities $Y$ involved for $\SI{5}{\TeV}$ primary beam energies we show in \cref{fig:initial_state_rapidities} the double-differential distribution of \cref{eq:double-diff-lumi} evaluated at two \ac{COM} energies $M = \{\num{1},\num{8}\}\SI{}{\TeV}$. These rapidity distributions are symmetric in $Y$ for initial states that match the primary beams. Asymmetries appear for initial states relying on secondary beam particles, for example $e^-e^-$ initial states from an $e^+e^-$ primary beam. 
We see very strong features at the end-points of the $Y$ distribution, which shift inwards as $M$ is increased. At these peaks, one particle has not radiated and therefore carries the full \SI{5}{\TeV} beam energy, while the second particle has much lower energy, thus generating an event that is boosted along the beam axis. 
Experimentally, this means there is a strong correlation between the boost of the event relative to the lab frame and the \ac{COM} energy of the collision. 
We also see that the detector should have a rapidity acceptance to roughly $|Y|\approx 3$, to make full use of the kinematic range of the colliders. 
At the LHC detectors, this degree of forward coverage has already been achieved, however the extent to which it would be compatible with the beam delivery system at a wakefield collider is yet unknown. 
For the $\gamma\gamma$ collider, the additional laser systems may further restrict the rapidity acceptance.

\section{Benchmark models\label{sec:model}}
\subsection{Motivation}
With the discovery of the Higgs boson, we now have all the building blocks to construct a minimal, self-consistent theory for the electroweak interactions. 
The LHC, however, cannot establish conclusively whether this minimal option is realized in Nature, or whether additional particles near the electroweak scale remain to be discovered.
There are compelling reasons to suspect that this may be the case: In \ac{SM} extensions where quantum corrections to the Higgs vacuum expectation value can be computed, these corrections are often large, unless there exist new particles with mass around \SI{1}{\TeV}. 
Supersymmetry is the best-known example; it predicts a partner particle for each \ac{SM} particle with identical charges (see \cite{Martin:1997ns} for a pedagogical review).
After accounting for the observed Higgs mass at \SI{125}{\GeV}, this theory predicts scalar partners for the top quark in the 1 to \SI{10}{\TeV} mass range, along with several other particles with similar mass \cite{PardoVega:2015eno}.
Similar arguments exist in composite Higgs theories, which typically require a fermion partner for the top quark with mass in the multi-TeV regime. While the spin and charges of these particles vary, they all interact under the electroweak interaction and are therefore produced with comparable cross sections through interactions with the $\gamma, W$ and/or $Z$ bosons.

Heavy electroweak particles can also be excellent dark matter candidates, as the weak force could be responsible for setting the correct relic density of the dark matter, the so-called ``WIMP miracle''.
In supersymmetry, the dark matter is typically the neutral component of a mixture of several such electroweak states, and its relic density can be achieved through co-annihilation with other superpartners. 
In the framework of ``minimal dark matter'', this picture is simplified by assuming that dark matter resides in a single electroweak multiplet~\cite{Cirelli:2005uq}. This assumption is highly predictive, as the mass of the new particle is the only free parameter in the model, up to discrete choices for the electroweak representation in which the dark matter is embedded. 
This mass parameter can be fixed by matching the relic density of annihilation freeze-out with the observed dark matter abundance (see \cite{Bottaro:2021snn,Bottaro:2022one} for recent calculations).
The simplest examples are fermionic triplet and doublet representations (see \cref{sec:modeldefinition}), which produce the correct abundance with masses of $m_\chi=\SI{2.86}{\TeV}$ and $m_\chi=\SI{1.08}{\TeV}$ respectively.  The triplet is currently disfavored as a dark matter candidate by observations of the $\gamma$-ray spectrum in the sky~\cite{Cohen:2013ama,Fan:2013faa,Safdi:2025sfs}. In contrast, the doublet dark matter scenario, often referred to as the ``thermal higgsino'', is currently unprobed and could be discovered by the Cherenkov Telescope Array Observatory (CTAO) in the next decade \cite{Rodd:2024qsi}. 
Such a discovery would greatly strengthen the case for a multi-TeV lepton or $\gamma\gamma$ collider. 
It would also immediately clarify what energy and luminosity such a collider should have in order to confirm the discovery.

Finally, electroweak symmetry breaking may be non-minimal, in the sense that a second field contributes to the masses of the $W$ and $Z$ bosons. 
If this is the case, there must exist additional heavy scalar fields charged under the electroweak force. Some of these fields must carry electric charge, and can therefore also be produced directly in $e^+e^-$ and $\gamma\gamma$ collisions.

In summary, new electroweak particles are a common feature of \ac{SM} extensions.. There are a large number of possibilities with different spins, masses and charges under the known \ac{SM} forces. Several of these particles may mix with one another or with the SM fermions, which further affects their collider phenomenology. 
For a given mass, their cross sections typically differ only by $\mathcal{O}(1)$. To compare production rates across future colliders, we therefore focus on two representative benchmark models.
However, the space of possible decay modes is much broader.
This is important, because the decay mode determines the size of the \ac{SM} background, which in turn impacts how much luminosity is needed for a discovery. 
For this reason we consider several different decay modes, which were chosen to represent low, intermediate and high background scenarios, as explained in the remainder of this section.

\subsection{Model definition \label{sec:modeldefinition}}
As benchmark models we choose a fermionic electroweak triplet and doublet, the so-called `wino' and `higgsino' in the parlance of supersymmetry. The triplet contains three components with electric charges $-1$, 0 and $+1$. The $+1$ and $-1$ charged particles combine into a Dirac fermion $\chi^\pm$ (``chargino''), while the neutral component is a Majorana fermion $\chi^0$ (``neutralino''). The doublet contains analogous charged states, but its neutral component $\chi^0$ is a Dirac fermion.
In both cases, $\chi^\pm$ and $\chi^0$ are nearly mass degenerate, with a small splitting of $\Delta m_\chi\equiv m_{\chi^\pm}-m_{\chi^0}\approx \SI{164}{\MeV}$ and $\Delta m_\chi\approx \SI{340}{\MeV}$ respectively for the triplet and doublet.
This splitting arises from \ac{SM} radiative corrections \cite{Pierce:1996zz,Cheng:1998hc,Gherghetta:1999sw} but can be modified by the effective operators\footnote{In models where no such operators are present, the triplet and doublet are often referred to as ``pure''.}
\begin{align}
    \Delta \mathcal{L}_{\text{triplet}} &\supset \frac{1}{\Lambda_\text{UV}^3} (H^\dagger \chi H) (H^\dagger \chi H)\,,\\
        \Delta \mathcal{L}_{\text{doublet}} &\supset \frac{1}{\Lambda_\text{UV}} (H \chi) ( \chi H)\,.
\end{align}
These operators can be generated from integrating out an additional doublet and singlet fermion or just a singlet fermion for respectively the triplet and doublet cases.
Both options can be realized in the \ac{MSSM} \cite{Gherghetta:1999sw,Feng:1999fu}.
Alternatively, an additional singlet fermion may be lighter than the triplet or doublet and still mix with its neutral component. 
This opens up the decay mode of the $\chi^\pm$ to this neutral singlet accompanied by a $W^\pm$.
In what follows, we denote the lightest neutral fermion as $\chi^0$, regardless of which of the two aforementioned scenarios we find ourselves in. This allows us to freely vary $m_{\chi^0}$ with respect to $m_{\chi^\pm}$, as this captures a larger range of scenarios beyond the pure triplet and doublet cases. 
For the purpose of our analysis, the phenomenology of the doublet and triplet is very similar, except for an $\mathcal{O}(1)$ difference in their cross section and natural mass splitting.

\subsection{Production rates\label{sec:productionrates}}

We will consider the following production modes:
\begin{align*}
\begin{array}{ll}
\text{Drell-Yan}\quad &e^+ e^- \to \chi^+ \chi^-  \\
\text{Photon fusion}\quad &\gamma \gamma \to \chi^+ \chi^- \\
\text{Associated production}\quad\quad & e^\pm \gamma \to e^\pm\chi^+\chi^- \quad \text{and }\quad
 e^\pm \gamma \to e^\pm\chi^0\chi^0\\
\end{array}
\end{align*}
We also calculated the rates for production through $WW$-fusion ($e^+ e^- \to \chi^+ \chi^- \nu_e\bar\nu_e$) and $ZZ$-fusion ($e^- e^- \to \chi^+ \chi^- e^- e^-$) and found them to be negligible compared to both Drell-Yan and photon fusion. 
For unpolarized beams, the doublet and triplet cross sections differ only by the rescaling factors
\begin{align}
\frac{\sigma(e^+e^- \to \chi^+ \chi^-)_{\text{doublet}}}{\sigma(e^+e^- \to \chi^+ \chi^-)_{\text{triplet}}} &= \frac{1+4\sin^4\theta_W}{4\cos^4\theta_W}=0.51\label{eq:DYrescale}\,,\\
\frac{\sigma(\gamma\gamma \to \chi^+ \chi^-)_{\text{doublet}}}{\sigma(\gamma\gamma \to \chi^+ \chi^-)_{\text{triplet}}} &= 1\,,\label{eq:photonrescale}\\
\frac{\sigma(e^-\gamma \to e^-\chi^{+/0} \chi^{-/0})_{\text{doublet}}}{\sigma(e^-\gamma \to e^-\chi^+ \chi^-)_{\text{triplet}}} &= \frac{1-2\sin^2\theta_W+6\sin^4\theta_W}{2\cos^4\theta_W}=0.72\,,\label{eq:egammarescale}
\end{align}
with $\theta_W$ the \ac{SM} Weinberg angle. 
The $e^-\gamma \to e^-\chi^+ \chi^-$ channel is only relevant for the scenario where the $\chi^\pm$ cannot be reconstructed in the detector, as covered in \cref{sec:invisible}.
Since $\chi^\pm$ are effectively assumed to be invisible in this case, we add their rate to the rate for the truly invisible $\chi^0$ particles in \cref{eq:egammarescale}. 
For the triplet model, the rate for $e^-\gamma \to e^-\chi^0 \chi^0$ vanishes because the $\chi^0$ is a Majorana fermion.
For (partially) polarized beams there are additional differences between the doublet and triplet models in the angular distributions of the final states~\cite{Choi:1998ut}. 
We leave these effects for future studies and assume unpolarized beams here.

\begin{figure}
    \centering
    \includegraphics[width=\linewidth]{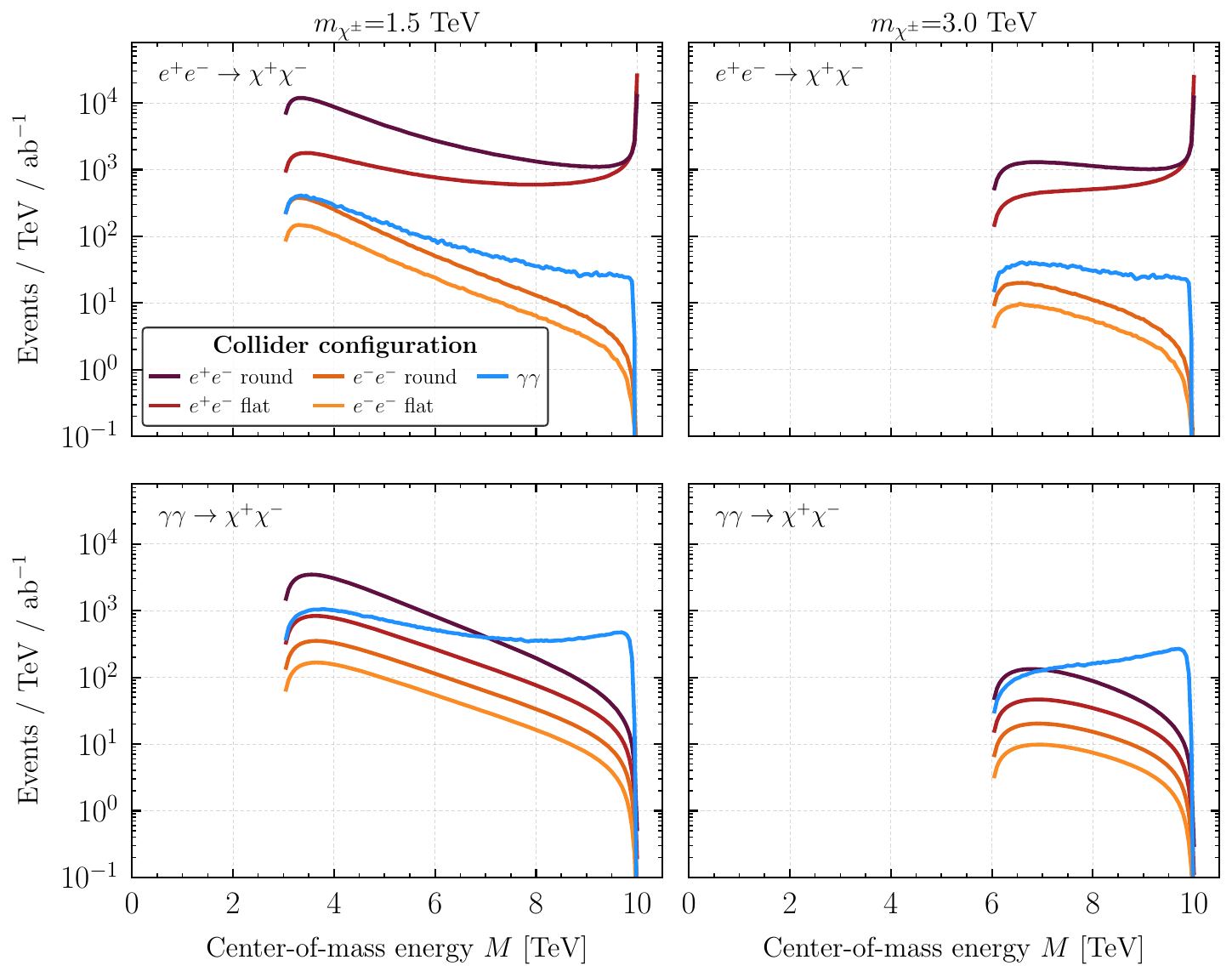}
    \caption{$\chi^\pm$ production rates in the triplet model as a function of the \ac{COM} energy $M$ of the collision. Top: Rate for Drell-Yan production for $m_\chi=\SI{1.5}{\TeV}$ and $m_\chi=\SI{3}{\TeV}$. Bottom: Rate for $\gamma\gamma \to \chi^+\chi^-$ for $m_\chi=\SI{1.5}{\TeV}$ and $m_\chi=\SI{3}{\TeV}$. 
    }
    \label{fig:differential}
\end{figure}

To obtain the particle production rates we fold the cross section as a function of the \ac{COM} energy, computed using \texttt{MadGraph5\_aMC@NLO\_v3} \cite{Alwall:2014hca,Frederix:2018nkq}, with the respective luminosity spectra in \cref{sec:wakefield}. 
In practice, we generate Monte Carlo events in discrete steps of \SI{50}{\GeV} in $M$ (with bin-centers $M_i$) as the cross section does not vary strongly for such a small interval. We then reweight these events with the integral over the appropriate luminosity spectrum over $[M_i-\SI{25}{\GeV},M_i+\SI{25}{\GeV}]$. The final bin ($M_i=\SI{10}{\TeV}$) of some of the spectra in \cref{fig:initial_state_lumis} is effectively a $\delta$-function, and is treated accordingly.  
This treatment with weighted events allows us to place analysis cuts on the events in \cref{sec:signatures} and efficiently carry out our studies for all five colliders at once.  
\cref{fig:differential} shows the resulting number of events for Drell-Yan and photon fusion as a function of the \ac{COM} energy for the triplet model for two masses, $m_{\chi^\pm} = \SI{1.5}{\TeV}$ and $m_{\chi^\pm} = \SI{3}{\TeV}$.
The rates for each collider are normalized to a geometric luminosity of $\SI{1}{\per\atto\barn}$.

For the $e^+e^-$ colliders, Drell-Yan production features a sharp peak at the endpoint of the distribution, which mirrors the peak in the luminosity spectra in \cref{fig:initial_state_lumis}.
We also find a prominent tail all the way down to the kinematic limit of $M=2m_{\chi^\pm}$.
This is because the Drell-Yan and photon fusion cross sections both scale as $\sigma \sim 1/M^2$, thus biasing the particle production rate towards lower $M$.
Concretely, we find that the fraction of $m_{\chi^\pm}=\SI{3}{\TeV}$  events with $M\in [\SI{9.9}{\TeV},\SI{10}{\TeV}]$ is respectively 41\% and 15\% for flat and round $e^+e^-$ colliders. For the $m_\chi = \SI{1.5}{\TeV}$ benchmark these fractions drop to 19\% and 3\%. 
This implies that the ``low energy'' tails are of qualitative importance, especially for round beams. 
We therefore cannot think of these machines as mono-energetic colliders, even if an $\mathcal{O}(1)$ fraction of their luminosity is contained in the endpoint peak. 
For the $e^-e^-$ and $\gamma\gamma$ colliders, the striking endpoint peak in Drell-Yan is absent, since the probability of finding a secondary $e^+$ carrying a large fraction of the beam energy is heavily suppressed. 
In photon fusion, the $\gamma\gamma$ collider has more support at high $M$, as expected from \cref{fig:initial_state_lumis}.

\begin{figure}
    \centering
    \includegraphics[width=\linewidth]{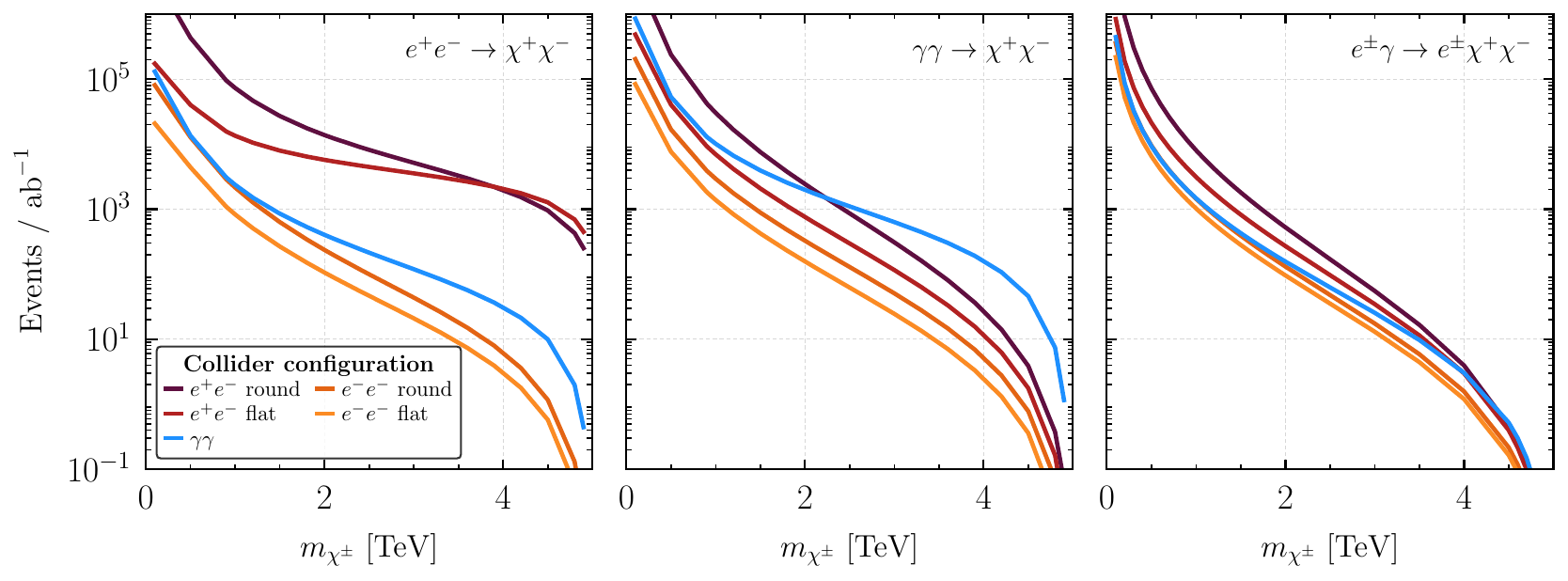}
    \caption{Total $\chi^\pm$ production rates in the triplet model as a function of the $m_{\chi^\pm}$ for Drell-Yan (left), photon fusion (middle) and $e^\pm$ associated production (right). For most scenarios, the sum of the Drell-Yan and photon fusion rates is the most relevant figure of merit. If both the $\chi$ particles are invisible to the detector, the rate for $e^\pm$ associated production is the right figure of merit. 
    }
    \label{fig:totalrate}
\end{figure}

To obtain the total number of events for a given production process and $\chi^\pm$ mass, we integrate the number of events over all \ac{COM} energies (\cref{fig:totalrate}). 
With the same geometric luminosity, we find that the $e^+e^-$ collider with round beams has the largest signal rate, except for $m_{\chi^\pm}\gtrsim \SI{4.5}{\TeV}$, where flat beams begin to out-perform the round beams.
The signal rate in the $e^-e^-$ colliders is roughly two orders of magnitude smaller at intermediate $m_{\chi^\pm}$, and suffers from an even greater suppression for high $m_{\chi^\pm}$. Nevertheless, for  $\SI{1}{\per\atto\barn}$ luminosity, these colliders could still produce between $10$ and $10^3$ $\chi^+\chi^-$ pairs, which could still be sufficient for a discovery, depending on how the $\chi^\pm$ behave after being produced.
As expected, the photon fusion process is very important for the $\gamma\gamma$ collider, to the extent that it always outperforms the two $e^-e^-$ colliders. 
This remains true for Drell-Yan production, as the photons from the laser stimulate the production of additional $e^+e^-$ pairs in the beam spot (see \cref{fig:initial_state_lumis}).

Although \cref{fig:totalrate} serves as a valuable initial figure-of-merit for comparing the colliders, its interpretation warrants caution:
First, we have assumed the same baseline geometric luminosity for all five colliders, even though the technological challenges they face on the accelerator side differ substantially.
Second, the ultimate sensitivity depends on the \ac{SM} backgrounds, which differ for each machine. These backgrounds also depend strongly on whether and how the $\chi^\pm$ decays.
To arrive at a comprehensive picture, we must therefore consider a range of possible decay modes, and analyze the corresponding backgrounds. 
The remainder of this paper is devoted to this analysis.

\subsection{Decay modes}
\begin{figure}
    \centering
    \includegraphics[width=0.8\linewidth]{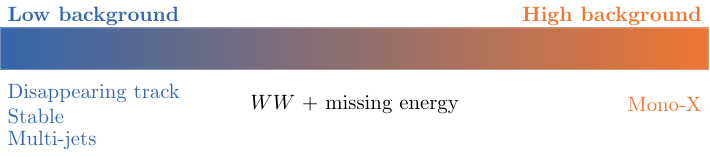}
    \caption{Schematic overview of the considered decay modes.
    }
    \label{fig:backgroundoverview}
\end{figure}

Certain $\chi^\pm$ decay scenarios are challenging to probe because of substantial \ac{SM} backgrounds, while others are effectively background-free.
This strongly affects the luminosity required for discovery, and a fair comparison between colliders therefore demands considering a range of scenarios, summarized schematically in \cref{fig:backgroundoverview}.
Our simplified model captures this by varying the mass of the lightest neutral state ($m_{\chi^0}$):
\begin{itemize}
    \item If $m_{\pi^\pm} < m_{\chi^\pm} - m_{\chi^0} \lesssim \SI{1}{\GeV}$, the dominant decay is $\chi^\pm \to \pi^\pm \chi^0$. With limited phase space, this decay can be slow enough for $\chi^\pm$ to travel several centimeters in the detector before decaying \cite{Chen:1996ap,Cheng:1998hc,Feng:1999fu,Gherghetta:1999sw}. The resulting track then terminates abruptly, and the low energy $\pi^\pm$ may not be reconstructed in the detector. There are essentially no \ac{SM} backgrounds for this ``disappearing track'' signature, apart from occasional misreconstructed particles. The $\chi^\pm$ lifetime in the lab frame also depends on the luminosity spectrum through the Lorentz boost of the \ac{COM} frame relative to the lab (see \cref{fig:initial_state_rapidities}), which in turn informs detector design. This case is discussed in \cref{sec:disappearingtrack}.

    \item If $m_{\chi^\pm} - m_{\chi^0} \gg m_W$, the dominant decay is $\chi^\pm \to W^\pm \chi^0$. This signature can be detected as two energetic $W$ bosons accompanied by missing energy. The background for this process is expected to be large; however, we find that suitable cuts can effectively suppress it, at least for the $e^+e^-$ and $\gamma\gamma$ colliders. This case is analyzed in \cref{sec:WWmet}.

    \item If \SI{1}{\GeV} $\lesssim m_{\chi^\pm} - m_{\chi^0} \lesssim \SI{10}{\GeV}$, the $\chi^\pm$ decays promptly to $\chi^0$ accompanied by soft leptons or hadrons. These decay products may in principle be reconstructed, but once the high-rate background from $\gamma\gamma \to$ hadrons is overlaid, they may not provide a useful handle for distinguishing signal from background.
    We therefore conservatively treat both $\chi^\pm$ and $\chi^0$ as effectively invisible to the detector. In order to record the event, an additional SM particle is needed in the final state. This class of searches are often called ``mono-X'' searches, where X can stand for a photon, lepton, electroweak boson or a jet. All mono-X searches tend to suffer from large \ac{SM} backgrounds from processes with neutrinos in the final state. We discuss this case in \cref{sec:invisible}.

    \item If $m_{\chi^0} > m_{\chi^\pm}$, the $\chi^\pm$ are stable and traverse the detector as slow, highly ionizing particles. There are essentially no \ac{SM} backgrounds, apart from the rare misreconstruction of \ac{SM} tracks. We study this case in \cref{sec:HCSP}.

    \item For completeness, we also comment on the scenario where the $\chi^\pm$ decay to three hard \ac{SM} jets, as is common in supersymmetry with R-parity violation. The backgrounds for this case are expected to be negligible, as discussed in \cref{sec:decaytojets}.
\end{itemize}
\section{Search Strategies}
\label{sec:signatures}
\subsection{Disappearing tracks}
\label{sec:disappearingtrack}
If the $\chi^\pm$ can travel a macroscopic distance before decaying, they leave hits in the innermost layers of the tracking detector before decaying into a soft $\pi^\pm$ and the invisible $\chi^0$.
These hits can be reconstructed to a charged track that suddenly ``disappears'' in the detector \cite{ATLAS:2022rme,CMS:2023mny}.
We assume that four hits are needed to reconstruct a high quality track. Let $\ell_{T4}$ and $\ell_{T5}$ denote the radial distance from the beam axis to the $4^\text{th}$ and $5^\text{th}$ tracking layer. We assume that observing a disappearing track corresponds to requiring the charged particle to decay within the region between $\ell_{T4}$ and $\ell_{T5}$, such that no hit is recorded in the $5^\text{th}$ tracking layer.\footnote{In principle, one could also search for decays between the $5^\text{th}$ and $6^\text{th}$ layers, and so on. Since the precise detector layout is not yet known, we interpret $\ell_{T4}$ and $\ell_{T5}$ here as the inner and outer radial boundaries of the decay region.}
For instance, in the ATLAS search for disappearing tracks \cite{ATLAS:2022rme}, the values $\ell_{T4} = \SI{12}{cm}$ and $\ell_{T5} = \SI{30}{cm}$ are used, corresponding to the fourth pixel layer and the innermost microstrip tracker, respectively.
Disappearing track searches are thus most sensitive to charged particles with average decay lengths on the order of a few centimetres, typically $c\tau \sim \mathcal{O}(1)\,\si{cm}$, where $c$ and $\tau$ are the speed of light and the proper lifetime of the particle, respectively.

The $\chi^\pm$ is a well-motivated candidate for such charged, long-lived particle. 
Its lifetime depends on the mass splitting $\Delta m_\chi$ between the charged and neutral components, which is a function of $m_{\chi^{\pm}}$ in the most minimal scenario.
For $m_{\pi^\pm} < \Delta m_\chi \lesssim \SI{1}{\GeV}$ the dominant decay channel is into a neutral component and a charged pion with the proper decay length given by~\cite{Chen:1995yu,Thomas:1998wy} 
\begin{align}\label{eq:ctautriplet}
    c\tau_{\text{triplet}} \simeq \SI{3.1}{cm}
        \left(
            \frac{\Delta m_\chi}{\SI{164}{MeV}}
         \right)^{-3}
         \left[1-\frac{m_{\pi^\pm}^2}{\Delta m_\chi^2}
    \right]^{-\frac12},
\end{align}
for the triplet and
\begin{align}\label{eq:ctaudoublet}
    c\tau_{\text{doublet}} \simeq \SI{0.7}{cm} 
        \left(
            \frac{\Delta m_\chi}{\SI{340}{MeV}}
        \right)^{-3}
        \left[ 1-\frac{m_{\pi^\pm}^2}{\Delta m_\chi^2}
    \right]^{-\frac12}\,,
\end{align}
for the doublet.
In the limit $m_{\chi^\pm} \gg m_W$, the mass splitting for a pure triplet (doublet) asymptotically approaches $\Delta m_\chi \to \SI{164}{MeV}$ ($\SI{340}{MeV}$)~\cite{Ibe:2012sx}. Inserting these values of $\Delta m_{\chi}$ into \cref{eq:ctautriplet,eq:ctaudoublet} yields proper decay lengths of $c\tau \simeq \SI{5.8}{cm}$ and \SI{1.3}{cm}, respectively. 
These approximations are accurate for $m_{\chi^\pm} \gtrsim \SI{1}{\TeV}$ and are shown as black dotted lines in \cref{fig:disappearing_track_reach}.

\begin{figure}
    \centering
    \includegraphics[width=\textwidth]{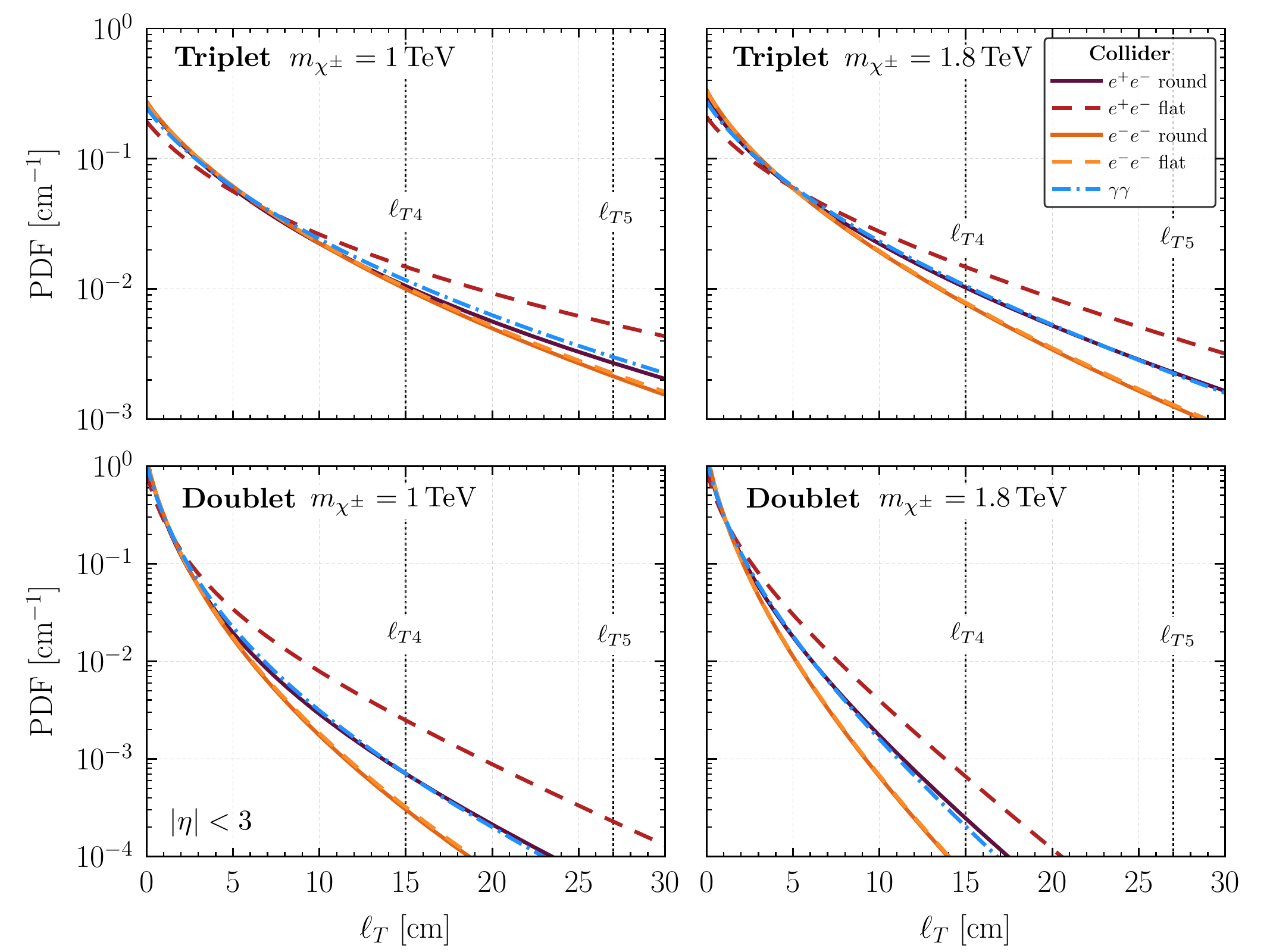}
    \caption{
        Lab-frame transverse flight-length distributions for the charged component of the triplet (top, $c\tau = \SI{5.8}{cm}$) and doublet (bottom, $c\tau = \SI{1.3}{cm}$) with $m_\chi = \SI{1}{\TeV}$ (left) and \SI{1.8}{\TeV} (right). 
        The $y$-axis is normalized such that the total number of events integrates to one in each distribution.
        The thin vertical lines at $\ell_{T4}=\SI{15}{cm}$ and $\ell_{T5}=\SI{27}{cm}$ indicate the fiducial cuts that were used to derive the discovery reach in \cref{fig:disappearing_track_reach}.
    }
    \label{fig:flight_length_transverse}
\end{figure}

\cref{fig:flight_length_transverse} shows the distribution of the $\chi^\pm$ flight length in the transverse plane  ($\ell_T$). 
The $y$-axis is normalized such that the total number of events integrates to one, allowing a comparison across setups with different cross sections.
All panels clearly illustrate the enhancement due to the Lorentz boost, as evidenced by a significant number of events with $\ell_T$ much larger than the intrinsic $c\tau$.
This effect is particularly pronounced for smaller $m_{\chi^\pm}$ and for $e^+e^-$ beams, which follow from higher collision energies in the Drell-Yan process.
It is largest for the flat $e^+e^-$ beams, as in this case a larger fraction of the signal events is produced at the maximum collision energy of \SI{10}{\TeV}.

\begin{figure}
    \centering
    \includegraphics[width=\textwidth]{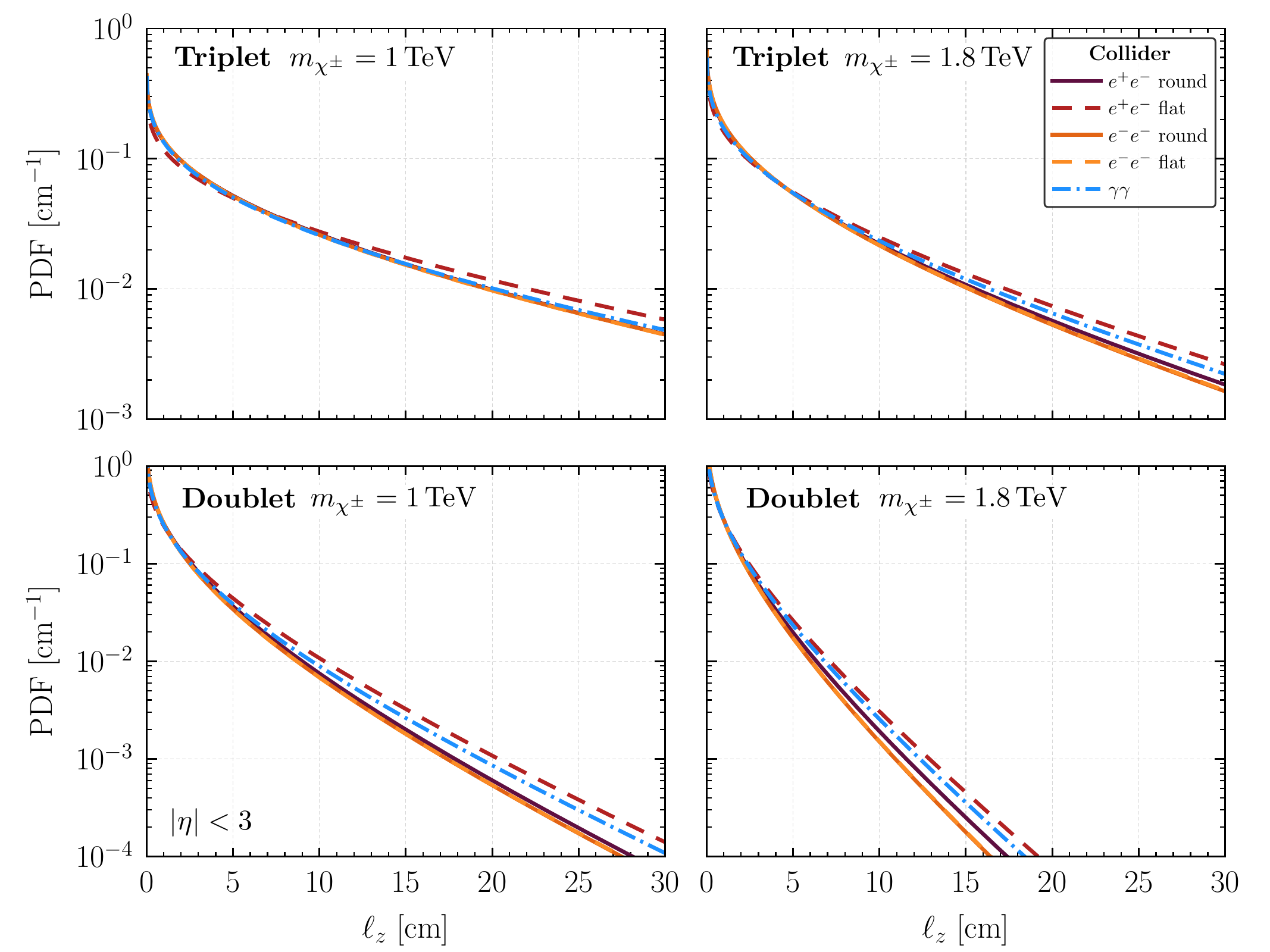}
    \caption{
        Lab-frame longitudinal flight-length distributions for charged triplets (top) and doublets (bottom).  
        The setup and conventions are identical to those in \cref{fig:flight_length_transverse}.
    }
    \label{fig:flight_length_longitudinal}
\end{figure}

\cref{fig:flight_length_longitudinal} shows the distribution of the longitudinal flight length $\ell_z$.
The longitudinal boost is largest if the initial state particles have very asymmetric energies.
Although the longitudinal flight length is not directly exploited in current disappearing track searches, these features may be useful in optimizing detector geometries for long-lived particle searches.
For example it may be possible to install dedicated forward tracking modules, along the lines of the LHCb VELO detector \cite{Bediaga:2013tje} and the ATLAS and CMS phase II forward tracking detectors \cite{ATLAS:2017svb,CMS:2017lum}.

\begin{figure}
  \centering
  \includegraphics[width=\textwidth]{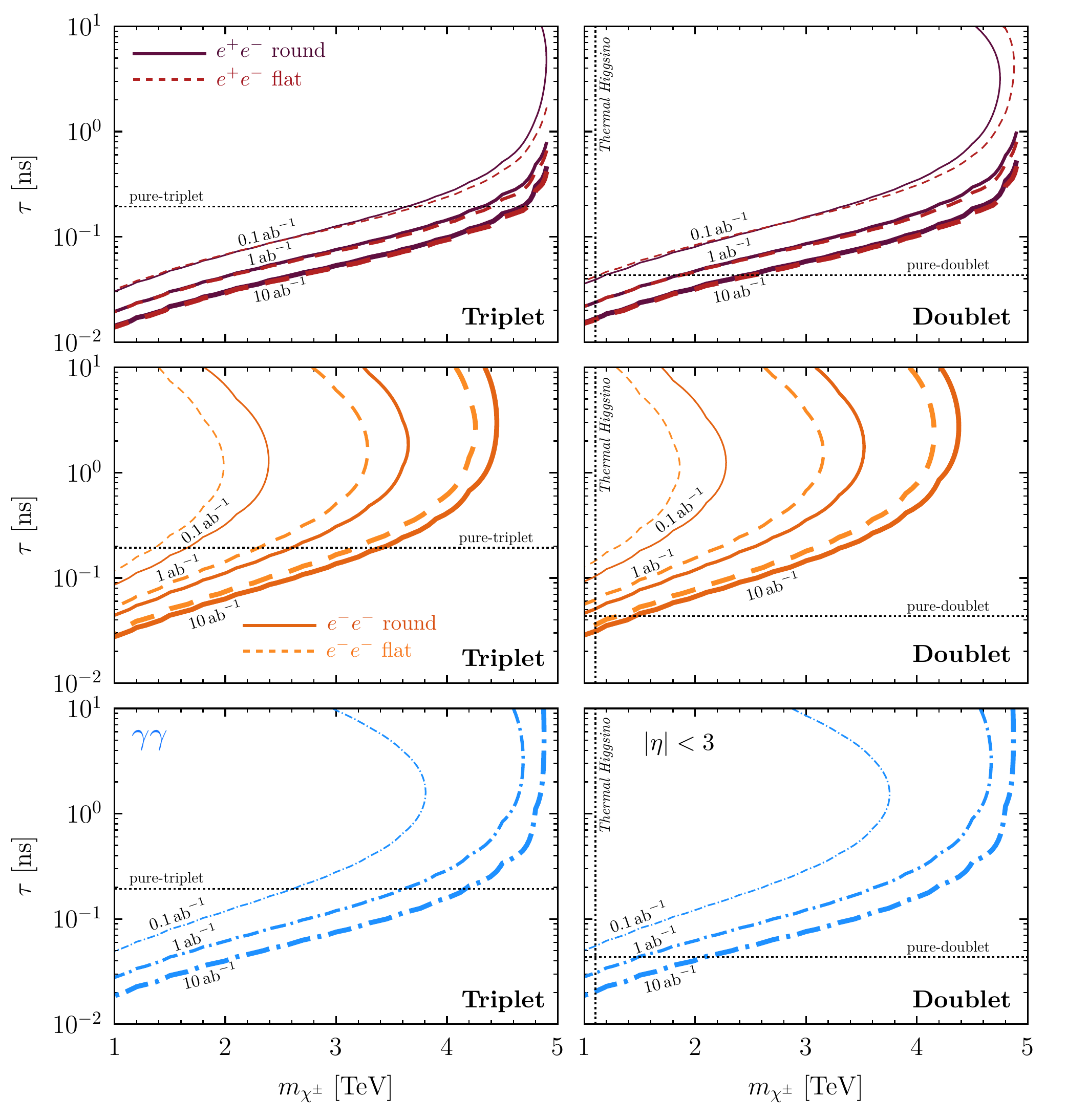}
  \caption{
    Reach of the disappearing track search for charged triplets (left) and doublet (right) with the detector setup of $\ell_{T4}=\SI{15}{cm}$ and $\ell_{T5}=\SI{27}{cm}$.
    Contours of the $10$ signal events are shown with different colors representing different integrated luminosities.
    The black dotted lines represent the lifetime of the pure triplet and doublet, in \eqref{eq:ctautriplet} and \eqref{eq:ctaudoublet}, as well as the mass of the thermal higgsino. Top to bottom are the $e^+e^-$ collider with round (solid) and flat (dashed) beams, the $e^-e^-$ collider with round (solid) and flat (dashed) beams, and the $\gamma\gamma$ collider.
  }
  \label{fig:disappearing_track_reach}
\end{figure}

In \cref{fig:disappearing_track_reach}, we present the projected sensitivity of the disappearing track search for the more general case where we allow $c\tau$ to be a free parameter.
We assume fiducial selection with $\SI{15}{cm} \leq \ell_{T} \leq \SI{27}{cm}$.
Only the dominant production channels are considered: $e^+e^- \to \chi^+\chi^-$ for the $e^+e^-$ collider and both $e^+e^- \to \chi^+\chi^-$ and $\gamma\gamma \to \chi^+\chi^-$ for the $e^-e^-$ and $\gamma\gamma$ colliders.
For the given geometric luminosity, the corresponding contour indicates regions where $10$ or more signal events are expected, which we assume would be sufficient to claim a discovery. For the doublet case (right panel), the black dotted line indicates the predicted $m_{\chi^\pm}$ for which the $\chi^0$ makes up all of the dark matter (``thermal higgsino'' scenario).

A comparison between the round and flat $e^+e^-$ beams reveals no significant differences in sensitivity. 
The signal production rate is highest in round beams, but the signal efficiency for a cut of $\ell_{T}\gtrsim \SI{15}{cm}$ is greater for flat beams (see \cref{fig:flight_length_transverse}). 
Both effects more or less compensate each other. 
Since this search is assumed to be free of irreducible SM background, a discovery can be made with very low luminosity in most of the parameter space.
As $c\tau$ decreases, the gain in sensitivity is roughly logarithmic in the luminosity, because the signal efficiency depends exponentially on $c\tau$. Both round and flat $e^+e^-$ beams can probe the thermal higgsino with a geometric luminosity of $\SI{0.1}{\per\atto\barn}$.

The $e^-e^-$ collider can also probe an interesting part of the parameter space, including the thermal higgsino, but will require more luminosity to do so. Here the round beams are always the better choice as we benefit from maximizing the luminosity of the $\gamma\gamma$ and $e^+e^-$ initial states. 
The discovery potential of the $\gamma\gamma$ collider is qualitatively similar to the one of the $e^+e^-$ machines, though slightly worse at high $m_{\chi^\pm}$. 
The $\gamma\gamma$ collider however maximizes its discovery potential at higher luminosity than the $e^+e^-$ options, as can be seen by the greater separation between the contours in the bottom panels of \cref{fig:disappearing_track_reach}.

\begin{figure}
    \centering
    \includegraphics[width=\textwidth,trim={0.4cm 8.cm 0.1cm 4.8cm}, clip]{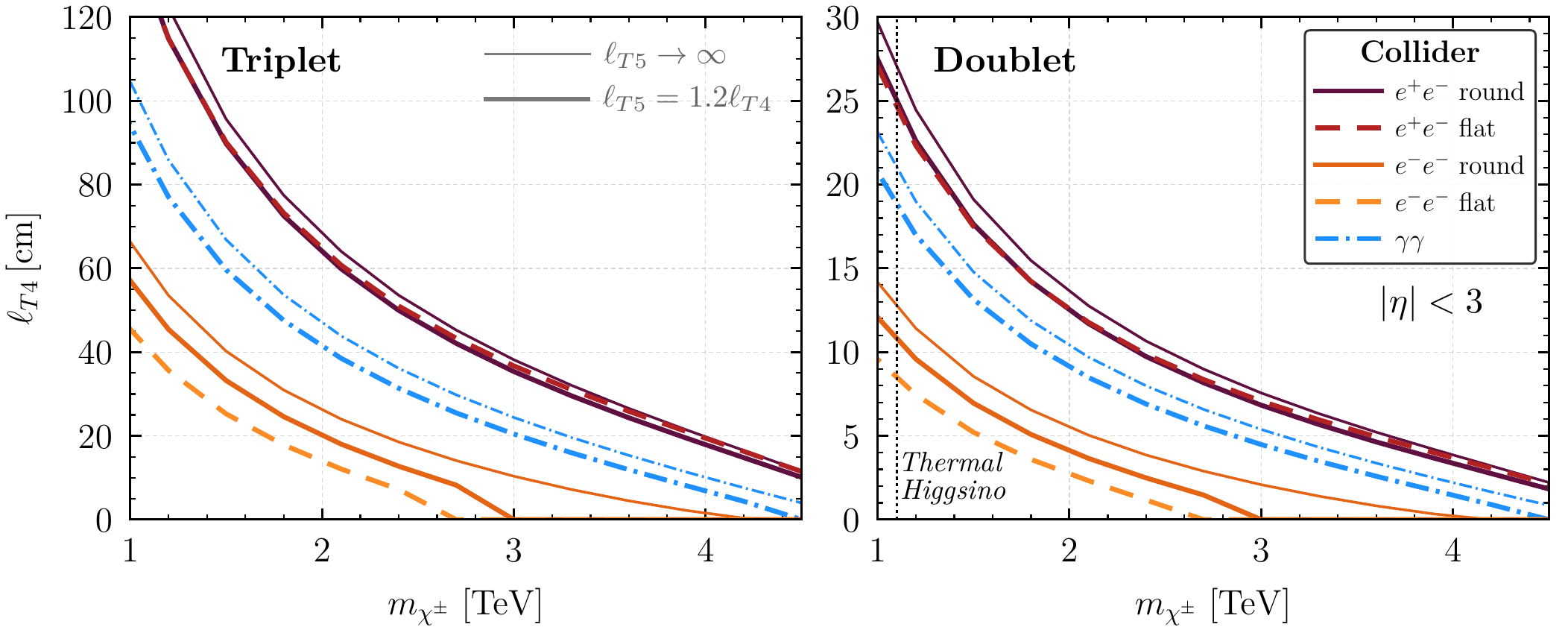}
    \caption{
        Required detector geometry parameter $\ell_{T4}$ to observe ten signal events as a function of $m_{\chi^\pm}$ for the triplet (left) and doublet (right) models, assuming a geometric luminosity of $\SI{1}{ab^{-1}}$.  
        The proper lifetimes are the same as those assumed in \cref{fig:flight_length_transverse}. 
        The vertical line on the right-hand panel indicates the value for $m_{\chi^\pm}$ for which $\chi^0$ constitutes 100\% of the dark matter relic density (``thermal higgsino''). 
        For clarity the $\ell_{T5}\to \infty$ lines were omitted for the flat beams.
    }
    \label{fig:ell_T4}
\end{figure}

Finally, we briefly comment on how the discovery potential depends on the detector geometry. In \cref{fig:ell_T4}, we show the most relevant detector geometry parameter, $\ell_{T4}$, needed to observe ten signal events for the triplet (left panel) and doublet (right panel), assuming a geometric luminosity of \SI{1}{\per\atto\barn}. The fifth layer position is fixed to $\ell_{T5} = 1.2\ell_{T4}$. To indicate that this latter assumption does not have much of an impact, we also show the case with $\ell_{T5} \to \infty$ for the round beams and the $\gamma\gamma$ collider.
This figure illustrates to what extent a lower signal rate in $e^-e^-$ or $\gamma\gamma$ collider, relative to the $e^+e^-$ colliders, can be compensated for by a more aggressive detector design.
\FloatBarrier
\subsection{$WW$ + missing momentum\label{sec:WWmet}}
We next consider the case where the $\chi^\pm$ decays promptly to a $W^\pm$ boson and a $\chi^0$ ($\chi^\pm \to W^\pm \chi^0$). 
We assume $\chi^0$ is stable and invisible, so the signature is a pair of $W$ bosons plus large missing momentum.
Missing momentum signals are generally well-motivated in the context of models of supersymmetry and dark matter.
The $WW$ + missing momentum signature is a particularly interesting test case to compare colliders with flat and round beams. This is because we expect large \ac{SM} backgrounds, which will be much more severe for round beams. 
Moreover, the \ac{COM} energy is not known on an event-by-event basis due to the luminosity spectra, unless all final states in the event are visible and reconstructed. 
This means that we can only rely on missing transverse momentum (MET), which is defined as the magnitude of the transverse component of the missing momentum vector.
For the case at hand this is
\begin{equation}
    \text{MET}\equiv \sqrt{(p_{W^+,x}+p_{W^-,x})^2+(p_{W^+,y}+p_{W^-,y})^2}\,,
\end{equation}
with $p_{W^+,x}$ the $x$-component of the $W^+$ momentum vector etc.
This is to be contrasted with a mono-energetic lepton-collider, e.g.~LEP, FCC-ee or muon colliders, where the full missing momentum vector and missing energy can be reconstructed for each event. 
This additional kinematic information is valuable for discriminating signal from background for models which contain invisible particles. 

At first glance, both features appear to favor flat beams: The backgrounds would be smaller and the collider would be closer to being mono-energetic, potentially helping with background discrimination. On the other hand, ATLAS and CMS routinely search for invisible particles using only transverse kinematics, despite much more challenging backgrounds. 
We will deploy techniques used at the LHC in the context of wakefield colliders and quantify the luminosity needed to make a 5$\sigma$ discovery for each collider configuration.
We use a straightforward cut-and-count strategy to ensure the results are conservative and can be reproduced easily. 
A more sophisticated strategy, e.g.~deploying machine learning, should therefore outperform our sensitivity estimates.

For the signal, we choose three benchmark points with $m_{\chi^\pm} =\SI{1.5}{\TeV}$, \SI{3}{\TeV} and \SI{4.5}{\TeV}. The $\chi^0$ mass is fixed to be $m_{\chi^0} = m_{\chi^\pm} / 3$. 
These choices place the model well outside the reach of the high-luminosity LHC (HL-LHC), and thus ensures we probe new parameter space. We also avoid the ``compressed'' regime where $(m_{\chi^\pm}-m_{\chi^0})/ m_{\chi^\pm} \ll 1$, thus guaranteeing a large amount of missing energy. The compressed case will be covered in \cref{sec:invisible}.

The dominant backgrounds are processes with hard neutrinos
\begin{align}
e^+ e^- &\to W^+ W^- Z, Z\to \nu\bar\nu \,, \label{eq:bgeetoWWZ}\\
\gamma \gamma &\to W^+ W^- Z, Z\to \nu\bar\nu \,, \label{eq:bgggtoWWZ}\\
e^+ e^- &\to W^+ W^- \nu \bar\nu \,, \label{eq:bgeptoWWvv}\\
e^- e^- &\to W^- W^- \nu \nu.\label{eq:bgeetoWWvv}
\end{align}
We consider only hadronic $W$-boson decays and therefore assume that the $W$ charges cannot be reconstructed.\footnote{In principle, identifying the charges of the $W$'s could help with background suppression. 
This is possible in the semi-leptonic channel, where we select for the $W^+\to \mu^+ \bar \nu$ and $W^-\to $ hadrons decay modes.
While such a selection would fully remove the $e^- e^- \to W^- W^- \nu\nu$ background, we find that it is too costly on the signal rate. This is because the $W^+\to \mu^+ \bar \nu$ branching ratio is only 11\%, and we estimate that the signal rate for the \SI{4.5}{\TeV} and \SI{3}{\TeV} benchmarks would be only $\sim\!0.1$ events per \SI{10}{\per\atto\barn} and $\sim\!5$ events per \SI{10}{\per\atto\barn}, respectively.}
For $e^+e^-$ colliders, the process in \eqref{eq:bgeptoWWvv} dominates over the processes in \eqref{eq:bgeetoWWZ} and \eqref{eq:bgggtoWWZ} by about an order of magnitude, depending on the cuts and beam configuration. 
For $e^-e^-$ colliders, the background from \eqref{eq:bgeetoWWvv} is about two orders of magnitude larger than \eqref{eq:bgggtoWWZ}, while \eqref{eq:bgeetoWWZ} is always negligible.
When referring to ``background'' in the remainder of this section, we mean the sum of all four processes, weighted by the relevant luminosity spectra, as laid out in \cref{sec:wakefield}. 
For example, the $\gamma\gamma$ collider has a large amount of luminosity in residual $e^-e^-$ collisions, to the extent that \eqref{eq:bgeetoWWvv} is the dominant background in this machine. 

We assume that the trigger will be close to 100\% efficient for the benchmark signal we consider, and thus neglect any losses of that nature.
The $W$'s are also expected to be highly energetic and well separated, and as such we also neglect any signal losses from the reconstruction efficiency associated with the hadronic $W$'s. 
Lastly, because the signature is inherently non-resonant, we do not expect our conclusions to be qualitatively impacted by the detector's energy resolution.
We therefore work with the truth-level four-vectors of the $W^\pm$. 
We further assume that the detector has a rapidity acceptance of $|\eta|<3$, which corresponds to $174^\circ\lesssim \theta \lesssim 6^\circ$ for the angle between the detector edge and the beamline. 
We verified that the $WW+$MET signature is insensitive to this choice, and a somewhat smaller detector acceptance would not qualitatively impact our conclusions.

\begin{figure}
    \centering
    \includegraphics[width=\linewidth]{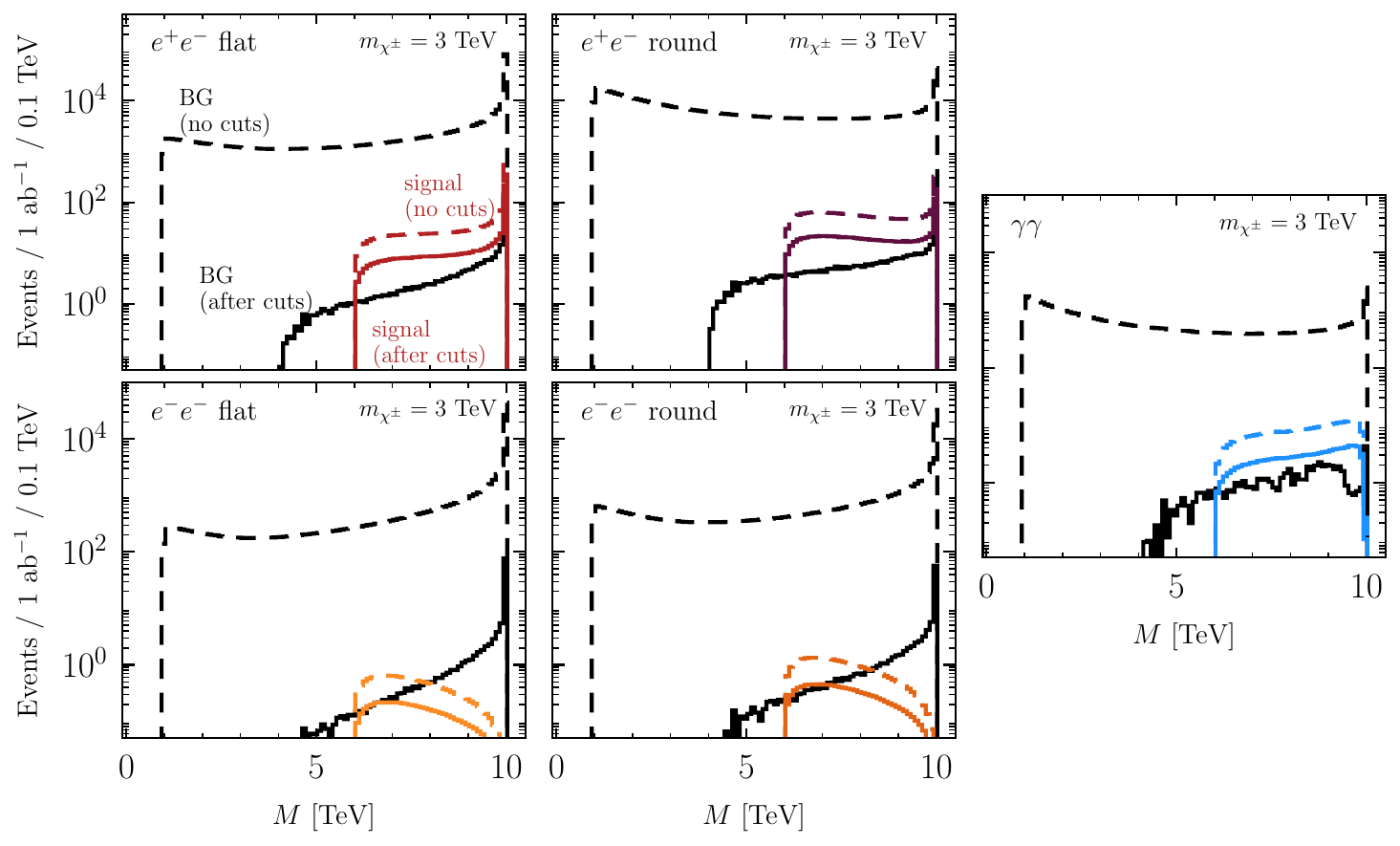}
    \caption{Signal and background distributions for the triplet model as function of the \ac{COM} energy ($M$), for the $m_{\chi^\pm}=\SI{3}{\TeV}$ benchmark, before and after imposing the cuts in \cref{tab:WWMETcuts}. $M$ itself cannot be reconstructed on an event-by-event basis and is therefore not available as a variable to help separate the signal from the background.  }
    \label{fig:WWmet_com_plot}
\end{figure}

To obtain the number of expected signal and background events, we reweight Madgraph Monte Carlo events with the luminosity spectra, as explained in \cref{sec:productionrates}.
The results are shown for the $m_\chi^\pm=\SI{3}{\TeV}$ triplet benchmark by the dashed curves in \cref{fig:WWmet_com_plot}. These lines are the same as the curves in \cref{fig:differential}, up to the detector fiducial efficiency and a rescaling factor that accounts for the $W\to~$~hadrons branching ratio. 
Comparing the $e^+ e^-$ beams first, we see that the signal rate is slightly higher for round beams, though the background is also higher by several orders of magnitude.
For $e^-e^-$ beams, the signal rates drop by about two orders of magnitude, because the necessary $e^+e^-$ or $\gamma\gamma$ initial state must be generated from the beam--beam interactions.
The $e^-e^- \to W^- W^- \nu \nu$ background process is unsuppressed in comparison, which explains why the $e^-e^-$ colliders are expected to be less effective.
In all cases, the signal is several orders of magnitude smaller than the background.

We must therefore impose additional selection cuts on the data, to suppress the background relative to the signal (solid lines in \cref{fig:WWmet_com_plot}). 
Due to the missing momentum, the \ac{COM} energy ($M$) of the collisions cannot be measured on an event-by-event basis. 
Following standard LHC practice, we instead define the $H_T$ variable as the scalar sum of the transverse momentum of the objects in the event:
\begin{equation}
    H_T \equiv p_{T,W^+} + p_{T,W^-} + \text{MET}
\end{equation}
with $p_{T,W^\pm}$ the magnitudes of the transverse momentum vectors of the $W^\pm$
\begin{equation}
 p_{T,W^\pm}\equiv \sqrt{p_{W^\pm,x}^2+p_{W^\pm,y}^2}\,.
\end{equation}
We further define a missing mass variable $M_\text{miss}$ by assuming that the \ac{COM} energy of the collision was $M=\SI{10}{\TeV}$. Under this assumption one could reconstruct the invariant mass of the full missing momentum four-vector: 
\begin{equation}\label{eq:missingmassWWMET}
M_\text{miss} \equiv \sqrt{(\SI{10}{\TeV} - E_{W^+}- E_{W^-})^2-|\vec p_{W^+}+\vec p_{W^-}|^2}
\end{equation}
with $E_{W^\pm}$ and $\vec p_{W^\pm}$ respectively the energy and three-momentum vectors of the $W^\pm$.

If the $M=\SI{10}{\TeV}$ hypothesis is correct for a particular event, this variable reconstructs the $Z$ mass ($M_\text{miss} = m_Z$) for the backgrounds in \cref{eq:bgeetoWWZ,eq:bgggtoWWZ}. For the signal on the other hand, one always finds $M_\text{miss} \gg m_Z$, and thus a mild cut on $M_\text{miss}$ provides an effective handle to suppress \cref{eq:bgeetoWWZ,eq:bgggtoWWZ} at no cost to the signal efficiency.
Finally, we use the variable $m_{T2}$, which has been developed for searches for supersymmetry at the LHC. 
Its definition is more involved than $H_T$ and $M_\text{miss}$, and we refer to Ref.~\cite{Lester:1999tx} for details. For our analysis we make use of the python implementation provided in Ref.~\cite{Lester:2014yga}.
To improve sensitivity, we loosely optimize the cuts on these variables for each signal benchmark, as shown in \cref{tab:WWMETcuts}. 
The tables with the full cut-flow for all benchmarks and collider configurations are provided in \cref{app:WWMET}.
\begin{table}[t]
    \centering
\begin{tabular}{ C{1.5cm} | C{2cm}  C{2cm}  C{2cm}  C{5.5cm}} \toprule       
$m_{\chi^\pm}$   & ${\text{MET}}$       & ${H_T}$              & $M_\text{miss}$        & $m_{T2}$ \\ \midrule
\SI{4.5}{\TeV}   & $>\SI{750}{\GeV}$    & $>\SI{5}{\TeV}$      & $>\SI{3}{\TeV}$        & $\SI{3}{\TeV}<m_{T2}< \SI{4.5}{\TeV}$     \\
\SI{3.0}{\TeV}   & $>\SI{750}{\GeV}$    & $>\SI{4}{\TeV}$      & $>\SI{3}{\TeV}$        & $\SI{2}{\TeV}<m_{T2}< \SI{3}{\TeV}\phantom{.5}$     \\
\SI{1.5}{\TeV}   & $>\SI{750}{\GeV}$    & $>\SI{2}{\TeV}$      & $>\SI{2}{\TeV}$        & $\SI{1}{\TeV}<m_{T2}<\SI{1.5}{\TeV}$     \\ \toprule
\end{tabular}
\caption{Cuts for the $WW$+MET signature, for each of the three signal benchmark points. \label{tab:WWMETcuts}}
\end{table}

\begin{figure}
    \centering
    \includegraphics[width=\linewidth]{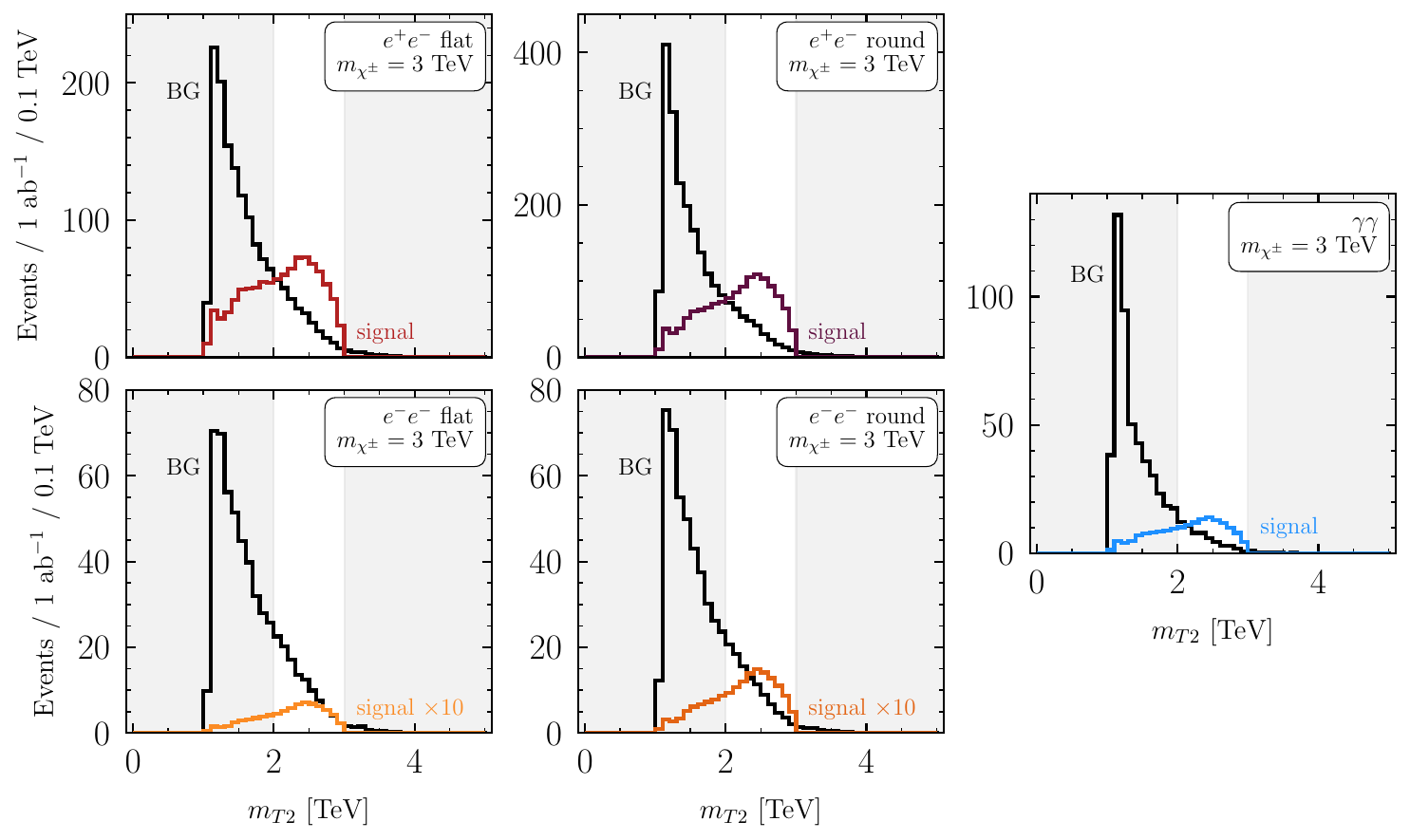}
    \caption{Distributions of the $m_{T2}$ variable for signal and background (BG) for the $m_{\chi^\pm}=\SI{3}{\TeV}$ benchmark. All cuts in \cref{tab:WWMETcuts} have been applied, except for the cut on $m_{T2}$ itself. The $m_{T2}$ cuts are indicated by the gray shading, where the signal region is the unshaded area. For clarity, the signal rate for both $e^-e^-$ colliders was multiplied $\times 10$. } 
    \label{fig:mt2ep_3TeV}
\end{figure}

\cref{fig:mt2ep_3TeV} shows the $m_{T2}$ distributions for the $m_{\chi^\pm}=\SI{3}{\TeV}$ triplet benchmark, after the MET, $H_T$ and $M_{\text{miss}}$ cuts in \cref{tab:WWMETcuts} have been applied. For both $e^+e^-$ colliders and the $\gamma\gamma$ collider we achieve excellent signal to background discrimination for a signal efficiency between 30\% and 40\%, depending on the beam configuration. This also holds for the $m_{\chi^\pm}=\SI{1.5}{\TeV}$ and $m_{\chi^\pm}=\SI{4.5}{\TeV}$ benchmarks; see \cref{fig:mt2ep_1p5TeV} and \cref{fig:mt2ep_4p5TeV} in \cref{app:WWMET}.
For $e^-e^-$ beams the picture is less favorable, as the signal rate is suppressed by several orders of magnitudes, due to the suppressed $\gamma\gamma$ and $e^+e^-$ luminosity spectra (\cref{fig:initial_state_lumis}). 

Finally, we estimate the geometric luminosity needed to make a convincing discovery.
To this end, we calculate the median expected significance as a function of the geometric luminosity assuming a single-bin Poisson likelihood, and solve for $5\sigma$ significance. 
For simplicity, we use the same set of selection cuts in \cref{tab:WWMETcuts} for both the triplet and doublet models, and rescale the signal rates using \eqref{eq:DYrescale} and \eqref{eq:photonrescale}.
The results are summarized in \cref{tab:WWMETsignificance} for all benchmarks and beam configurations. Given the effective background discrimination in this analysis, we find that the dominant driver of the sensitivity is ultimately the production rate, and colliders with round beams outperform those with flat ones except in the highest mass benchmark.
For the $e^+e^-$ configurations, a comparatively low geometric luminosity of $\sim\!\SI{300}{\text{fb}^{-1}}$ is sufficient to discover all benchmarks. Roughly 200 $\text{fb}^{-1}$ would be enough to discover the \SI{1.5}{\TeV} and \SI{3}{\TeV} benchmarks with the $\gamma\gamma$ collider, while the \SI{4.5}{\TeV} benchmark would require roughly \SI{3}{\per\atto\barn}.
Finally, very high luminosities would be needed for the $e^-e^-$ colliders, such that only the \SI{1.5}{\TeV} and \SI{3}{\TeV} benchmarks can be discovered with a geometric luminosity of the order of $\mathcal{O}(\SI{10}{\per\atto\barn})$. 
Though more sophisticated analysis strategies could improve the situation, it appears unlikely that the \SI{4.5}{\TeV} benchmarks could be accessible in an $e^-e^-$ collider with a realistic luminosity.

\begin{table*}
\centering
\begin{tabular}{L{2cm} | C{3.7cm}  C{3.7cm}  C{3.7cm}  } \toprule
Collider & $m_{\chi^\pm}=\SI{1.5}{\TeV}$ & $m_{\chi^\pm}=\SI{3}{\TeV}$ & $m_{\chi^\pm}=\SI{4.5}{\TeV}$\\\hline
$e^+e^-$ round 	& 6 (15) 	& 21 (60) 	& 93 (291) \\
$e^+e^-$ flat 	& 42 (107) 	& 34 (104) 	& 50 (148) \\
$e^-e^-$ round 	& 989 ($1.3 \times 10^3$) 	& $2.1 \times 10^4$ ($3.0 \times 10^4$) 	& $6.5 \times 10^6$ ($1.1 \times 10^7$) \\ 
$e^-e^-$ flat 	& $4.7 \times 10^3$ ($6.0 \times 10^3$) 	& $9.9 \times 10^4$ ($1.4 \times 10^5$) 	& $3.2 \times 10^7$ ($5.3 \times 10^7$) \\ 
$\gamma\gamma$ 	& 51 (56)	& 188 (202) 	& $2.7 \times 10^3$ ($3.0 \times 10^3$)\\
 \toprule  
\end{tabular}
\caption{Approximate geometric luminosity (in $\mathrm{fb}^{-1}$) needed for a $5\sigma$ discovery via the $WW$+MET search for the triplet (doublet) model for the given $m_{\chi^\pm}$. }
\label{tab:WWMETsignificance}
\end{table*}

\subsection{Mono-$X$ searches}
\label{sec:invisible}
In \cref{sec:disappearingtrack} and \cref{sec:WWmet} we assumed $m_\pi<\Delta m_\chi \lesssim \SI{1}{\GeV}$ and $m_W~\ll ~\Delta m_\chi$ respectively, which produced the disappearing track and $WW\,+$\,missing energy signatures.
In both cases, suitable cuts can suppress the backgrounds below the signal rate.
In this section we consider intermediate mass splittings, specifically $\SI{1}{\GeV}\lesssim \Delta m_\chi \lesssim \SI{10}{\GeV}$. This is the most challenging case for any future collider, since there are large SM backgrounds that cannot be fully suppressed.
The only discovery path is therefore to collect a large amount of luminosity and search for small deviations on top of the predicted SM background.

The difficulty lies in the fact that the $\chi^\pm$ decay promptly to a few low energy hadrons and the invisible $\chi^0$.
It is likely feasible for the detector to reconstruct some of these soft hadrons, however it remains unclear to what degree this would help separate the signal from background.
This is because every event in a high energy collider will come with various amounts of low energy hadronic activity. 
This is most severe at hadron colliders, where the huge soft QCD cross section and the high pile-up conditions inject a large amount of soft hadronic activity into the detector with every collision, completely overwhelming any particles from the $\chi^\pm$ decays.
In muon and electron/photon colliders, direct QCD production is not a factor. There are however other challenges that must be mitigated:
In muon colliders we expect a sizable amount of hadronic activity from beam-induced backgrounds \cite{Collamati:2021sbv}, while in high energy electron/photon colliders the $\gamma\gamma \to$ hadrons process is expected to produce low energy hadrons in each bunch crossing \cite{Barklow:1443518}.
Both effects can likely be mitigated to a degree, but this requires dedicated studies that include more detailed assumptions with respect to the detector design and modeling of the beam--beam interactions. 
Such a study for wakefield colliders is underway by detector and accelerator experts at LBNL \cite{detectorpaper}. \ATLAS{Please advise on how this sentence should be worded, and what you want in the citation in terms of authors, titles, etc.}
In this work we therefore adopt a maximally conservative approach, and assume the decay products of the $\chi^\pm$ are not energetic enough to be reconstructed at a \SI{10}{\TeV} wakefield collider. Towards the end of this section we will comment on how including the soft hadrons may improve the sensitivity.

If $\chi^\pm$ are effectively invisible, $e^+e^-\to \chi^+\chi^-$ and $\gamma\gamma\to \chi^+\chi^-$ no longer provide usable triggers, as there is no visible energy in the event to trigger the detectors. 
One must therefore consider subleading production modes, with one visible particle in the final state in addition to the invisible $\chi\chi$ pair. 
We have studied the following possibilities:
\begin{align}
\text{mono-$\gamma$:\phantom{ZWe}}\quad &e^+ e^- \to \gamma \chi \chi  \\
\text{mono-$Z$:\phantom{$\gamma$We}}\quad &e^+ e^- \to Z \chi \chi \\
\text{mono-$W$:\phantom{Z$\gamma$e}}\quad &e^+ e^- \to W^\pm \chi^{\mp} \chi^{0}\\
\text{mono-$e$:\phantom{ZW$\gamma$}}\quad &\gamma e^- \to e^- \chi \chi\label{eq:monoeproduction}\\
& e^+ e^- \to e^\pm\bar\nu \chi^{\mp} \chi^{0}
\end{align}
where the $\chi\chi$ final state represents both $\chi^{+} \chi^{-}$ and $\chi^{0} \chi^{0}$.
We find that the mono-$e$ channel always performs best and we will therefore exclusively focus on this case.
A similar conclusion was reached for muon colliders, where the mono-$\mu$ final state was identified as the most promising \cite{Han:2020uak}.
The signal rate for the $\gamma e^- \to e^- \chi \chi$ process at all five colliders was shown in \cref{fig:totalrate}. For the $e^+e^-$ colliders both $\gamma e^- \to e^- \chi \chi$ and $\gamma e^+ \to e^+ \chi \chi$ are included, while the latter is negligible for the $e^-e^-$ and $\gamma\gamma$ colliders.
The contribution from $e^+ e^- \to e^\pm\bar\nu \chi^{\mp} \chi^{0}$ is included but always subdominant and numerically relevant only for $e^+e^-$ colliders. 

Due to the large beam--beam interactions in a plasma wakefield collider, there are two key differences from the muon collider analogue of this search: Firstly, in a muon collider the initial state $\mu^\pm$ nearly always carries the full \SI{5}{\TeV} of energy, which is not always the case in the plasma wakefield colliders. 
This gives the muon collider an additional handle to discriminate the signal from the background. 
Second, in a wakefield collider, the initial state $\gamma$ arises from beamstrahlung or, in the $\gamma\gamma$ configuration, from the laser system. 
In the muon collider, the initial state photon is virtual and arises from the muon's parton distribution function.
Numerically, this implies larger signal and background rates in a plasma wakefield collider, compared to a muon collider with similar energy and luminosity (see \cref{app:pdf} for details).

\begin{figure}
    \centering
    \includegraphics[width=\linewidth]{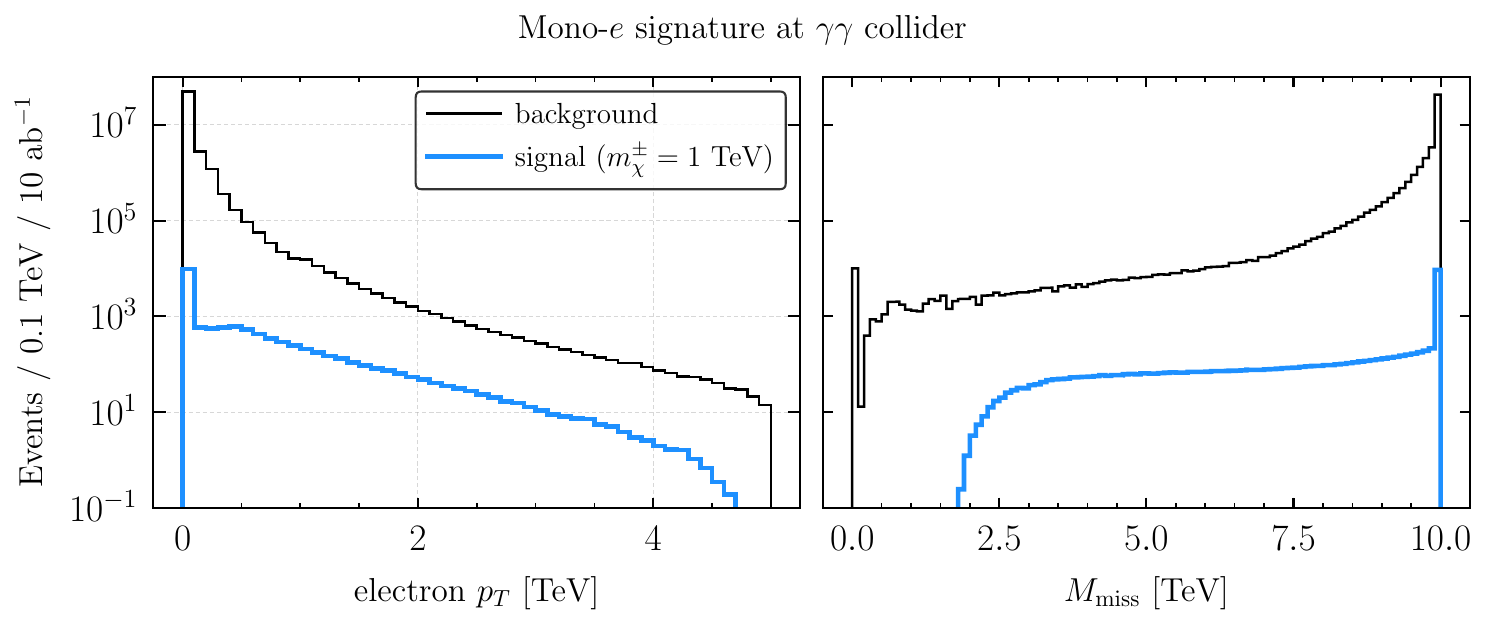}
    \caption{Signal and background distributions for the $e^-\gamma \to e^- \chi\chi$ channel, in  the $p_T$ of the $e^-$ (left) and missing mass (right).  The $\chi^\pm$ and $\chi^0$ are both assumed to be invisible. The luminosity spectra are those from the $\gamma\gamma$ collider; the distribution for all other colliders are qualitatively similar. }
    \label{fig:monoe_distributions}
\end{figure}

The dominant backgrounds are
\begin{align}
e^-\gamma &\to e^- \nu \bar\nu\,,\\
e^+e^- &\to e^+e^- \nu \bar\nu\,,\\
e^- \gamma & \to e^- \gamma\,, \label{eq:forwardphotonbackground}
\end{align}
where in the latter two cases the $e^+$ or $\gamma$ must fall outside the detector acceptance.
Both on-shell and off-shell contributions from $Z\to \nu\bar \nu$ and $W^-\to e^- \nu$ are included in our calculation of these backgrounds. 
The rate for \eqref{eq:forwardphotonbackground} is particularly large, though for this process the $p_T$ of the electron is bounded by
\begin{equation}
p_T \leq \frac{M}{2 \cosh(\eta_{\text{det}}-\log(\SI{10}{\TeV}/M))} \lesssim \SI{500}{\GeV}\,,
\end{equation}
with $\eta_{\text{det}}$ the rapidity acceptance of the detector, which we take to be $\eta_{\text{det}}=3$.
This background is therefore easily suppressed with a moderate $p_T$ cut. 

We assume access only to the four-momenta of the electron $(E_e,\vec p_e)$, which restricts our ability to separate the signal from the background. The most promising variables are the electron $p_T$ and the missing mass variable
\begin{equation}
M_\text{miss} = \sqrt{(\SI{10}{\TeV} - E_{e})^2-|\vec p_{e}|^2}\,,
\end{equation}
analogous to \cref{eq:missingmassWWMET}. 
The signal and background distributions for both variables are shown in \cref{fig:monoe_distributions} for the $\gamma\gamma$ collider.
The distributions for the other collider configurations are similar.
To estimate the sensitivity, we construct a binned 2D likelihood in $p_T$ and $M_\text{miss}$, with ten same-size bins in each variable.
The associated significance is defined by 
\begin{equation}\label{eq:emetsignificance}
    \text{significance}= \sqrt{\sum_i \frac{S_i^2}{B_i + (\epsilon_B B_i)^2}}\,,
\end{equation}
with $S_i$ and $B_i$ the number of signal and background events in the $i^{\text{th}}$ bin. Here, $\epsilon_B$ is the expected systematic uncertainty on the background distribution.
The luminosity spectra, missing higher-order electroweak corrections and the uncertainty on the electron reconstruction efficiency all contribute to $\epsilon_B$.
The luminosity spectra can likely be calibrated to the sub-$0.1\%$ level using large cross-section processes such as $e^-\gamma\to e^-\gamma$.
The uncertainty from higher order electroweak corrections to the background cross sections will likely be negligible in comparison. 
Overall, we expect the dominant uncertainty to be due to the uncertainty on the electron reconstruction efficiency. 
ATLAS has already achieved a total uncertainty of $\sim 0.2\%$ on their reconstruction efficiency of high $p_T$ electrons \cite{ATLAS:2023dxj}.
Here, we assume a total uncertainty of $\epsilon_B=0.5\%$, which ought to be conservative in light of the considerations above.
This is small enough that the significance for TeV-scale particle discovery remains statistically limited if one assumes a geometric luminosity of \SI{10}{\per\atto\barn}.

\begin{figure}
    \centering
\includegraphics[width=\linewidth]{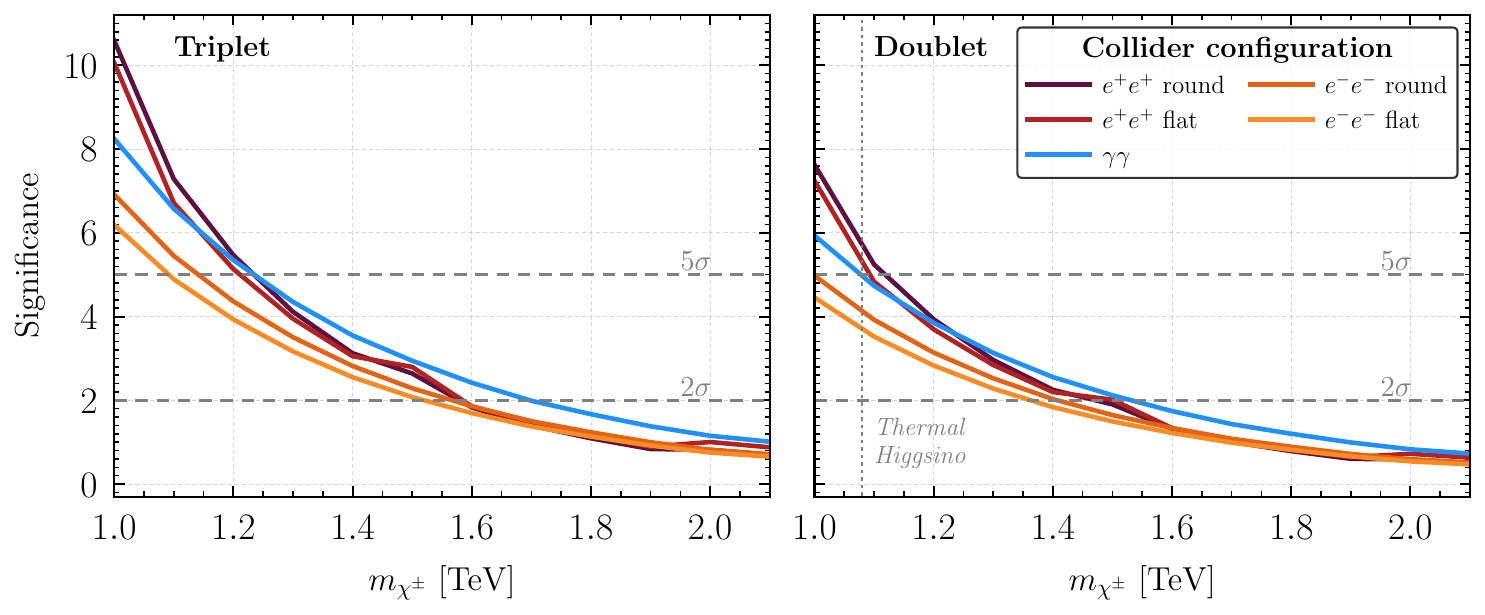}
    \caption{Expected significance of the mono-$e$ search for a geometric luminosity of \SI{10}{\per\atto\barn}. The systematic uncertainty on the background is assumed to be 0.5\%. The vertical line on the right-hand panel indicates the value for $m_{\chi^\pm}$ for which $\chi^0$ constitutes 100\% of the dark matter relic density (``thermal higgsino'').}
    \label{fig:monoX}
\end{figure}

\begin{table*}\centering
\begin{tabular}{L{2.9cm} | C{2.4cm}  C{2.4cm}  C{2.4cm} C{2.4cm} } \toprule
Collider & Triplet ($5\sigma$) & Triplet ($2\sigma$) &Doublet ($5\sigma$) & Doublet ($2\sigma$)\\\hline
$e^+e^-$ round	 &	1.23	 &	1.58	 &	1.12	 &	1.47 \\
$e^+e^-$ flat	 &	1.21	 &	1.58	 &	1.09	 &	1.5 \\
$e^-e^-$ round	 &	1.14	 &	1.57	 &	1.0	 &	1.41 \\
$e^-e^-$ flat	 &	1.09	 &	1.52	 &	0.94	 &	1.36 \\
$\gamma\gamma$	 &	1.23	 &	1.7	 &	1.08	 &	1.53 \\
\midrule
$\mu^+\mu^-$ \SI{10}{\TeV}	\cite{Han:2020uak} &  1.2	& 1.7  &  0.9 & 1.3\\
$pp$ \SI{100}{\TeV} \cite{Han:2018wus}	&  0.55	& 1.47  & 0.2 & 0.84\\
 \toprule  
\end{tabular}
\caption{Reach on $m_{\chi^\pm}$ (in TeV) for discovery ($5\sigma$) and exclusion ($2\sigma$) for the triplet and doublet model in a mono-$e$ search, assuming an integrated geometric luminosity of \SI{10}{\per\atto\barn} and a systematic uncertainty on the background of 0.5\%. For comparison, we show the analogous estimates for a \SI{10}{\TeV} muon collider (mono-$\mu$ search) with a luminosity of \SI{10}{\per\atto\barn} \cite{Han:2020uak} and a \SI{100}{\TeV} $pp$-collider  (mono-jet search) with a luminosity of \SI{30}{\per\atto\barn} \cite{Han:2018wus}. 
The assumed systematic uncertainties on the background modeling in these cases were respectively $0.1\%$ and $2\%$.
\label{tab:invisiblemassreach} }
\end{table*}

The significance defined in \cref{eq:emetsignificance} is plotted in \cref{fig:monoX} as a function of $m_{\chi^\pm}$, assuming a geometric luminosity of \SI{10}{\per\atto\barn} for each collider. 
The corresponding discovery ($5\sigma$) and exclusion ($2\sigma$) potential is shown in \cref{tab:invisiblemassreach}.
The signal production process in \eqref{eq:monoeproduction} does not require positrons, which explains why the performance of the $e^-e^-$ colliders is qualitatively similar to that of the $e^+e^-$ and $\gamma\gamma$ colliders.
In \cref{tab:invisiblemassreach} we also compare with the reach of a mono-$\mu$ search at muon collider \cite{Han:2020uak} and a mono-jet search at a \SI{100}{\TeV} $pp$ collider \cite{Low:2014cba,Cirelli:2014dsa,Han:2018wus}.
All wakefield colliders perform comparable to the muon collider, and significantly outperform the \SI{100}{\TeV} $pp$ collider. To discover the thermal higgsino, we estimate that the $e^+e^-$ collider with round (flat) beams would need a geometric luminosity of \SI{7.6}{\per\atto\barn} (\SI{8.6}{\per\atto\barn}). For the $e^-e^-$ collider with round (flat) beams \SI{15}{\per\atto\barn} (\SI{19}{\per\atto\barn}) is needed, compared to \SI{10}{\per\atto\barn} for the $\gamma\gamma$ collider.

\begin{figure}
    \centering
    \includegraphics[width=\textwidth,trim={0.4cm 8.cm 0.1cm 4.8cm}, clip]{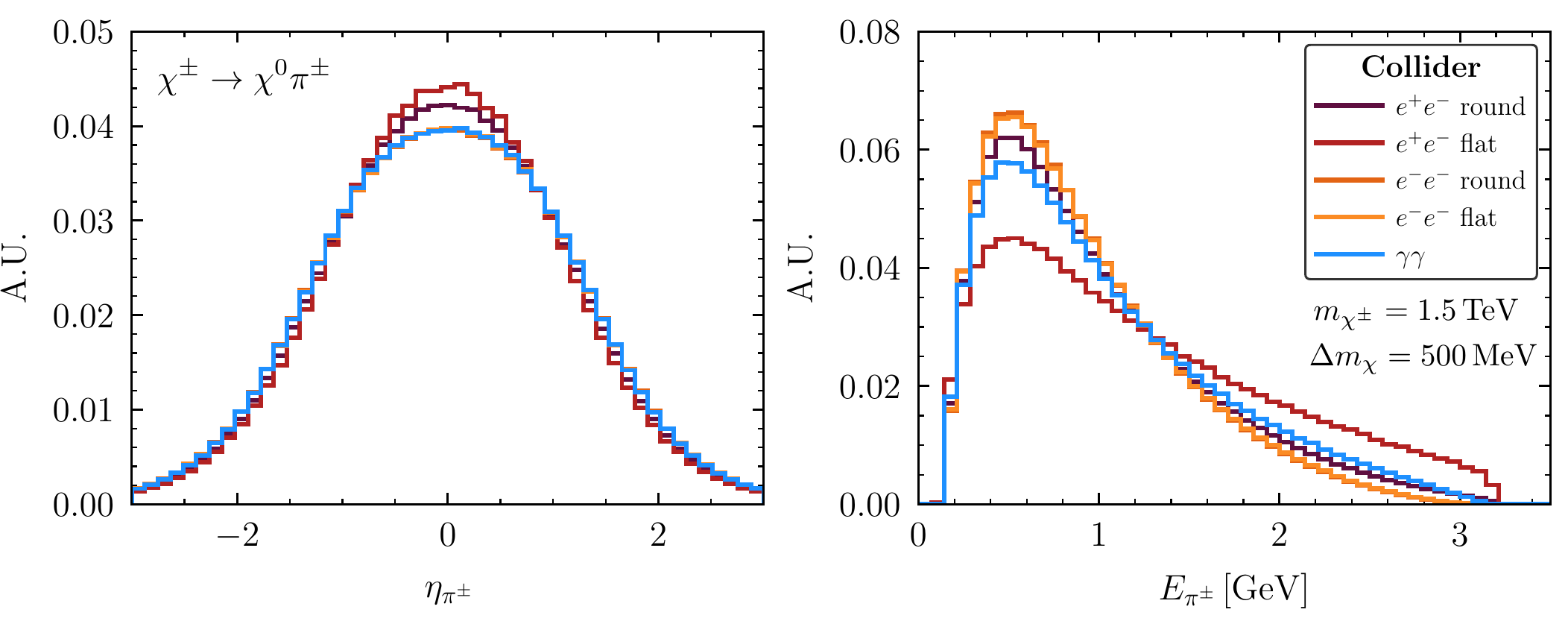}
    \caption{
        Distributions of the pseudorapidity (left) and energy (right) of the charged pion from the triplet-like $\chi^\pm$ decay $\chi^\pm \to \chi^0 \pi^\pm$, assuming $m_{\chi^\pm} = \SI{1.5}{\TeV}$ and $\Delta m_\chi = \SI{500}{\MeV}$.
    \label{fig:soft_charged_search}}
\end{figure}

To conclude, we comment briefly on ways to further improve the sensitivity to this scenario. Firstly, one could attempt to utilize the soft tracks produced in the $\chi^\pm$ decays. 
This has been studied for the LHC \cite{CMS:2025twk} and muon colliders \cite{Capdevilla:2024bwt}. \ATLAS{Simone, Fig 4 in this paper seems sketchy. We are inclined to cite it anyways, since we have no hard evidence that it is wrong. But perhaps this has already been scrutinized in the muon collider community?} 
A conclusive study requires carefully modeling low energy backgrounds, which is outside the scope of our current paper.  
As a first step, we can however quantify the typical kinematics of the soft signal tracks. 
As an example, we consider the two-body decay $\chi^\pm \to \chi^0 \pi^\pm$ with fixed masses: $m_{\chi^\pm} = \SI{1.5}{\TeV}$ and $m_{\chi^0} = m_{\chi^\pm} - \Delta m_\chi$, where $\Delta m_\chi = \SI{500}{MeV}$.  
Given the hierarchy $m_{\chi^0} \gg m_{\pi^\pm}$, the charged pion acquires an approximately mono-energetic energy $\sim \Delta m_\chi$ in the rest frame of $\chi^\pm$, which is then boosted in the lab frame.
\cref{fig:soft_charged_search} shows the distributions of the $\pi^\pm$ pseudorapidity $\eta$ (left panel) and energy $E$ (right panel) of the $\pi^\pm$ in the lab frame.

The right panel of \cref{fig:soft_charged_search} illustrates the potential for significant Lorentz boosts, resulting in a sizable fraction of events where the charged pion carries a few~\si{GeV} of energy.  
The endpoint of the energy distribution can be estimated as $\sim 2( \SI{10}{\TeV}/2m_{\chi^\pm})\Delta m_\chi$, where the factor of two accounts for the case in which $\chi^\pm$ and $\pi^\pm$ are collinear.  
A larger proportion of charged pions attain energies near this endpoint in $e^+e^-$ collisions with flat beams, owing to the more sharply peaked energy spectrum  of the beam particles.
For these energies and rapidities, we do not expect major obstructions designing detectors that can reconstruct the hard part of the spectrum in \cref{fig:soft_charged_search}.
Whether or not this is sufficient to improve the sensitivity depends on the background rate for a hard $e^-$ plus missing energy and one or two $\mathcal{O}$(GeV) pions. 

A second avenue of improvement would be to consider the virtual contributions of the $\chi$-particles to a precision measurement of Bhabha scattering and/or $\gamma\gamma\to W^+W^-$ cross sections. The former was shown to be promising for muon colliders \cite{Franceschini:2022sxc,Okabe:2023esr,Fukuda:2023yui}, and we expect both measurements to provide complementary sensitivity to the mono-$e$ search. We leave this and the full soft track analysis for future studies.

\subsection{Heavy stable charged particles \label{sec:HCSP}}
It is possible that $m_{\chi^\pm}- m_{\chi^0}< m_\pi$, which would imply that the decay length of the $\chi^\pm$ is much longer than the expected radius of the detector. 
In the context of the doublet and triplet model we considered, such a small splitting would require a minor amount of fine-tuning.
However, one may also imagine an alternative scenario with a charged particle that does not come with a neutral component and is exactly stable.
In our simplified models this corresponds to assuming $m_{\chi^\pm}< m_{\chi^0}$.
In both cases, the charged particles would behave as \acp{HSCP} in the detector, a very striking signature \cite{ATLAS:2022pib,CMS:2024nhn}.

The $\chi^\pm$ are most often produced near their kinematic threshold (see \cref{fig:differential}) and therefore tend to travel at an $\mathcal{O}(1)$ fraction of the speed of light. 
All detector-stable SM particles ($\mu^\pm,\pi^\pm,\cdots$) are very light and travel at the speed of light.
This allows signal and background to be separated using time-of-flight.
We assume that a timing detector, similar to the \ac{MIP} timing detector designed for the phase-2 CMS upgrade~\cite{CMS:2667167}, will be placed at a radial distance of $R=\SI{5}{\meter}$ away from the beam axis. 

The time-delay of each $\chi^\pm$ in the event is the difference between its arrival time at the detector and the arrival time of a particle traveling at the speed of light along the same trajectory
\begin{align}
    \Delta t_{\chi} & = \frac{R}{c\sin\theta}\left(\frac{1}{\beta}-1\right) = \frac{m_{\chi^\pm}}{p_T}\left(\xi-\sqrt{\xi^2-1}\right)\,, &
    \xi&\equiv \sqrt{1+\frac{p^2_T}{m^2_{\chi^\pm}}}\cosh{y}\,,
\end{align}
where $y$ is the particle's rapidity in the lab frame, $c$ is the speed of light and $\theta$ is the polar angle with respect to the beam-axis
The time delay of the event $\Delta t$ is the shortest of the time delays of the charged pair. 
The time delay is a monotonically-decreasing function of $\xi$, and thus of $\cosh{y} = \cosh\left(y^{\rm COM}+Y\right)$. 
Here $y^{\rm COM}$ is the rapidity of the particle in the \ac{COM} frame. 
Requiring that $\Delta t$ is longer than some threshold $\Delta t_{\rm{min}}$ then translates to a cut on the collision's lab-frame rapidity $Y$ which depends on $m_{\chi^{\pm}}$ and the event's kinematics in the \ac{COM}-frame. 
We are therefore able to implement a time-delay cut analytically, by reweighting each MadGraph event, as explained in \cref{sec:productionrates}, by $\frac{\mathrm{d}\Lumint}{\mathrm{d}M}$ calculated according to \cref{eq:dL_dM}.
The $Y$-integral limits are set by satisfying $\Delta t \geq \Delta t_{\rm{min}}$ in addition to a lab-frame $\eta$ cut, which also implies a similar cut on~$Y$.

\begin{figure}
    \centering
    \includegraphics[scale=0.5]{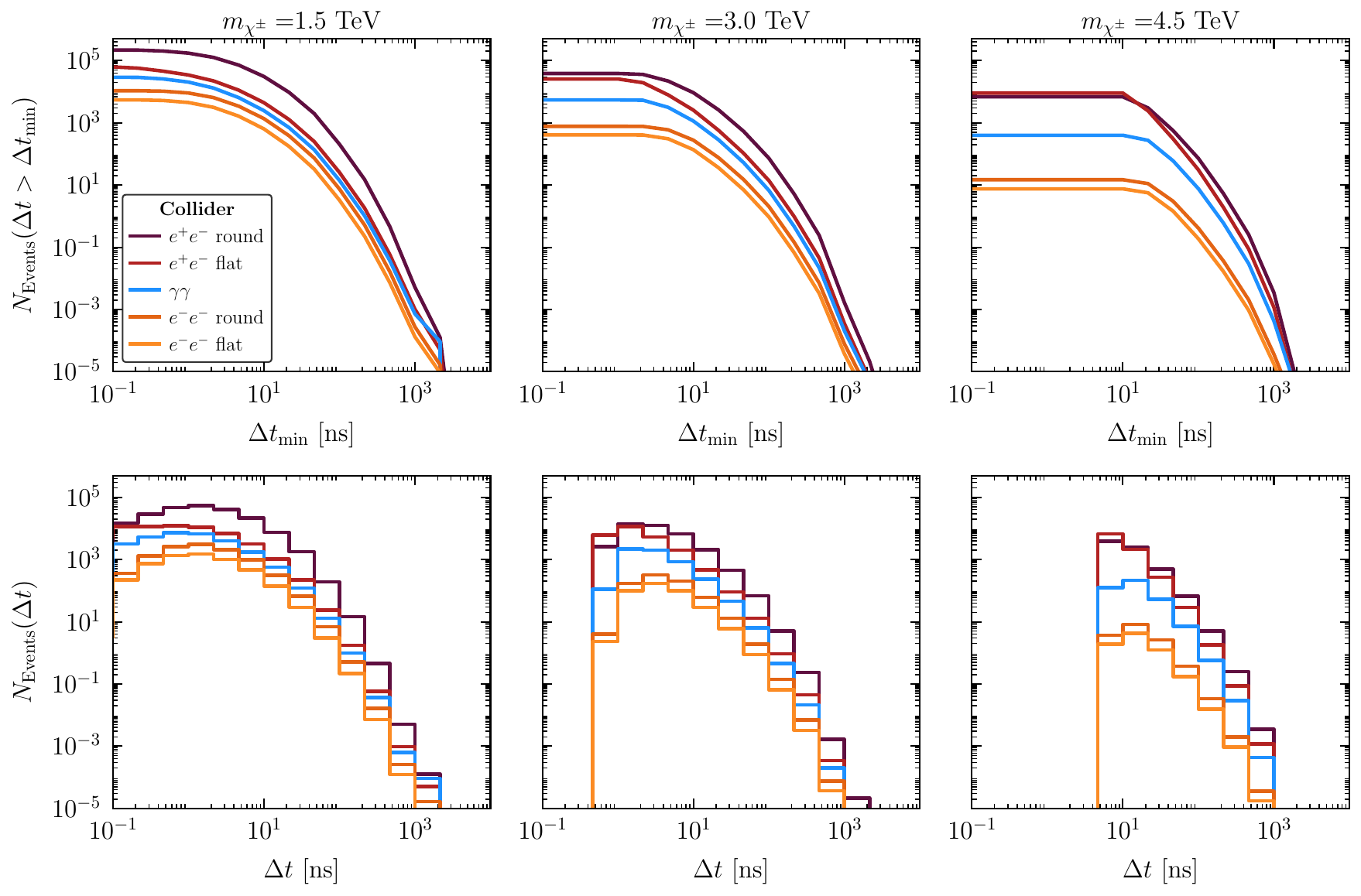}
    \caption{Number of events as a function of time delay $\Delta t$ for $\chi^\pm$ that are pair-produced and do not decay within the detector for a geometric luminosity of $\SI{1}{\per\atto\barn}$. Plotted are masses $m_{\chi^\pm} = \SI{1.5}{\TeV}$ (left), $m_{\chi^\pm} = \SI{3}{\TeV}$ (middle), $m_{\chi^\pm} = \SI{4.5}{\TeV}$ (right). Top: the number of events with time delay longer than $\Delta t_{\rm min}$, bottom: number of events with a time delay $\Delta t$.}
    \label{fig:HCSP}
\end{figure}

The distribution of time delays $\Delta t$ and the number of events with time delays larger than $\Delta t_{\rm min}$ are plotted in \cref{fig:HCSP} for the triplet model, assuming a geometric luminosity of $\SI{1}{\rm{ab}^{-1}}$. We only include events in which two charged $\chi$ particles are produced and reach the timing detector. 
The only additional cut applied for this analysis is $|{\eta}|\leq 3$ on both $\chi^\pm$. 

\ATLAS{Does the paragraph below seem reasonable from an experimental point of view?}
We see that most of the signal events arrive with a delay $\Delta t \gtrsim 1$ ns. 
A timing detector with resolution $\mathcal{O}(\text{ns})$ should therefore suffice to reject most backgrounds while maintaining an $\mathcal{O}(1)$ signal efficiency.
This implicitly assumes that the backgrounds are gaussian to a very high degree in the timing resolution.
At the LHC, this assumption would not be justified, given the out-of-time contributions from beam-induced backgrounds and subsequent events \cite{CMS:2019qjk}. 
The repetition rate of a wakefield collider, however, would only be $\mathcal{O}(\SI{10}{\kilo\Hz})$, as compared to the \SI{40}{\mega\Hz} collisions at the LHC.
In other words, out-of-time backgrounds from neighboring collisions are simply not a factor at a plasma wakefield collider, and the gaussian assumption appears to be justified at this stage.
In summary, we assume that with a suitable timing detector it will be possible to reject nearly all backgrounds at modest cost in signal efficiency.
In our summary plots in \cref{sec:discuss}, we will therefore assume that 10 signal events suffices to claim a discovery. 

\subsection{Hadronic decays\label{sec:decaytojets}}
Finally, for completeness we consider the scenario where the $\chi^0$ decays to jets instead of being invisible. 
This occurs in supersymmetric models with baryon-number-violating R-parity violation (RPV) via the $\lambda''_{ijk}\bar{U}_i\bar{D}_j \bar{D}_k$ operator in the superpotential. In particular, by allowing a nonzero $\lambda''_{ijk}$ coupling (with indices chosen to satisfy flavor constraints \cite{Dercks:2017lfq}, e.g., $\lambda''_{112}$ or similar), the $\chi^0$, 
which is assumed to be lighter than the $\chi^\pm$, can decay promptly into three quarks rather than contributing to missing energy. For example, the full decay chain may be
\begin{align}
\chi^+\chi^- \to (\chi^0 + \pi^+)(\chi^0 + \pi^-) \to \pi^++\pi^- + 6 \;\text{jets}\,.
\end{align}
Depending on $m_{\chi^\pm}-m_{\chi^0}$, the $\pi^\pm$ may be replaced with $W^\pm$. 
Either way, the key point is that these scenarios naturally yield six or more jets in the final state. However, reliably estimating the \ac{SM} background for$e^+e^-\to 6$ jets is challenging. 
Instead, we calculate the rate for four jets with Madgraph, which provides a very conservative upper bound on the rate for the 6 jets final state.
A $p_T$ cut of a few hundred GeV is already sufficient to suppress the 4 jet rate well below the $\chi^\pm$ pair-production rate for $m_{\chi^\pm} = \SI{2.5}{\TeV}$. Consequently, the combination of RPV-induced fully visible decays plus stringent kinematic cuts provides a clean avenue for probing both the triplet and doublet in plasma-wakefield accelerator experiments

\section{Discussion}
\label{sec:discuss}

For brevity, we present only the triplet-model summary results when comparing the different wakefield collider configurations, while our comparison of wakefield colliders with other future collider options covers both the doublet and triplet models.

\ATAP{We can certainly discuss how much of the conclusions should be here, vs reserved for the short paper that we all write together. Feedback welcome, we want everyone to be satisfied. }

\ATLAS{We can certainly discuss how much of the conclusions should be here, vs reserved for the short paper that we all write together. Feedback welcome, we want everyone to be satisfied. }

\subsection{Comparing wakefield colliders}
\begin{figure}
    \centering
    \includegraphics[width=\textwidth]{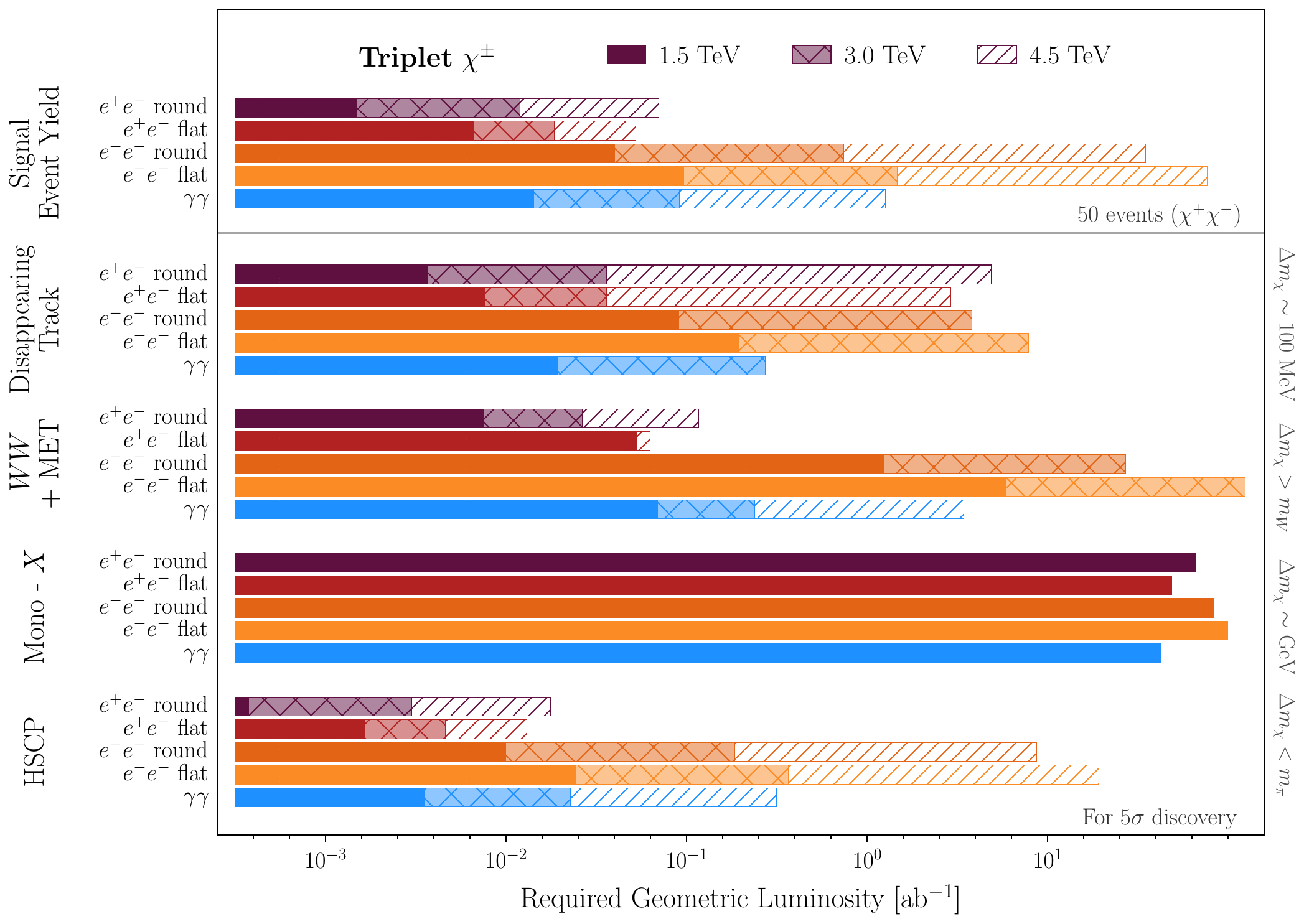}
    \caption{The amount of integrated geometric luminosity necessary for each of the collider options studied in this work to discover charged triplet particles with masses $1.5, 3, \SI{4.5}{\TeV}$. We show the luminosity needed for pair production of 50 $\chi^+\chi^-$ events (via Drell-Yan and photon fusion), and subsequently the requirements for each of the strategies discussed above. For the \ac{HSCP} and disappearing track signatures we assume the searches are effectively background free and 10 events are sufficient for a discovery; in the latter case the results are  quoted for the pure triplet with $\Delta m_\chi = \SI{164}{\MeV}$. No bar was included if the required geometric luminosity exceeds $\sim\!\SI{100}{\per\atto\barn}$. 
    }
    \label{fig:barchart_lumi}
\end{figure}

In \cref{fig:barchart_lumi} we compare the geometric luminosity that is needed to make a discovery in each of the five wakefield colliders.
This is \emph{not} meant to imply that the collider that requires the lowest luminosity is the ``best'' collider, since all options face very different obstacles on the accelerator side (see \cref{sec:wakefield}).
Rather, \cref{fig:barchart_lumi} is meant to inform the accelerator community about what specifications each collider should achieve to discover multi-TeV electroweak particles. 
This is intended as input to the R\&D decision making process, alongside the relevant technological challenges.

The uppermost panel of Fig.~\ref{fig:barchart_lumi} shows the luminosity required to produce 50 $\chi^+\chi^-$ pairs for three different benchmark mass points $m_{\chi^\pm}=1.5, 3, \SI{4.5}{\TeV}$, while the lower panels show the required geometric luminosities for a $5\sigma$ discovery for the searches discussed in Sections~\ref{sec:disappearingtrack},~\ref{sec:WWmet},~\ref{sec:invisible} and~\ref{sec:HCSP}, together with the $m_{\chi^\pm}-m_{\chi^0}$ mass splitting for which they are valid.  
For the mono-$e$ and $WW+$MET searches, the $\chi^\pm$ are assumed to decay promptly, while for the \ac{HSCP} search its proper decay length is assumed to exceed the size of the detector.
For the disappearing track projections the lifetime  
assumption is appropriate for the pure triplet, through \eqref{eq:ctautriplet}.  

Since $\chi^\pm$ pair production at these colliders proceeds predominantly via Drell–Yan and only subdominantly through photon fusion (see \cref{fig:totalrate}), it is unsurprising that the $e^+e^-$ configurations are more efficient in terms of signal event yield. By contrast, the $e^-e^-$ options must rely on secondary particles in the initial state and therefore require a higher geometric luminosity to achieve a comparable number of signal events.
On this metric, the photon collider outperforms both $e^-e^-$ configurations  but underperforms against both $e^+e^-$ options. 
Though the detailed numbers depend on the beam geometry and $m_{\chi^{\pm}}$, in general a factor of $\sim$10 (100) times more geometric luminosity is needed for a potential $\gamma\gamma$ ($e^-e^-$) collider to achieve a comparable signal yield to the $e^+ e^-$ counterpart, with this factor becoming more severe for larger $m_{\chi^{\pm}}$. 
At parton-level, $e^+e^- \to \chi^+\chi^-$ and $\gamma\gamma \to \chi^+\chi^-$ cross sections are comparable. However, the Compton backscattering of the laser light in the photon collider is only about $\sim10\%$ efficient at converting $e^-$ energy into high energy $\gamma$'s \cite{acceleratorpaper}.
This is why the $\gamma\gamma$ luminosity in the $\gamma\gamma$ collider is lower than the $e^+e^-$ luminosity in the two $e^+e^-$ configurations (see \cref{fig:initial_state_lumis}), which explains the lower signal event yield for the same geometric luminosity.
Given the challenges of positron acceleration, it is however plausible that in a fully realistic wakefield accelerator design, the signal yield would be highest in a $\gamma\gamma$ collider.

Both the Drell-Yan and photon fusion cross sections are largest for $M\sim 2 m_{\chi^\pm}$.
The additional beamstrahlung for round $e^+e^-$ beams moreover provides a higher luminosity spectrum for any $M$ below the maximum \ac{COM} energy of the collider (see \cref{tab:beam-parameters}).
Together, these effects imply higher signal yields for round $e^+e^-$ beams than for flat beams, except when $2m_{\chi^\pm}\approx \SI{10}{\TeV}$.
 
That the particle production occurs preferentially on threshold contributes further to the lower signal yield of $e^-e^-$ beams compared to their  $e^+e^-$ counterparts, as the anti-pinch leads to a larger fraction of the luminosity remaining in the endpoint. Within $e^-e^-$ colliders, the signal event yield is in fact set by the beamstrahlung-generated secondary $e^+$ flux, so round beams systematically outperform flat ones.

To get a more complete picture, we must compare the discovery potential of the colliders rather than the signal yield. Starting with the low background \ac{HSCP} and disappearing track channels, we find the same qualitative patterns as in the raw signal yields. For the disappearing track, the signal reconstruction efficiency is not 100\%, which is why additional luminosity is needed to claim a discovery (see \cref{sec:disappearingtrack} for details). 
For the triplet model, the $m_{\chi^\pm}=\SI{3}{\TeV}$ benchmark can be discovered with \SI{10}{\per\atto\barn} for all machines considered, though once again an order of magnitude separates the required luminosity of $e^+e^-$ and $\gamma\gamma$, and $\gamma\gamma$ and $e^-e^-$ options, respectively. 
For the higher $m_{\chi^\pm}=\SI{4.5}{\TeV}$ benchmark, the number of particles produced with a long enough lab frame decay-length to yield a disappearing track is quite sensitive to the ($e^+e^-$ or $\gamma\gamma$ initial state) luminosity near the peak \ac{COM} energy. The required geometric luminosity for discovery is $\sim\!\SI{4}{\per\atto\barn}$ for an $e^+e^-$ collider, $\sim\!\SI{200}{\per\atto\barn}$ for a photon collider and as much as $\sim\!\SI{e+4}{\per\atto\barn}$ for the $e^-e^-$ colliders. This is four orders of magnitude more than needed by $e^+e^-$, and is decidedly out of reach. It is therefore not shown in \cref{fig:barchart_lumi}.
That said, we emphasize that for such high $m_{\chi^\pm}$ the discovery reach is exponentially sensitive to the detector design (see \cref{fig:ell_T4}). Rather than aiming for very high luminosities, this parameter space may be best accessed with a more ambitious detector design than what we have assumed here.

In the $WW$+MET analysis in \cref{sec:WWmet}, we found very large backgrounds, especially for round beams compared to flat beams.
However, these backgrounds can be suppressed effectively with suitable analysis cuts, at least for the $e^+e^-$ and $\gamma\gamma$ colliders. 
A $5\sigma$ discovery of the $m_{\chi^\pm}=\SI{4.5}{\TeV}$ benchmark is within reach for $e^+e^-$ and $\gamma\gamma$ machines, assuming geometric luminosities of respectively $\sim\!\!\SI{1}{\per\atto\barn}$ and $\sim\!\!\SI{10}{\per\atto\barn}$.
The round $e^+e^-$ beams outperform the flat $e^+e^-$ beams for $m_{\chi^\pm}=$ \SI{1.5}{\TeV} and $m_{\chi^\pm}=\SI{3}{\TeV}$, and both are comparable for $m_{\chi^\pm}=\SI{4.5}{\TeV}$ (see \cref{tab:WWMETsignificance} for details).
For the $e^-e^-$ machines, the $m_{\chi^\pm}=\SI{3}{\TeV}$ benchmark is plausibly discoverable if a large amount of luminosity can be collected, but the discovery of a $m_{\chi^\pm}=\SI{4.5}{\TeV}$ particle would require $\mathcal{O}(\SI{e+4}{\per\atto\barn})$. Similarly to the disappearing tracks search, since such a high luminosity is likely unattainable, the $m_{\chi^\pm}=\SI{4.5}{\TeV}$ scenario is not shown in Fig.~\ref{fig:barchart_lumi} for $e^- e^-$ beams.

Finally, in the mono-$e$ analysis the five options perform similarly. This is perhaps surprising, given that this analysis suffers from the most backgrounds, and one might have expected a similar pattern to the $WW$+MET analysis.
The reason is that the primary production channel, $e^\pm \gamma \to e^\pm \chi^\pm \chi^\mp$ does not favor $e^+$ over $e^-$, and as such there is no significant advantage for the $e^+e^-$ colliders. In this case, the discovery reach is limited to the 1-\SI{1.5}{\TeV} range for all colliders. This is however comparable with or better than the reach of other future colliders, as we will see in the next section.

\subsection{Comparison with other 10~TeV pCM colliders}
\begin{figure}
    \centering
    \includegraphics[width=\textwidth]{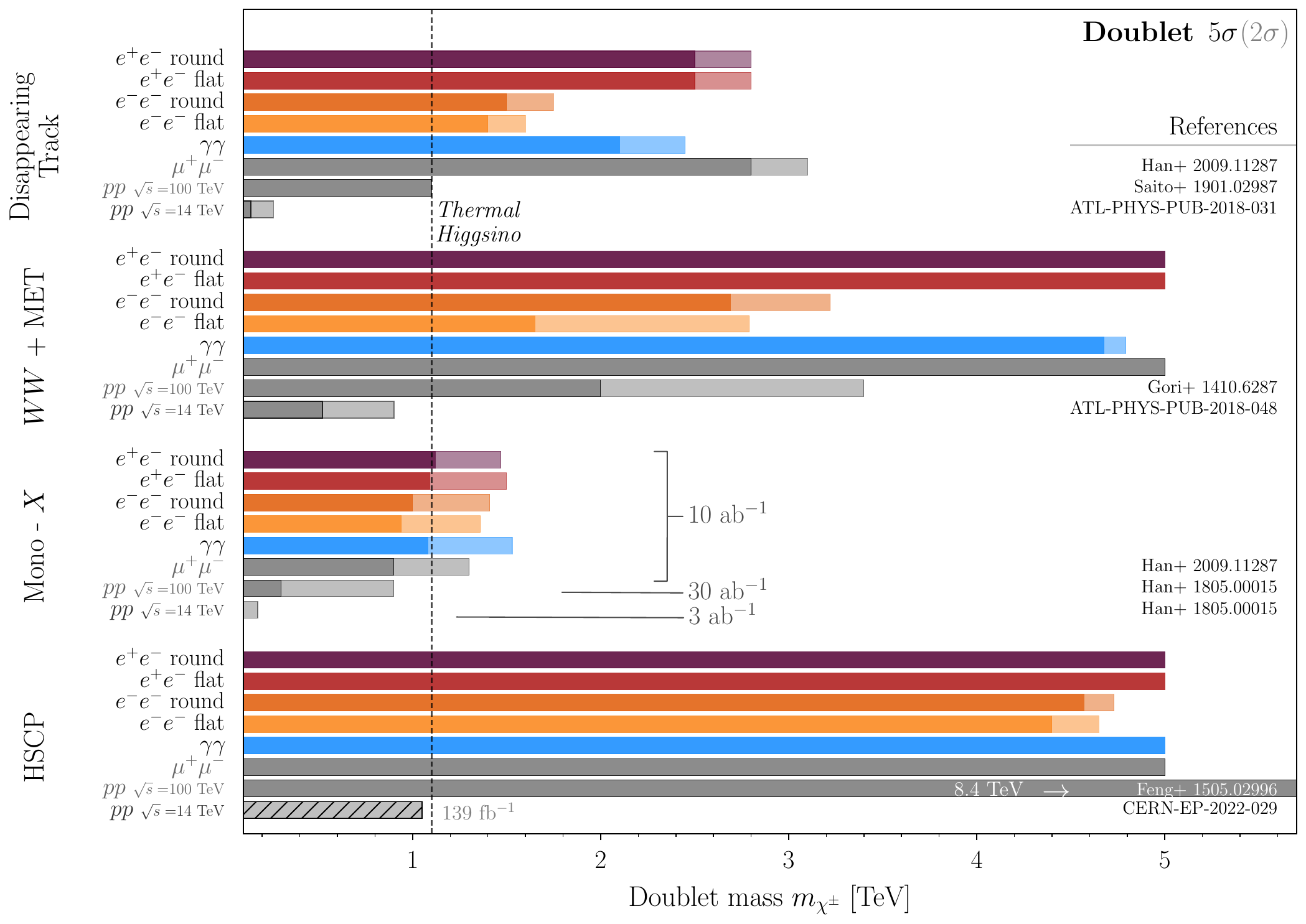}
    \caption{The comparative 5$\sigma$ discovery potential of key \SI{10}{\TeV} \ac{pCM} collider options for a heavy electroweak doublet. The colored bands represent the 5 wakefield beam options studied in this work, compared with a \SI{10}{\TeV} muon collider and \SI{100}{\TeV} $pp$ collider; shown also is the projected reach of HL-LHC. Where literature results are  available and the kinematic limit is not saturated, the projected 2$\sigma$ exclusion projections are shown in lighter colors. We assume a total luminosity of \SI{10}{\per\atto\barn} is collected for the lepton and photon colliders, \SI{30}{\per\atto\barn} for the $pp$ choice, and \SI{3}{\per\atto\barn} for HL-LHC; the exception is indicated with a hatched texture, where high luminosity projections could not be found.   
    The vertical line corresponds to the thermal higgsino prediction, which can be discovered with $e^+ e^-$ and $\gamma\gamma$ wakefield machines in all channels. Caution should be used in the interpretation of this figure, as referenced works make different assumptions about detector configuration, background control, and underlying UV physics. For example, the discovery potential for the disappearing track signature is more sensitive to the assumed detector configuration than to the luminosity of the collider. See text for further details. 
    }
    \label{fig:barchart_mass_higgsino}
\end{figure}

\begin{figure}
    \centering
    \includegraphics[width=\textwidth]{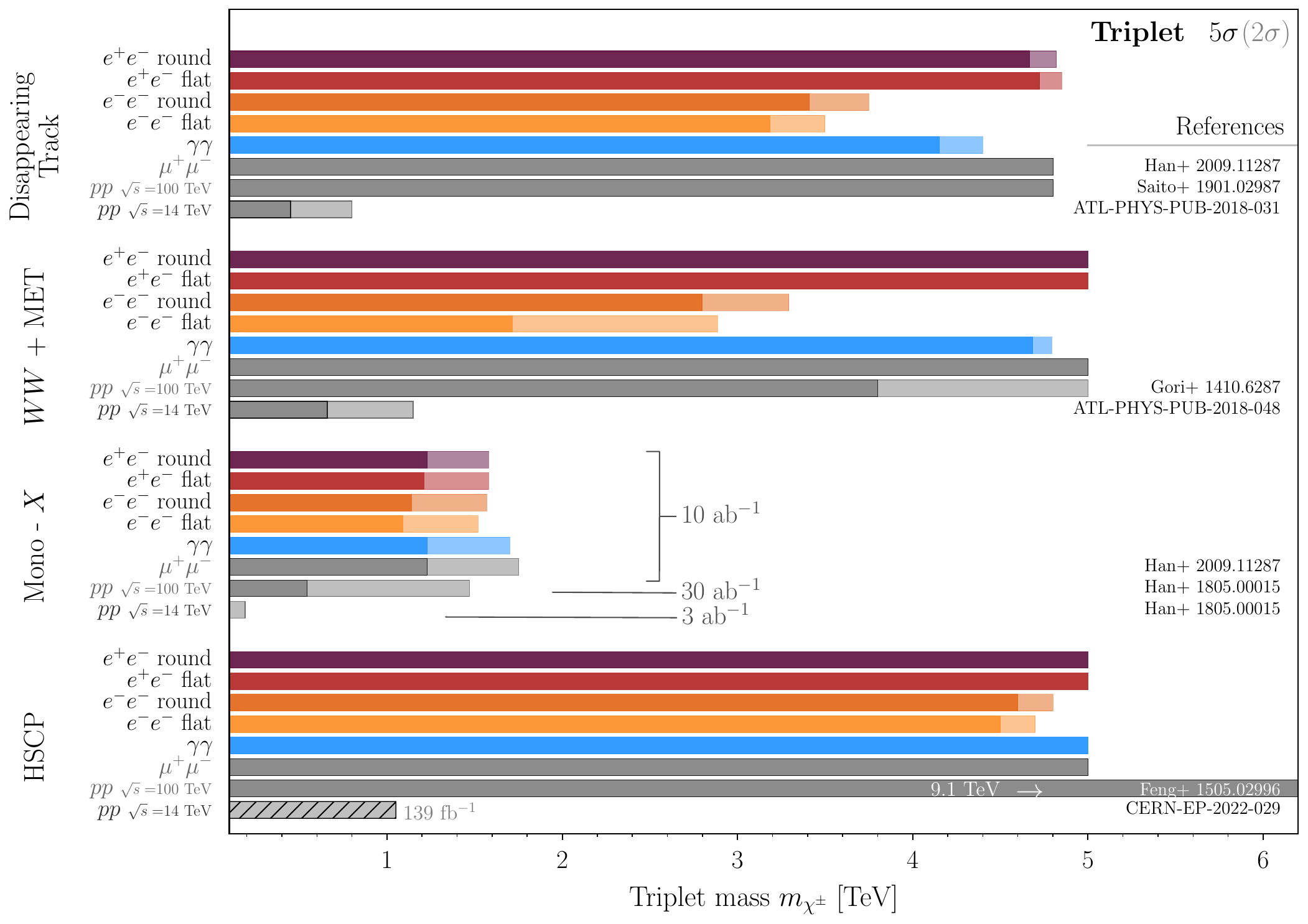}
    \caption{ 
    Similar to \cref{fig:barchart_mass_higgsino} for the electroweak triplet.}
    \label{fig:barchart_mass_wino}
\end{figure}

To complement \cref{fig:barchart_lumi} we fix the geometric luminosity to a benchmark value and study the discovery potential as a function of $m_{\chi^\pm}$.
This allows us to compare wakefield colliders directly with proposed future muon and hadron machines.
Any such comparison should be treated with caution, since it relies on assumptions about energy and luminosity that ultimately depend on how these accelerator R\&D programs evolve.
In \cref{fig:barchart_mass_higgsino} and \cref{fig:barchart_mass_wino} (for the doublet and triplet models, respectively) we assume an integrated geometric luminosity of $\SI{10}{\per\atto\barn}$ for all wakefield colliders, to be compared with a \SI{10}{\TeV} muon collider (\SI{10}{\per\atto\barn}) and a \SI{100}{\TeV} $pp$ collider (\SI{30}{\per\atto\barn}).\footnote{The recent design update for the FCC-hh has revised its expected energy to 80-\SI{90}{\TeV} \cite{Benedikt:2928941}. Most studies in the literature still assume \SI{100}{\TeV}, which we adopt here for consistency.} We also include the reach of the HL-LHC, assuming a luminosity of \SI{3}{\per\atto\barn}.

Starting with the \SI{100}{\TeV} $pp$-collider, the sensitivity of a disappearing track search is strongly dependent on how close to the \ac{IP} one is able to place tracking layers.  Ref.~\cite{Saito:2019rtg} showed that the default configuration \cite{Mangano:2022ukr} has limited discovery potential, in particular being unable to probe the \SI{1}{\TeV} doublet benchmark. A more optimal detector configuration has however been proposed in~\cite{Saito:2019rtg}; we use the corresponding improved discovery potential in \cref{fig:barchart_mass_higgsino} and \cref{fig:barchart_mass_wino}.  
In a \SI{100}{\TeV} collider this design requirement must be balanced with the expected radiation damage to the tracking layers, which is still an ongoing subject of study. \ATLAS{Anything more specific we should say here?}

To carry out the analogue of the $WW$+MET analysis in \cref{sec:WWmet} at a high energy hadron collider, one must rely on the leptonic branching ratios of the $W$ and $Z$ bosons. This produces a distinctive 3 lepton signature\footnote{This signature assumes the presence of an additional neutral particle close to the $m_{\chi^\pm}$ scale, which decays to the lighter neutral state via $Z$ emission. For details on the model see the references cited.} \cite{Acharya:2014pua,Gori:2014oua}, but reduces the overall signal rate due to the relatively small leptonic branching ratios of the $W$ and $Z$.
The reach is also very sensitive to $m_{\chi^\pm}-m_{\chi^0}$, and a moderate amount of compression in the spectrum is sufficient to hide the $\chi^\pm$.
For these models, the \SI{100}{\TeV} $pp$-collider's discovery potential is also very sensitive to additional assumptions about the ultraviolet physics \cite{Gori:2014oua}. 
Assuming the most optimistic parameter choices in \cite{Gori:2014oua}, the triplet (doublet) model can be discovered with \SI{30}{\per\atto\barn}
if $m_{\chi^\pm}\lesssim \SI{3.8}{\TeV}$ ($m_{\chi^\pm}\lesssim \SI{2}{\TeV}$), assuming an uncompressed spectrum.\footnote{The original study in \cite{Gori:2014oua} assumed a luminosity of \SI{3}{\per\atto\barn}, which we rescaled to \SI{30}{\per\atto\barn}, assuming an improvement on the cross section reach of $\sim\sqrt{\SI{3}{\per\atto\barn}/\SI{30}{\per\atto\barn}}$.
We use the cross section computation in \cite{Acharya:2014pua} to convert the reach on the rescaled cross sections to a reach in terms of $m_{\chi^\pm}$.}  
If the spectrum is compressed and $\chi^\pm$ is effectively invisible in the detector, a \SI{100}{\TeV} $pp$-collider would rely on the mono-jet signature \cite{Low:2014cba,Cirelli:2014dsa,Han:2018wus} as the counterpart of the mono-$e$ signature in \cref{sec:invisible}.
As our comparison point, we use the projections in \cite{Han:2018wus} that assume a 2\% systematic uncertainty on the \ac{SM} background. The discovery reach would then extend to $m_{\chi^\pm}\leq \SI{545}{\GeV}$ ($m_{\chi^\pm}\leq \SI{199}{\GeV}$) for the triplet (doublet) model. 
The discovery potential for the \ac{HSCP} signature was calculated for slepton models, assuming $\SI{3}{\per\atto\barn}$~\cite{Feng:2015wqa}. 
We rescale this result to $\SI{30}{\per\atto\barn}$ assuming negligible backgrounds and correct for the higher cross sections of a fermionic triplet and doublet, using the calculations in~\cite{Acharya:2014pua}.
With these assumptions, \SI{100}{\TeV} $pp$ collider can discover the triplet (doublet) model up to $m_{\chi^\pm}\lesssim \SI{9.1}{\TeV}$ ($m_{\chi^\pm}\lesssim \SI{8.4}{\TeV}$)

To compare with the performance of a muon collider, we assume a \SI{10}{\TeV} machine with \SI{10}{\per\atto\barn} of integrated luminosity. 
For its discovery potential in the disappearing track search we follow \cite{Han:2020uak}. Assuming one reconstructed track and 10 events needed for discovery, it was found that a muon collider could discover the triplet (doublet) model for $m_{\chi^\pm} \leq \SI{4.8}{\TeV}\,(\SI{2.8}{\TeV})$.
Similar to the $e^+e^-$ colliders, any muon collider with moderately high luminosity should be able to probe the $WW$+MET signature all the way to the kinematic limit of the collider ($m_{\chi^\pm} = \SI{5}{\TeV}$). The discovery potential of a stable $\chi^\pm$ should likewise almost saturate this limit.
Finally, the mono-$\mu$ channel dominates the sensitivity if the $\chi^\pm$ are effectively invisible, with a discovery reach of \SI{1.2}{\TeV} (\SI{0.9}{\TeV}) for the triplet (doublet) model~\cite{Han:2020uak}.

Lastly, we show the expected reach of collecting \SI{3}{\per\atto\barn} of luminosity at the LHC. 
The HL-LHC will be able to exclude the disappearing track signature up to $m_{\chi^\pm}\leq \SI{0.85}{\TeV}$ and $m_{\chi^\pm}\leq \SI{0.26}{\TeV}$ for the triplet and doublet models, respectively \cite{ATLAS:2018jjf}.
The projected exclusion limits for the analogue of the $WW$+MET signature at HL-LHC are subject to the same caveats as at the \SI{100}{\TeV} $pp$ collider discussed above. 
The expected exclusion reach is $m_{\chi^\pm}\leq\SI{1.15}{\TeV}$ and $m_{\chi^\pm}\leq \SI{0.9}{\TeV}$ for the uncompressed triplet and doublet models respectively \cite{ATLAS:2018diz}.\footnote{The doublet case was obtained by rescaling the bound on triplet with the respectively cross sections.}
For scenarios where the mass spectrum is compressed, the monojet search is expected to exclude triplet (doublet) masses of $m_{\chi^\pm} \leq \SI{191}{\GeV}$ (\SI{175}{\GeV}), again assuming 2\% uncertainties on the background~\cite{Han:2018wus}. 
To our knowledge, there is currently no public projection for the \ac{HSCP} signature for the HL-LHC. We therefore instead show the existing LHC exclusion bound for this case \cite{ATLAS:2022pib}.   

To conclude this section, we summarize the main lessons from \cref{fig:barchart_mass_higgsino} and \cref{fig:barchart_mass_wino}. 
An $e^+e^-$ or $\gamma\gamma$ wakefield collider, irrespective of whether the beams are round or flat, achieves a similar reach to an equal–energy muon collider for charged electroweak states in all the channels we studied.
The $e^-e^-$ configurations are less favorable, but can still  probe parameter space well beyond the HL-LHC if a geometric luminosity $\sim\!\SI{10}{\per\atto\barn}$ can be achieved.
The $e^+e^-$ and $\gamma\gamma$ colliders perform similarly or better than a \SI{100}{\TeV} $pp$-collider in all channels except in the low-background HSCP scenario. 
The analogue of the $WW$+MET signature at a \SI{100}{\TeV} $pp$-collider is moreover subject to additional model-dependent assumptions, and alternative assumptions could significantly reduce the sensitivity compared to those discussed above. 

Of particular interest is the difficult-to-access case of the thermal higgsino at $m_{\chi^\pm} = \SI{1.08}{\TeV}$, shown with a dashed line in \cref{fig:barchart_mass_higgsino}.
The $e^+e^-$ and $\gamma\gamma$ colliders could discover or exclude this important case in both the disappearing track and the mono-$e$ channel, 
while the $e^-e^-$ colliders could discover or exclude it in the disappearing track channel and exclude it with mono-$e$ channel.
A disappearing track search at a \SI{100}{\TeV} $pp$ collider could discover this scenario as well, but a more aggressive detector design relative to the baseline design would be needed~\cite{Saito:2019rtg}.
This is because the lifetime of $\chi^\pm$ component of the doublet is expected to be rather short, and thus one needs to construct tracking layers as close as possible to the \ac{IP}. 
The radiation environment in a lepton collider is expected to be less severe in comparison, and additional small-radius tracking layers may therefore be less challenging to realize. 
The mono-jet search at a \SI{100}{\TeV} $pp$ collider could likely not discover this case, though it could exclude it if it can achieve systematic uncertainties on the background model at or below the 1\% level \cite{Han:2018wus}.

\section{Conclusions}
\label{sec:conclusion}
\ATAP{We can certainly discuss how much of the conclusions should be here, vs reserved for the short paper that we all write together. Feedback welcome, we want everyone to be satisfied. }

\ATLAS{We can certainly discuss how much of the conclusions should be here, vs reserved for the short paper that we all write together. Feedback welcome, we want everyone to be satisfied. }

We combined preliminary accelerator optimization \cite{acceleratorpaper} with detailed beam--beam simulations \cite{simulationpaper} to assess the discovery potential of a \SI{10}{\TeV} plasma wakefield collider for multi-TeV electroweak states.
We compared five configurations: $e^+e^-$ and $e^-e^-$ machines with round and flat beams, and a $\gamma\gamma$ option based on laser Compton backscattering (see \cref{tab:beam-parameters}). 
Using realistic luminosity spectra, we evaluated the geometric luminosity required for discovery across representative electroweak multiplet scenarios and decay patterns, and summarized the resulting requirements in \cref{fig:barchart_lumi}.

A central message of this study is that beam--beam effects do not undermine the electroweak physics case at multi-TeV plasma wakefield colliders.
For the benchmarks considered here, production is often dominated by near-threshold kinematics, so the broadened luminosity spectrum characteristic of a round-beam $e^+e^-$ collider can be neutral or even beneficial.
Consequently, we find no clear, universal physics advantage of a flat-beam $e^+e^-$ collider over a round-beam collider for heavy electroweak pair production.
This shows that achieving robust reach for electroweak particles does not require solving the full flat-beam challenge before a compelling energy-frontier program becomes viable.

The $e^-e^-$ configurations remain interesting but are systematically less effective, since the dominant production modes are only accessible through beam-induced secondary initial states.
For several signatures closing this gap would require unrealistically large geometric luminosities (notably in high-background channels; see \cref{tab:WWMETsignificance} and \cref{fig:barchart_lumi}).
By contrast, the $\gamma\gamma$ option always outperforms both $e^-e^-$ colliders and offers a credible alternative pathway if positron acceleration proves too challenging at the multi-TeV scale.  
In view of the different technological challenges across configurations, our results help quantify which accelerator trade-offs may be acceptable while preserving decisive sensitivity to key electroweak targets.

Overall, we find that the $e^+e^-$ and $\gamma\gamma$ colliders perform qualitatively similar to a muon collider with equal energy and  luminosity, and demonstrate strong discovery potential for new multi-TeV electroweak states. 
Of particular note is that both configurations are expected to be sensitive to the thermal higgsino benchmark. This remains true even in the most pessimistic invisible decay scenario, if a geometric luminosity of $\sim\!\SI{10}{\per\atto\barn}$ can be achieved. In this case, wakefield colliders outperform both the \SI{10}{\TeV} muon and \SI{100}{\TeV} proton collider options.

More broadly, this work highlights the need for sustained, tightly coupled progress across accelerator design, beam--beam modeling, detector concepts, and theory. 
Realistic luminosity spectra and beam-induced initial states qualitatively reshape search strategies compared to idealized lepton-collider assumptions, and must therefore be treated as core inputs to both physics projections and R\&D prioritization. 
The analysis presented here is designed to be updated as the wakefield design study matures, and forthcoming companion papers will address additional physics targets, the evolving accelerator parameter space, improved simulations, and detector-driven refinements of the search program.

\acknowledgments
This work would not have been possible without close collaboration with members of the LBNL Accelerator Technology \& Applied Physics Division and the LBNL ATLAS group.
In particular, we are indebted to Stepan Bulanov, Arianna Formenti, Remi Lehe,  Jens Osterhoff, Carl Schroeder and Jean-Luc Vay for providing their preliminary results on the accelerator parameters and beam--beam simulations.
We thank Simone Pagan Griso and Angira Rastogi for their input on the detector-related assumptions. 
We are also grateful to all of the above for comments on the manuscript and the regular discussions which clarified important concepts for us and helped focus and refine our goals and conclusions. 
We further thank Daniel Downey for helping us make progress towards a public-facing version of our analysis code.
In addition, we are thankful to 
Wolfgang~Altmanshofer, 
Tim~Barklow, 
Innes~Bigaran, 
Massimo~Cipressi, 
Nathaniel~Craig, 
Katie~Fraser, 
Spencer~Gessner, 
Kevin~Langhoff, 
Zoltan~Ligeti, 
David~Marzocca, 
Pankaj~Munbodh, 
Michael~Peskin, 
Niki\v{s}a~Ple\v{s}ec, 
Nick~Rodd, 
Robert Szafron, 
and Jure~Zupan
for useful comments, discussions and/or collaboration on related work.
The work of SC is supported by the Simons Foundation.
The work of SC is supported by the U.S. Department of Energy, Office of Science, the BNL C2QA award under grant Contract Number DESC0012704 (SUBK\#390034).
The work of SK, CS and IS is supported by the Office of High Energy Physics of the U.S.\ Department of Energy under contract DE-AC02-05CH11231. IS also acknowledges support by the Weizmann Institute of Science Women’s Postdoctoral Career Development Award. The work of WLX was supported by the Kavli
Institute for Particle Astrophysics and Cosmology, and
by the U.S. Department of Energy under contract DE-AC02-76SF00515. In addition this work was performed in part at
the Aspen Center for Physics, which is supported by National Science Foundation grant PHY-2210452. 
This research used resources of the National Energy Research Scientific Computing Center (NERSC), a Department of Energy Office of Science User Facility using NERSC award HEP-ERCAP0031191.

\appendix
\section{Effect of parton distribution functions}
\label{app:pdf}

Throughout this work, we have assumed that beam--beam interactions are the dominant source of lower-energy secondary particles. In this appendix we compute the contribution from EW parton distribution functions (PDF) and quantify their size relative to beam--beam effects.
To distinguish the PDF contributions to the luminosity spectra introduced in \cref{sec:wakefield}, we will refer to the latter as the ``machine luminosity spectra'' in this appendix.
In complete generality, under collinear factorization and assuming factorization with the beam--beam dynamics, the PDF contributions to the effective luminosities are given by
\begin{align}\label{eq:pdf-general}
    \frac{\diff \mathscr{L}_{\text{int},\,ab}^\text{PDF}}{\diff M} &= \sum_{a^\prime b^\prime}\bigg[\int_M^{\sqrt{s}} \diff M^\prime   \frac{2 M}{(M^\prime)^2} \int_\tau^1  \frac{\diff x}{x}f_{a^\prime\to a}(x,Q)f_{b^\prime\to b}(\tau/x,Q)\frac{\diff \mathscr{L}_{\text{int},\,a^\prime b^\prime }}{\diff M^\prime} \\  &  \hspace{8.5cm} + (a^\prime \leftrightarrow b^\prime \text{ if $a^\prime \neq b^\prime$})\bigg] \notag\,,
\end{align}
where $f_{a^\prime\to a}(x,Q)$ is the PDF for the particle $a$ with energy fraction $x = E/M^\prime$ in a beam of $a^\prime$ particles. The sum over $a^\prime b^\prime$ indicates all possible contributions leading to the desired state $ab$. The luminosity of the primary and secondary beam particles at the \ac{IP} is described by the integrated luminosity $\diff \mathscr{L}_{\text{int},\,a^\prime b^\prime }/\diff M^\prime$, see \cref{subsec:beam-parameters-lumi}. In the above $\tau \equiv (M/M^\prime)^2$ and we choose the scale $Q = M/2$. Note that under the replacement
\begin{align}
    \frac{\diff \mathscr{L}_{\text{int},\,a^\prime b^\prime }}{\diff M^\prime} \to \mathscr{L}_\text{geom}  \delta(M^\prime - \sqrt{s})\,,
\end{align}
\cref{eq:pdf-general} reduces to the case of a standard monochromatic-energy collider with beams of particles $a^\prime b^\prime$.
In the evaluation of \cref{eq:pdf-general} we make the following approximations:
\begin{enumerate}
    \item We neglect contributions from secondary beam species in $\diff\mathscr{L}_{\text{int},\,a'b'}/\diff M'$ and include only the machine’s primary beams $a'b'\in\{e^+e^-,\,e^-e^-\}$, which depend on the collider configuration.  
    \item We use a bin integrated expression for the machine luminosity spectra, namely
    \begin{align}
        A_j^{a^\prime b^\prime} = \int_{M_j - \Delta M/2}^{M_j + \Delta M/2}  \diff M^\prime \frac{\diff \mathscr{L}_{\text{int},\,a^\prime b^\prime }}{\diff M^\prime}\,,
    \end{align}
    with bin centers $M_i$ and bin-widths $\Delta M = \SI{50}{\GeV}$. Consequently we also show the resulting PDF luminosities on the same grid. 
\end{enumerate}
Under these assumptions, the PDF–induced contribution to the effective luminosity in the destination bin centered at $M_i$ reads
\begin{align}\label{eq:pdf-binned}
  \left.\frac{\diff \mathscr{L}_{\text{int},\,ab}^\text{PDF}}{\diff M}  \right|_{M_i} &\simeq \sum_j A_j^{a^\prime b^\prime} \frac{2 M_i}{M_j^2} \int_{\tau_{ij}}^1 \frac{\diff x}{x} f_{a^\prime\to a}(x,M_i/2)f_{b^\prime\to b}(\tau_{ij}/x,M_i/2) \,,
\end{align}
where $\tau_{ij} \equiv (M_i/M_j)^2$. We now turn to the structure of the \acp{PDF}. The lepton \acp{PDF} have a pronounced endpoint at $x\to1$. In pure QED, the DGLAP solution can be organized as (see Eq.~(2.4) of Ref.~\cite{Garosi:2023bvq})
\begin{align}
  f_{e^\pm}(x,Q) = f_{e^\pm}^{\text{cont}}(x,Q)\;+\; w_{e^\pm}(Q)\,\delta(1-x)\,,
\end{align}
where $f^{\text{cont}}$ is the smooth continuum part and 
$w_{e^\pm}(Q)$ is the no-emission (Sudakov) weight at scale $Q$, fixed by momentum and lepton number conservation
$\int_0^1\! f_{e^\pm}^{\text{cont}}(x,Q)\diff x + w_{e^\pm}(Q)=1$.\footnote{This expression is approximate, since closure under QED also requires the inclusion of positron contributions which are however extremely small in comparison. In the case of EW \acp{PDF} the neutrinos must also be included, see Refs.~\cite{Garosi:2023bvq,Han:2020uid,Han:2021kes}.}
At the level of the convolution, this endpoint structure modifies \cref{eq:pdf-binned}. For example in the case of $e^-e^-$ from $e^-e^-$ primary beams we obtain
\begin{figure}
    \centering
    \includegraphics[width=0.496\textwidth,trim={3cm 0.4cm 4cm 0.25cm}, clip]{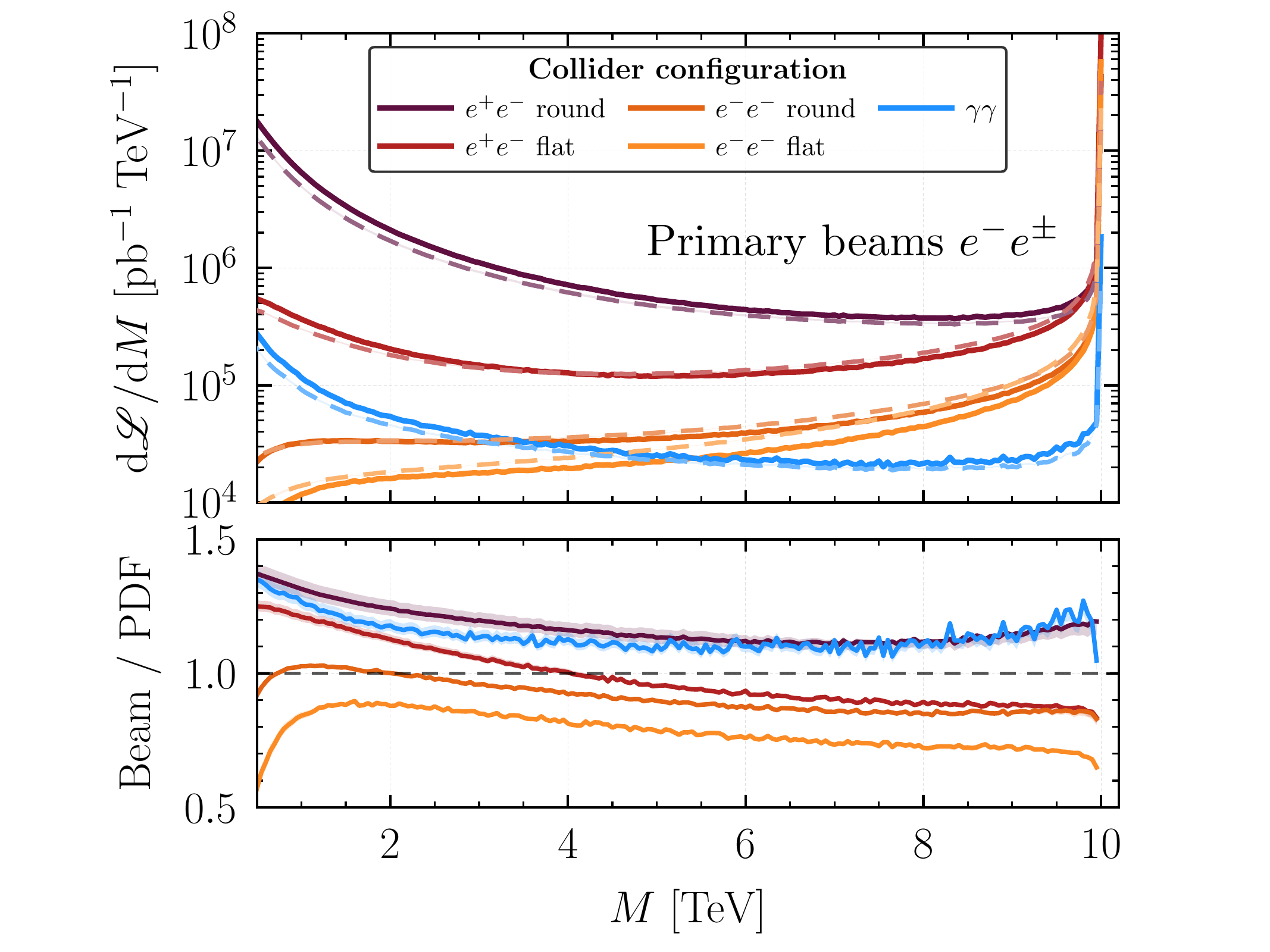} \hfill
    \includegraphics[width=0.496\textwidth,trim={3cm 0.4cm 4cm 0.25cm}, clip]{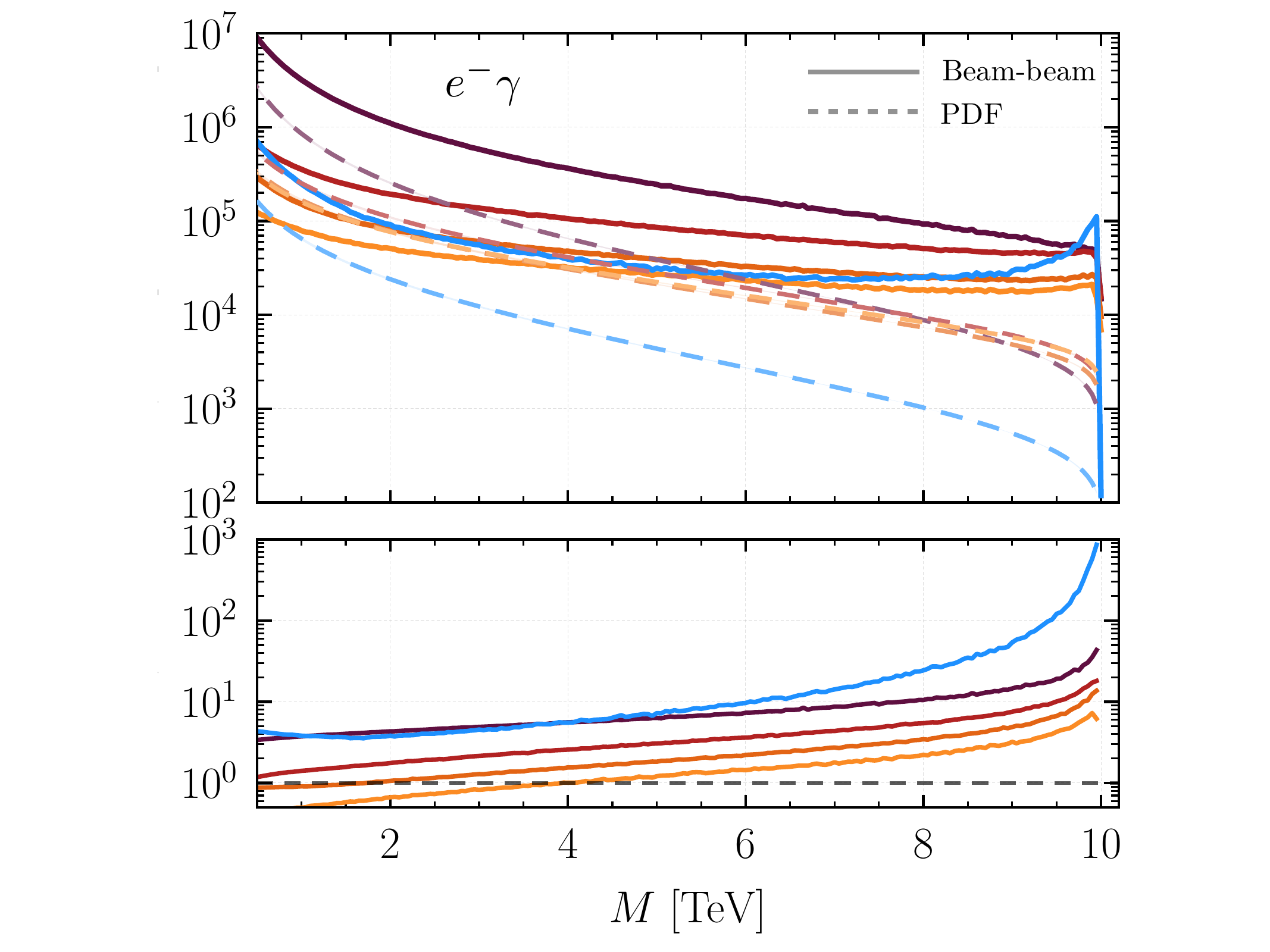} 
    \caption{Electroweak PDF contributions to the differential luminosity functions compared with the corresponding machine (beam–beam) spectra used in \cref{fig:initial_state_lumis}. 
    In each subplot, the lower panel shows ratio of the solid divided by the dashed curves of the corresponding panel, as well as an uncertainty band from varying the factorization scale $Q=\kappa M/2$ with $\kappa\in\{0.5,2\}$.
    Left: Primary lepton–lepton channels ($e^-e^-$ and $e^+e^-$). The dashed (solid) lines indicate the effective luminosity with (without) including the PDF corrections. The PDF curves include the $\delta\times\delta$ term of the lepton structure functions, which reduces the luminosity in some cases. The impact of the correction is at most $\mathcal{O}(1)$. Right: $e^-\gamma$ channel. 
    Here the dashed lines correspond to the PDF correction alone, while the solid lines show the machine (beam--beam) spectra without the inclusion of the PDFs.
    The true effective luminosity spectra in this case is therefore the sum of both curves, which would be the analogue of the dashed curve in the left-hand panel.
In this channel the PDF correction is negligible, except when $M < \SI{4}{\TeV}$ in $e^-e^-$ colliders.
     }
    \label{fig:pdf-effects} 
\end{figure}
\begin{align}\label{eq:pdf-discrete}
    \left.\frac{\diff \mathscr{L}_{\text{int},\,e^-e^-}^\text{PDF}}{\diff M}  \right|_{M_i} &=  A_i^{e^-e^-} \frac{w_{e^-}^2(M_i/2)}{\Delta M} +   \sum_j A_j^{e^-e^-} \frac{2 M_i}{M_j^2} \Bigg[2 f^\text{cont}_{e^-}(\tau_{ij}, M_i/2) w_{e^-}(M_i/2)\notag\\
    &\qquad \qquad +\int_{\tau_{ij}} ^1 \frac{\diff x}{x} f^\text{cont}_{e^-}(x,M_i/2)f^\text{cont}_{e^-}(\tau_{ij}/x,M_i/2)\Bigg]\,.
\end{align}
Note that as we are considering the \acp{PDF} for lepton-beams only, the photon contributions carry no $\delta(1-x)$ term and as such only the integral piece above would contribute to channels such as $e^\pm \gamma$ and $\gamma \gamma$.
In \cref{fig:pdf-effects} we compare electroweak PDF contributions to the machine (beam–beam) luminosity spectra as introduced in \cref{sec:wakefield}. For these numerical results we utilize the electroweak \acp{PDF} from Ref.~\cite{Garosi:2023bvq}. The left panel shows the primary lepton–lepton channels ($e^+e^-$ and $e^-e^-$), evaluated using \cref{eq:pdf-discrete}. 
For these channels the PDF curve contains a contribution proportional to $w_{e^\pm}^2$ that is exactly aligned with the original machine spectrum (the discrete $\delta\times\delta$ term, corresponding to events where neither beam radiates), plus additional terms from the continuum parts of the \acp{PDF}. 
The net impact at low $M$ can be a mild enhancement or depletion, depending on the level of beamstrahlung. For round $e^+e^-$ beams the shape is essentially unchanged but the normalization is reduced (a net decrease).
For the other configurations, the PDF piece tends to be larger near the endpoint, while the machine spectrum dominates or becomes relatively more important at lower $M$. 
Since these effects amount to order-one shifts in the overall normalization of the primary lepton–lepton luminosity, they are largely degenerate with modest changes in machine parameters (such as the electrons per bunch and bunch dimensions) and do not qualitatively alter our physics reach. For this reason we do not fold them into the main reach projections and instead consider them as a (controllable) theoretical systematic on the absolute rates.

The right panel of \cref{fig:pdf-effects} shows the $e^-\gamma$ channel. Here the PDF curves  for producing a photon contain no $\delta$ spike and must therefore be \emph{added} to the corresponding machine channels to obtain the total luminosity. In these cases the PDF contributions are negligible over most of the range and only become comparable at small $M$. For $e^-e^-$ flat beams---and to a lesser extent for round beams---parity is reached around $M \lesssim \SI{4}{\TeV}$, while above this the machine contribution clearly dominates. This conclusion also holds for the $\gamma\gamma$ channel (not shown). Neglecting electroweak \acp{PDF} therefore has two main consequences: (i) it introduces an $\mathcal{O}(1)$ change in the absolute normalization of the primary lepton–lepton luminosities, and (ii) it slightly underestimates the $e\gamma$ and $\gamma\gamma$ fluxes at low invariant mass. Since our reach estimates are driven by electroweak particle production near threshold and by relative comparisons between collider options, these effects are numerically subleading compared to the spread between different machine scenarios.

\section{Details on $WW$ + missing momentum\label{app:WWMET}}
\begin{figure}[h]
    \centering
    \includegraphics[width=\linewidth]{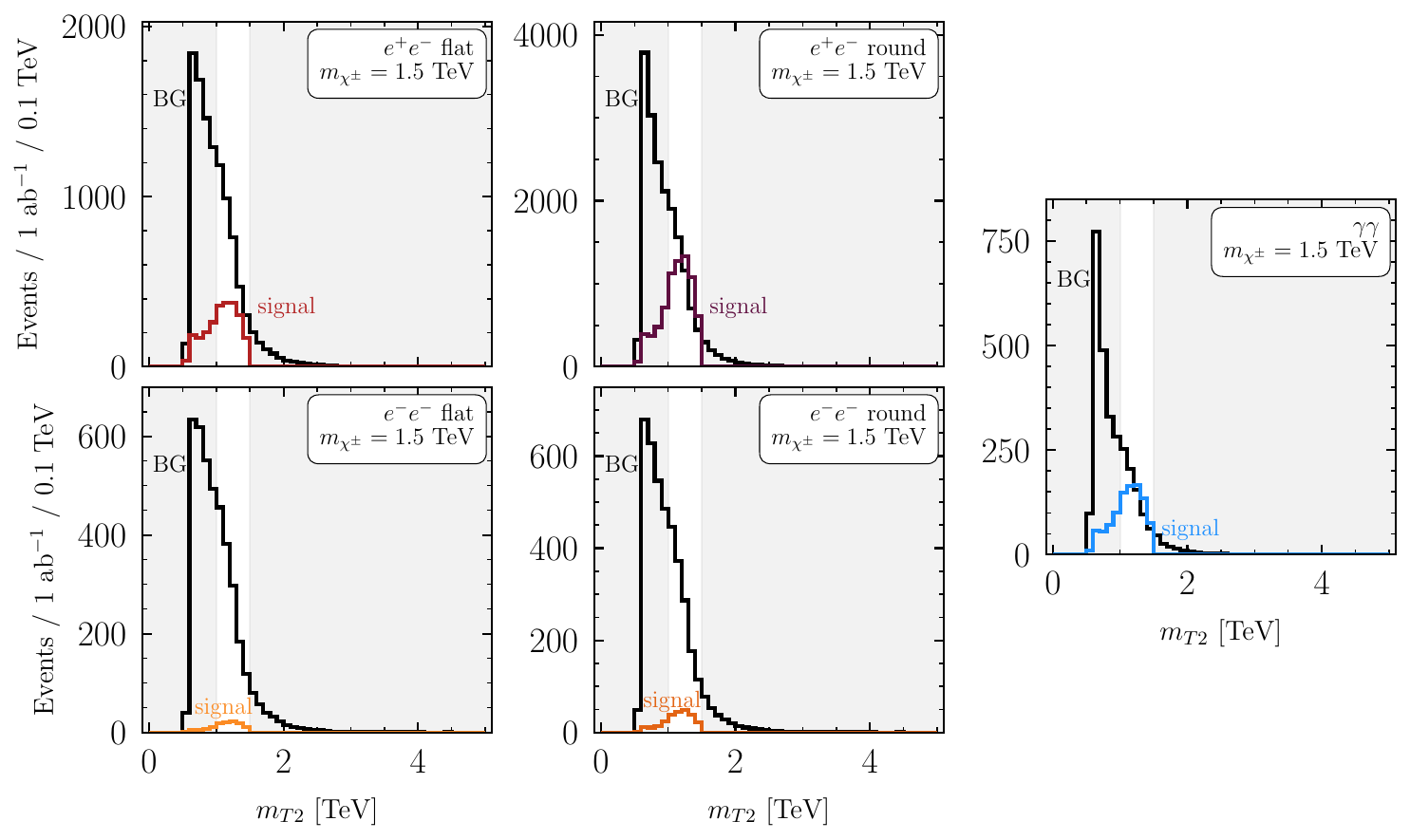}
    \caption{Distributions of the $m_{T2}$ variable for signal and background (BG) for the $m_{\chi^\pm}=\SI{1.5}{\TeV}$ benchmark. All cuts in \cref{tab:WWMETcuts} have been applied, except for the cut on $m_{T2}$ itself. The $m_{T2}$ cuts are indicated by the gray shading, where the signal region is the unshaded area.}
    \label{fig:mt2ep_1p5TeV}
\end{figure}

\begin{figure}[h]
    \centering
    \includegraphics[width=\linewidth]{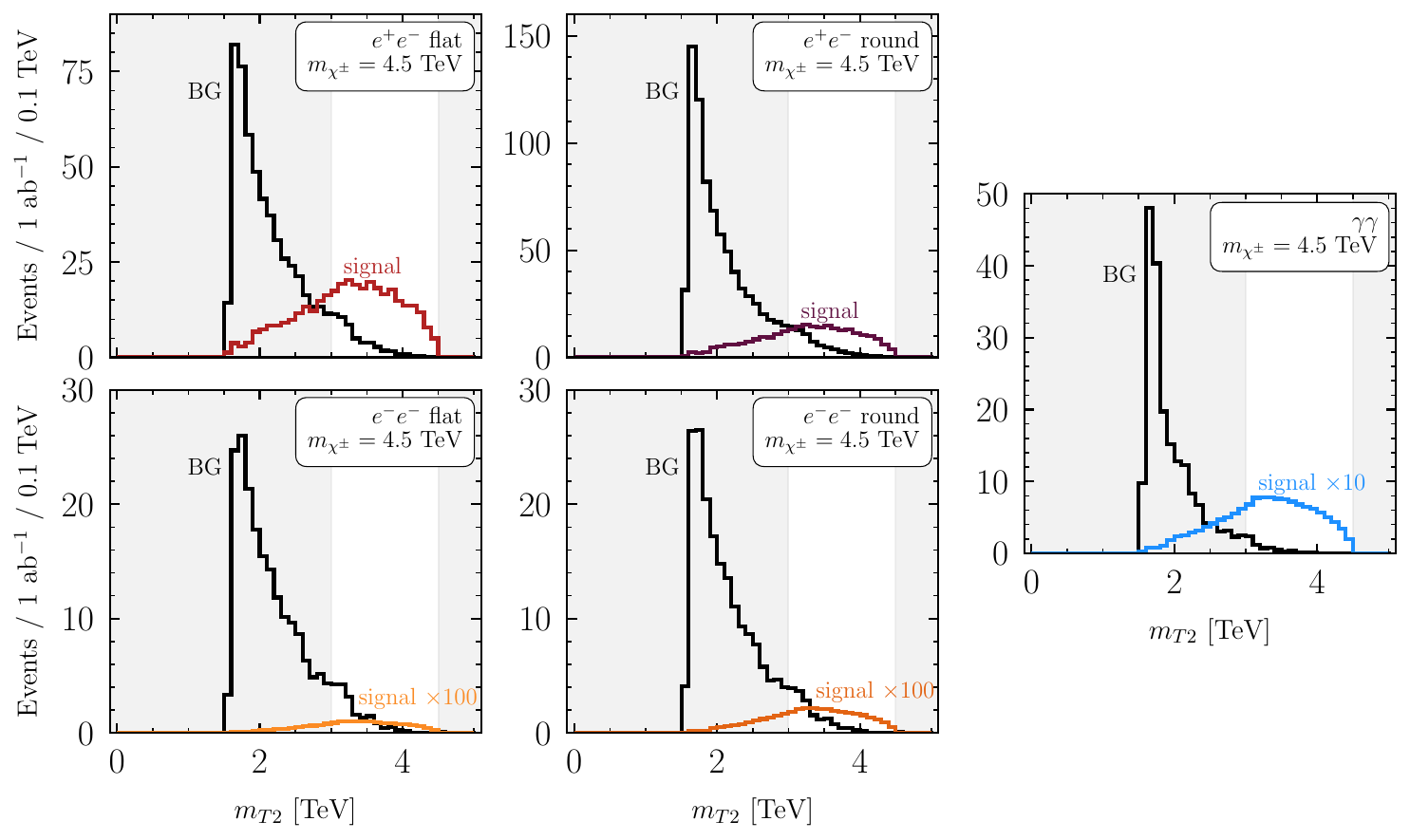}
    \caption{
    Same as in \cref{fig:mt2ep_1p5TeV}, but for $m_{\chi^\pm}=\SI{4.5}{\TeV}$.
    For clarity, the signal rate for both $e^-e^-$ colliders was multiplied $\times 100$. The signal rate for the $\gamma\gamma$ collider was multiplied $\times 10$.}
    \label{fig:mt2ep_4p5TeV}
\end{figure}

In this appendix we provide some additional, quantitative details for the $WW$ + missing momentum \cref{sec:WWmet}. \cref{fig:mt2ep_1p5TeV} and \cref{fig:mt2ep_4p5TeV} show the $M_{T2}$ distributions for the $m_{\chi^\pm}=\SI{1.5}{\TeV}$ and $m_{\chi^\pm}=\SI{4.5}{\TeV}$, in analogy with \cref{fig:mt2ep_3TeV}. \cref{tab:WWMETcutflow} shows the consecutive and total efficiencies of all the cuts we imposed in \cref{tab:WWMETcuts}. It also shows the total number of signal and background events after cuts, assuming a geometric luminosity of \SI{10}{\per\atto\barn}.


\renewcommand{\arraystretch}{0.9} 
\begin{table*}[p]\centering
\begin{tabular}{p{2.1cm} | C{1.1cm} | p{1.4cm}  p{1.4cm}  p{1.4cm}  p{1.4cm} p{1.4cm} | p{1.4cm}} \toprule
\multicolumn{7}{c}{$e^+e^-$ flat beams}\\\midrule
& $m_{\chi^\pm}$ & $\epsilon_{\text{MET}}$ & $\epsilon_{H_T}$ & $\epsilon_{m_\text{miss}}$ & $\epsilon_{m_{T2}}$ & $\epsilon_\text{tot}$ & $N_\text{events}$\\\midrule
 signal		& 4.5 & 0.907    &0.706      &1.0        &0.643     &0.412     & 2333.0 \\
 background	& 4.5 & 0.025    &0.047      &0.714      &0.092     &0.001     & 522.1 \\
 signal		& 3.0 & 0.841    &0.718      &1.0        &0.586     &0.354     & 5769.3 \\
 background	& 3.0 & 0.025    &0.102      &0.884      &0.193     &0.006     & 2927.4 \\
 signal		& 1.5 & 0.608    &0.911      &0.999        &0.649     &0.36     & 15872.3 \\
 background	& 1.5 & 0.025    &0.668      &0.972      &0.341     &0.082     & 37180.0 \\\midrule
\multicolumn{7}{c}{$e^+e^-$ round beams}\\\midrule
& $m_{\chi^\pm}$ & $\epsilon_{\text{MET}}$ & $\epsilon_{H_T}$ & $\epsilon_{m_\text{miss}}$ & $\epsilon_{m_{T2}}$ & $\epsilon_\text{tot}$ & $N_\text{events}$\\\midrule
 signal		& 4.5 & 0.906    &0.697      &1.0        &0.652     &0.412     & 1751.7 \\
 background	& 4.5 & 0.02    &0.035      &0.806      &0.083     &0.0     & 680.5 \\
 signal		& 3.0 & 0.834    &0.666      &1.0        &0.629     &0.349     & 8480.8 \\
 background	& 3.0 & 0.02    &0.082      &0.94      &0.166     &0.002     & 3709.5 \\
 signal		& 1.5 & 0.562    &0.872      &1.0        &0.728     &0.357     & 54240.4 \\
 background	& 1.5 & 0.02    &0.644      &0.988      &0.312     &0.03     & 57831.4 \\\midrule
\multicolumn{7}{c}{$e^-e^-$ flat beams}\\\midrule
& $m_{\chi^\pm}$ & $\epsilon_{\text{MET}}$ & $\epsilon_{H_T}$ & $\epsilon_{m_\text{miss}}$ & $\epsilon_{m_{T2}}$ & $\epsilon_\text{tot}$ & $N_\text{events}$\\\midrule
 signal		& 4.5 & 0.902    &0.662      &1.0        &0.69     &0.411     & 1.2 \\
 background	& 4.5 & 0.025    &0.038      &0.872      &0.092     &0.002     & 186.1 \\
 signal		& 3.0 & 0.827    &0.608      &1.0        &0.684     &0.344     & 54.3 \\
 background	& 3.0 & 0.025    &0.091      &0.973      &0.209     &0.01     & 1148.9 \\
 signal		& 1.5 & 0.548    &0.861      &1.0        &0.758     &0.358     & 887.8 \\
 background	& 1.5 & 0.025    &0.664      &0.991      &0.354     &0.126     & 14410.9 \\\midrule
\multicolumn{7}{c}{$e^-e^-$ round beams}\\\midrule
& $m_{\chi^\pm}$ & $\epsilon_{\text{MET}}$ & $\epsilon_{H_T}$ & $\epsilon_{m_\text{miss}}$ & $\epsilon_{m_{T2}}$ & $\epsilon_\text{tot}$ & $N_\text{events}$\\\midrule
 signal		& 4.5 & 0.902    &0.663      &1.0        &0.688     &0.411     & 2.5 \\
 background	& 4.5 & 0.025    &0.036      &0.882      &0.085     &0.001     & 166.7 \\
 signal		& 3.0 & 0.827    &0.608      &1.0        &0.684     &0.344     & 112.5 \\
 background	& 3.0 & 0.025    &0.089      &0.977      &0.196     &0.006     & 1047.4 \\
 signal		& 1.5 & 0.547    &0.859      &1.0        &0.761     &0.357     & 1921.2 \\
 background	& 1.5 & 0.025    &0.663      &0.992      &0.344     &0.08     & 13971.9 \\\midrule
\multicolumn{7}{c}{$\gamma\gamma$}\\\midrule
& $m_{\chi^\pm}$ & $\epsilon_{\text{MET}}$ & $\epsilon_{H_T}$ & $\epsilon_{m_\text{miss}}$ & $\epsilon_{m_{T2}}$ & $\epsilon_\text{tot}$ & $N_\text{events}$\\\midrule
 signal		& 4.5 & 0.903    &0.668      &1.0        &0.672     &0.405     & 92.2 \\
 background	& 4.5 & 0.044    &0.076      &0.614      &0.033     &0.0     & 66.3 \\
 signal		& 3.0 & 0.835    &0.663      &1.0        &0.626     &0.347     & 1076.4 \\
 background	& 3.0 & 0.044    &0.152      &0.843      &0.106     &0.002     & 574.5 \\
 signal		& 1.5 & 0.571    &0.883      &1.0        &0.701     &0.353     & 6899.9 \\
 background	& 1.5 & 0.044    &0.75      &0.905      &0.267     &0.021     & 7730.8 \\
\toprule  
\end{tabular}
\caption{Cut flow table for $WW$+MET search in $e^+e^-$, $e^-e^-$ and $\gamma\gamma$ colliders, with $m_{\chi^\pm}$ in units of TeV. The $\epsilon$'s are the efficiencies of the cuts in \cref{tab:WWMETcuts}, as executed in order from left to right. $\epsilon_\text{tot}$ is the efficiency of all cuts combined. For the number of events, $N_\text{events}$, a geometric luminosity of \SI{10}{\per\atto\barn} was assumed.\label{tab:WWMETcutflow}}
\end{table*}
\clearpage

\bibliographystyle{JHEP}
\bibliography{bibliography}
\end{document}